\documentclass[aps,preprint,nofootinbib]{revtex4}%
\usepackage{amsmath}
\usepackage{graphicx}
\usepackage{amsfonts}
\usepackage{amssymb}%
\providecommand{\U}[1]{\protect\rule{.1in}{.1in}}
\providecommand{\U}[1]{\protect\rule{.1in}{.1in}}
\providecommand{\U}[1]{\protect\rule{.1in}{.1in}}
\providecommand{\U}[1]{\protect\rule{.1in}{.1in}}
\providecommand{\U}[1]{\protect\rule{.1in}{.1in}}
\providecommand{\U}[1]{\protect\rule{.1in}{.1in}}
\providecommand{\U}[1]{\protect\rule{.1in}{.1in}}
\providecommand{\U}[1]{\protect\rule{.1in}{.1in}}

\begin{document}
\preprint{ }



\begin{center}

{\Large Complete Set of Homogeneous Isotropic Analytic Solutions }

{\Large in} {\Large Scalar-Tensor Cosmology with Radiation and Curvature }

{\vskip0.3cm}

\textbf{Itzhak Bars}$^{\ast}$\textbf{, Shih-Hung Chen}$^{\dagger}$,
\textbf{Paul J. Steinhardt}$^{\#}$\textbf{\ and Neil Turok}$^{\dagger}%
$\textbf{\ }

{\vskip0.2cm}

$^{\ast}$\textsl{Department of Physics, University of Southern
California,\ Los Angeles, CA 90089, USA}

$^{\dagger}$\textsl{Perimeter Institute for Theoretical Physics, Waterloo, ON
N2L 2Y5, Canada}

$^{\#}$\textsl{Department of Physics and Princeton Center for Theoretical
Physics}

\textsl{Princeton University, Princeton, NJ 08544, USA}

{\vskip0.0cm} \textbf{Abstract}
\end{center}

We study a model of a scalar field minimally coupled to gravity, with a
specific potential energy for the scalar field, and include curvature and
radiation as two additional parameters. Our goal is to obtain
 analytically the complete set of configurations of a
homogeneous and isotropic universe as a function of time.
This leads to a geodesically complete description of the universe, including
the passage through the cosmological singularities, at the classical level. We
give all the solutions  analytically without any restrictions on the
parameter space of the model or initial values of the fields. We find that
for generic solutions the universe goes through a singular (zero-size) bounce
by entering a period of antigravity at each big crunch and exiting
from it at the following big bang. This happens cyclically again and again
without violating the null energy condition. There is a special subset of
geodesically complete non-generic solutions which perform zero-size
bounces without ever entering the antigravity regime in all cycles. For
these, initial values of the fields are synchronized and quantized but the
parameters of the model are not restricted. There is also a subset of spatial
curvature-induced solutions that have finite-size bounces in the
gravity regime and never enter the antigravity phase. These exist only within
a small continuous domain of parameter space without fine tuning initial
conditions. To obtain these results, we identified 25 regions of a 6-parameter
space in which the complete set of analytic solutions are explicitly obtained.

{\small PACS: 98.80.-k, 98.80.Cq, 04.50.-h.}

{\small Keywords: big bang, inflation, cosmology, Weyl symmetry, gravity,
antigravity, 2T-physics.}

\newpage\newpage

\section{Introduction}

Scalar-tensor theory has been one of the most popular tools for building
models in cosmology. It is sufficiently simple while having a variety of
applications. One common application is to describe early universe inflation
\cite{inflation Guth,inflation Linde,inflation Steinhardt} where the scalar
plays a central role in driving a period of accelerated expansion that solves
the homogeneity (horizon) and the flatness problems, and generates the
primordial density perturbation that seeds the subsequent growth of the large
scale structure\cite{Mathiazhagan,extendedLa,hyperextendedAccetta,LindeBD}.
Alternatively, some ekpyrotic models for the early universe models
\cite{ekpyrotic1,ekpyrotic2} also utilize scalar-tensor theories to produce a
period of a slow contracting phase before a big crunch that eliminates the
horizon problem, and solves the flatness problem as well. With some specific
matching rules inspired by a colliding two branes picture, ekpyrotic models
can also generate scale invariant density perturbations as observed today.
Another application of scalar-tensor theories is to produce the late time
acceleration of the universe that is inferred from type IA supernova
observations \cite{SN}. Such models are called "quintessence" models where a
small nonzero dynamical vacuum value of a scalar potential replaces the
cosmological constant \cite{quint}. In addition, by using a conformal
transformation, it has been shown that modified gravity theories, such as
$f\left(  R\right)  $ gravity, are equivalent to scalar-tensor theory with a
specific scalar potential \cite{f(R)}.

Despite such broad applications of scalar-tensor theories, only a few isolated
examples of analytic solutions have been found so far. This is because the
coupled second order nonlinear differential equations are hard to solve
analytically.  Our goal in this paper is to provide a full set of analytic solutions that
give all possible configurations of a \textit{homogeneous} and
\textit{isotropic} universe as a function of time. This expands on our
previous work in \cite{inflationBC,cyclic BCT,cyclic IB} by including
additional degrees of freedom, in particular radiation. The effects of
anisotropy are discussed elsewhere \cite{BCSTletter,BCST}.

Our overall approach in this paper is in contrast to specific analytic,
approximate or numerical solutions that are usually fine tuned from the point
of view of initial conditions and/or the potential energy function $V\left(
\sigma\right)  $~for the scalar field, to force a solution in which the
universe has a particular desired behavior as motivated by prejudices and
observations. Instead, we would like to understand the global structure of
solution space that can emerge from a class of theories, so that we can gain a
better understanding of how the features of our own universe could emerge.
Obtaining the full set of classical solutions can provide some such insights.
Indeed through our solutions we gain new understanding about general generic
behavior as we will see below.

We can obtain the full set of analytic solutions of the scalar-tensor theory
for several forms of the potential energy function $V\left(  \sigma\right)
$~for the scalar field. In this paper we concentrate on the case:
\begin{equation}
V\left(  \sigma\right)  =\left(  \frac{6}{\kappa^{2}}\right)  ^{2}\left[
c\sinh^{4}\left(  \sqrt{\frac{\kappa^{2}}{6}}\sigma\right)  +b\cosh^{4}\left(
\sqrt{\frac{\kappa^{2}}{6}}\sigma\right)  \right]  , \label{V}%
\end{equation}
where the parameters $b$ and $c$ are dimensionless free parameters, and
$\kappa^{-1}$ is the reduced Planck mass $\kappa^{-1}=\sqrt{\frac{hc}{8\pi G}%
}=2.43\times10^{18}\frac{GeV}{c^{2}}.$ We note that this potential has
familiar features. For example, if $\left(  b+c\right)  >0,$ depending on the
various values and signs of $b,c$, $V(\sigma)$ has a single well or double
well with stable minima, similar to other potentials used in cosmological
applications. Because our analysis has a broad range of applications beyond
cosmology, we will classify all the solutions regardless of the physical
application. In other papers, including \cite{BCSTletter,BCST}, the role of
these solutions is discussed in a cosmological setting. Some of these cases
will be pointed out briefly later in the paper. In a separate paper we will
present the analytic solutions for the potentials $V\left(  \sigma\right)
=\frac{6^{2}}{\kappa^{4}}be^{2p\kappa\sigma/\sqrt{6}}$ for arbitrary $p$ and
$V\left(  \sigma\right)  =\frac{6^{2}}{\kappa^{4}}\left(  be^{-2\kappa
\sigma/\sqrt{6}}+ce^{-4\kappa\sigma/\sqrt{6}}\right)  $, where $b,c,p$ are
dimensionless real parameters.

For the potential \ref{V}, we can solve the Friedmann equations exactly for all time
intervals before or after the big bang. The method, which is based on
conformal symmetry, was introduced in previous papers \cite{inflationBC}%
-\cite{BCST}. It was applied to the cases of the flat and curved isotropic
Friedmann universes without radiation or matter \cite{cyclic BCT}. It was also
applied to the case of an anisotropic universe in the absence of the potential
energy, but with the inclusion of radiation \cite{BCST}. In this paper, we
further generalize our method for the isotropic universe to include both
curvature and radiation with the potential, where radiation is taken in the
form of a perfect fluid. The inclusion of radiation is a simple mathematical
exercise beyond our previous paper \cite{cyclic BCT}, but it describes richer
physics, and leads to more complicated analytic expressions for the solutions.
So, the reader may wish to first understand the previous work in a simpler
setting \cite{cyclic BCT}.

Including radiation, the model is defined by 4 parameters, namely $\left(
b,c\right)  $ in the potential, the curvature parameter $K$ in the metric,
and  $\rho_{r},$ the energy density of radiation when the scale factor is
$a=1.$ The two fields of interest are the time dependent scale factor
$a\left(  \tau\right)  $ and the scalar field $\sigma\left(  \tau\right)  .$
Their initial conditions $\left(  a\left(  \tau_{0}\right)  ,\dot{a}\left(
\tau_{0}\right)  ,\sigma\left(  \tau_{0}\right)  ,\dot{\sigma}\left(  \tau
_{0}\right)  \right)  $ introduce 4 more parameters which enter the general solution
to second order. However, the zero energy condition including
gravity (or Friedmann equation) eliminates one of the initial values, and one
other initial value can be absorbed into a redefinition of the initial time
$\tau_{0}$ by using the time translation symmetry of the differential
equations. Hence the complete set of solutions are described by 4+2=6
parameters given by $\left(  b,c,K,\rho_{r},E,a\left(  \tau_{0}\right)
\right)  ,$where a single energy parameter $E$ is used conveniently instead of
the two related initial velocities $\dot{a}\left(  \tau_{0}\right)
,\dot{\sigma}\left(  \tau_{0}\right)  $. We will give all the analytic
solutions without putting any conditions on these 6 parameters at any
$\tau_{0}$. It turns out that there are 25 distinct regions of this parameter
space in which the solutions take different analytic forms in terms of Jacobi
functions. The 25 regions and the corresponding solutions are given explicitly
in the Appendix. If an unstable potential with $(b+c)<0$ is of interest there
would be more regions and corresponding solutions; with appropriate
modifications these can be obtained from those available in the Appendix.

As in our previous work \cite{cyclic BCT,cyclic IB}, we find that in the
Einstein frame the generic solutions are geodesically incomplete at the
cosmological singularity (see Eqs. (3) and (4) in \cite{cyclic BCT}). We will
then find that the knowledge of the full set of solutions suggests how to
complete the space and make it geodesically complete for the generic solution.
This completion includes time intervals during which the gravitational field
effectively acts like a repulsive force (antigravity) rather than an
attractive force (gravity). So, the generic solution has alternating time
intervals of gravity and antigravity as graphically illustrated in
Fig.~\ref{figGeneric}. In this figure, each time the trajectory crosses the
45$^{o}$ lines, the universe transits from gravity to antigravity or
vice-versa. There is however a subset of non-generic solutions that are
geodesically complete in the Einstein frame without ever entering the
antigravity region. These are of two types: (i) singular zero-size bounces at
the cosmic singularity without violating the null energy condition as shown in
Fig.~\ref{CCfig2} and (ii) non-singular finite-size bounces as shown in
Figs.(23,24) in \cite{cyclic BCT}.

The zero-size bounces (i), which are classified in tables I, IIb and III, with
the related analytic expressions in the Appendix, are obtained by
synchronizing and/or quantizing some initial values. These tables provide the
most general parameter subspace (within the 6-parameter set) for which the
geodesically complete singular bounce occurs without antigravity. The
parameter subset consists of 4 continuous and one quantized parameter, which
is obviously smaller than the available continuous 6-parameter set. Despite
the fact that this subset of solutions may be considered a set of measure zero
as compared to the full set, it is distinguished as the only zero-size bounce
set of solutions that are geodesically complete in the Einstein frame, and do
not enter the antigravity region at any time in any cycle.

The finite-size-bounces (ii), which are classified in table IIa, with the
related analytic expressions in the Appendix, describe a universe that
contracts up to a minimum non-zero size, at which point the spatial curvature
causes the universe to bounce back into an expanding phase. This kind of
spatial curvature-induced bounce is already familiar in cosmology. Here we
provide analytic solutions for the finite-size bounces. As the universe turns
around to repeat such cycles, the minimum size is not necessarily the same in
each cycle, as this depends on the parameters. Such solutions occur when the
parameters satisfy, $\rho_{r}<K^{2}/16b$ and $\phi_{\text{min}}^{2}\left(
\tau_{0}\right)  >K/4b>s_{\text{max}}^{2}\left(  \tau_{0}\right)  $ (see Table
IIa). Note that there are still 6 parameters, so this is a continuous set, but
it is a restricted region of parameter space or initial values.

The solutions above -- the generic case, type (i) or type (ii) bounces -- are
the exact and complete set of solutions in the absence of anisotropy. Although
anisotropy can be neglected as the universe expands, it can grow to be a
dominant effect near the singularity. We do not consider those cases here;
they are described in \cite{BCSTletter,BCST} where it is proven that there is
an attractor mechanism that is independent of initial conditions. The
attractor distorts the zero-bounce solutions, both the generic or non-generic
type (i), to behave in a unique way such that these solutions must undergo a
big crunch/big bang transition by contracting to zero size, passing through a
brief antigravity phase, shrinking to zero size again, and re-emerging into an
expanding normal gravity phase.

This paper is organized as follows. In Sec.~\ref{model}, we introduce the
standard scalar-tensor theory, with a single minimally coupled scalar field
$\sigma\left(  x\right)  ,$ as the gauge fixed version of a locally scale
(Weyl) invariant reformulation of Einstein's theory of gravity that contains
two conformally coupled scalar fields $\phi\left(  x\right)  ,s\left(
x\right)  $. This \textquotedblleft Weyl lifted\textquotedblright\ version has
an extra scalar field as well as a local Weyl symmetry that compensate each
other, so that the physical degrees of freedom are the same number as in the
standard formulation of the theory. This model and method of solution emerged
directly from the 2T-physics formulation of gravity
\cite{2TPhaseSpace,2Tgravity,2TgravityGeometry}. The model was also inspired
in the context of braneworld notions \cite{HoravaWitten,RandallSundrum} that
led to the colliding brane scenario for cosmology
\cite{ekpyrotic1,ekpyrotic2,KOSST,branesMT}. Recently 'tHooft also motivated
the same Weyl invariant theory because of its ability to give a better
description of black and white holes in a convenient gauge \cite{tHooft}. Such
gauge choices, including the $E,\gamma,c$ and $s$ gauges discussed in this
paper, are just a small subset of examples of 3+1 dimensional \textquotedblleft
shadows\textquotedblright\ that 2T-physics yields as dual forms of the same
parent theory that unifies them in 4+2 dimensions.

The Weyl lifted version has no dimensionful constants, not even the
gravitational constant. The extra scalar field can be eliminated by gauge
fixing it to a dimensionful constant, thus reaching what we call the
\textquotedblleft Einstein gauge\textquotedblright\ ($E$-gauge), the standard
formulation of the theory in the Einstein frame with the usual gravitational
constant and the scalar field $\sigma$. The advantage of the Weyl lifted
version is that it allows us to choose another more convenient gauge that we
call the $\gamma$-gauge, in which the cosmological equations greatly simplify
and can be solved analytically. The full set of solutions is then mapped to
the Einstein frame by a gauge transformation from the $\gamma$-gauge to the
$E$-gauge, and verified that they are the solutions of the Friedmann
equations. In this process we only add gauge degrees of freedom to Einstein's
theory. But in the presence of the gauge degrees of freedom we find naturally
the geodesically complete space and understand much more clearly the nature of
the complete space of solutions. In particular we learn that geodesic
completeness requires an extension of the domain of the original fields in the
scalar-tensor theory in the Einstein frame, such that with this extension, the
same fields can describe also an antigravity region not captured at first sight.

In Sec.~\text{\ref{solve}} we show how the complete set of solutions of the
Friedmann equations, including curvature and radiation, are obtained
analytically without constraining the 6-parameter space. The complete set of
solutions is explicitly given in the Appendix, where in 25 different regions
of the parameter space the analytic expressions take different forms. These
solutions are all cyclic and geodesically complete in an enlarged domain as
described above.

The non-generic solutions of type (i) with zero-size bounces, and type (ii)
with finite-size bounces, which never enter the antigravity regime, are still
geodesically complete in the gravity domain. In Sec.~\ref{geoComplete} we
determine the constraints on parameter space and initial conditions that
distinguish the geodesically complete solutions in the restricted Einstein
frame. The corresponding parameter spaces are classified in Tables I, IIa,b
and III.

In Sec.~\text{\ref{antig}} we comment on the generic solutions that are
geodesically complete provided an anti-gravity regime is included.

In Sec.~\text{\ref{discussion}} we summarize our results. In an Appendix
$\left(  \ref{analytic}\right)  $ we list all the analytic cosmological
solutions predicted by our model introduced in Sec.~\text{\ref{model}}.

\section{The Weyl lifted Model}

\label{model}

Our approach begins with the standard action typically used in cosmological
models that describe a scalar field $\sigma\left(  x^{\mu}\right)  $ minimally
coupled to gravity
\begin{equation}
S_{gravity}=\int d^{4}x\sqrt{-g}\left\{
\begin{array}
[c]{c}%
\frac{1}{2\kappa^{2}}R\left(  g_{E}\right)  -\frac{1}{2}g_{E}^{\mu\nu}%
\partial_{\mu}\sigma\partial_{\nu}\sigma-V\left(  \sigma\right) \\
+radiation+matter
\end{array}
\right\}.  \label{action gravity}%
\end{equation}
We are able to solve for the complete set of homogeneous, isotropic,
cosmological solutions of this model when the potential $V\left(
\sigma\right)  $ is given as in Eq.~(\ref{V}). Through these solutions we
discover the geodesic completion of the space through the cosmological singularity.

To solve the equations and to understand the geodesic completion we will use a
device that we call ``Weyl extension'' in which the model is enlarged by
adding gauge degrees of freedom that are compensated with a local scaling
(Weyl) symmetry. The local scaling symmetry does not allow the usual
Einstein-Hilbert term $R\left(  g_{E}\right)  /2\kappa^{2}$, but allows
conformally coupled scalars. The following action, which will be shown to be
related to the one above by a gauge choice, contains two conformally coupled
scalars, $\phi,s,$ interacting with the curvature term with the coefficient
$\frac{1}{12}$ dictated by the gauge symmetry
\begin{equation}
S=\int d^{4}x\sqrt{-g}\left(  \frac{1}{2}g^{\mu\nu}\partial_{\mu}\phi
\partial_{\nu}\phi-\frac{1}{2}g^{\mu\nu}\partial_{\mu}s\partial_{\nu}%
s+\frac{1}{12}\left(  \phi^{2}-s^{2}\right)  R\left(  g\right)  -\phi
^{4}f\left(  \frac{s}{\phi}\right)  \right)  . \label{conformal action}%
\end{equation}
The function $f(z)$ is determined by the scalar field potential (see below)
and, in general, can be an arbitrary function of the gauge invariant ratio
$z\equiv\frac{s}{\phi}$ and still maintain the Weyl symmetry. The metric
$g_{\mu\nu}$ can differ from the metric $g_{\mu\nu}^{E}$ in
Eq.~(\ref{action gravity}) by an arbitrary conformal factor because of the
gauge symmetry under the following local transformation%
\[
\phi\rightarrow\Omega\phi,\;s\rightarrow\Omega s,\;g_{\mu\nu}\rightarrow
\Omega^{-2}g_{\mu\nu},
\]
where $\Omega\left(  x\right)  $ is an arbitrary function of spacetime.

This action was described as the \textit{conformal shadow} of the 2T-Gravity
action \cite{2Tgravity,2TgravityGeometry} in 4+2 dimensions. In this setting,
it was shown that the local scaling gauge symmetry is a remnant of general
coordinate transformations in the extra 1+1 dimensions.

We draw attention to the fact that $\phi$ has the wrong sign kinetic term
while $s$ has the correct sign. It appears as if $\phi$ is a ghost; however,
since the ghost disappears for some choices of gauge (e.g. the gauge that
restores the Einstein gravity form of the theory), there is no real problem
with ghosts or unitarity. In this connection, note also that there is no
gravitational constant; rather, the factor $\frac{1}{12}\left(  \phi^{2}%
-s^{2}\right)  $ effectively behaves like a gravitational parameter which
replaces the usual expression $\left(  16\pi G\right)  ^{-1}=\left(
2\kappa^{2}\right)  ^{-1}$ where $G$ is the Newton constant. If $\phi$ had the
opposite sign kinetic term, then this factor would become purely negative
$\frac{1}{12}\left(  -\phi^{2}-s^{2}\right)  $ in order to maintain the Weyl
symmetry, but then the gravitational parameter would have the wrong sign. This
is the reason why $\phi$ must be introduced initially with the
\textquotedblleft wrong\textquotedblright\ sign, so that the gravitational
parameter $\frac{1}{12}\left(  \phi^{2}-s^{2}\right)  $ is positive at least
in some regions of field space.

Fermionic and gauge fields, as well as more conformally coupled scalars, can
be added, to construct a completely Weyl invariant model, such as the Standard
Model of particle physics. The Higgs field must also couple as a conformal
scalar to preserve the Weyl symmetry. The Higgs mass term is not allowed by
the gauge symmetry, but it can be generated by coupling the Higgs doublet $H$
to the singlets $\phi,s$ in Weyl invariant quartic terms of the form
$H^{\dagger}H\left(  \alpha\phi^{2}+\beta s^{2}\right)  $. There are various
possible models for the effective Higgs mass by choosing the parameters
$\alpha,\beta$. One possibility is related to the fact that $\left(  \phi
_{E}^{2}-s_{E}^{2}\right)  $ gets fixed to a constant in the Einstein gauge of
Eq.~(\ref{Egauge}). Another possibility emerges from our solutions, by noting
that cosmologically $s_{E}=$ $\frac{\sqrt{6}}{\kappa}\sinh\left(  \frac
{\kappa\sigma\left(  \tau\right)  }{\sqrt{6}}\right)  $ evolves to a field
$s_{E}\sim\sigma\left(  \tau\right)  $ which is much smaller than the Planck
scale, and at late times behaves almost like a constant that can mimic the
Higgs mass term. This might also explain the mass hierarchy \cite{2Tsm,higgsCosmo}.

We discuss here four gauge choices that are useful for finding solutions and
interpreting them: Einstein gauge ($E$-gauge), the $\gamma$-gauge, the
supergravity gauge ($c$-gauge) and the string gauge ($s$-gauge). To
distinguish the fields in each gauge we denote them by the subscripts
$E,\gamma,c,s$ respectively. We also define the following Weyl gauge invariant
quantity $\chi$ which plays an important role in our discussion%
\begin{equation}
\chi\equiv\frac{\kappa^{2}}{6}(-g)^{\frac{1}{4}}(\phi^{2}-s^{2}).
\end{equation}
Another gauge invariant is the ratio $s/\phi.$

\noindent\textit{$E$-gauge:} The usual Einstein gravity in
Eq.~(\ref{action gravity}) is obtained in the $E$-gauge, in which we denote
the fields by $\phi_{E}\left(  x\right)  ,s_{E}\left(  x\right)  ,g_{E}%
^{\mu\nu}\left(  x\right)  $ with a subscript $E.$ In this gauge the
gravitational parameter is gauge fixed to the usual Newton constant for all
spacetime $x^{\mu}$%
\begin{equation}
\frac{1}{12}\left(  \phi_{E}^{2}\left(  x\right)  -s_{E}^{2}\left(  x\right)
\right)  =\left(  2\kappa^{2}\right)  ^{-1}.
\end{equation}
Then parameterizing this gauge with a single scalar field $\sigma$,
\begin{equation}
\phi_{E}\left(  x\right)  =\pm\frac{\sqrt{6}}{\kappa}\cosh\left(  \frac
{\kappa\sigma\left(  x\right)  }{\sqrt{6}}\right)  ,\;s_{E}\left(  x\right)
=\pm\frac{\sqrt{6}}{\kappa}\sinh\left(  \frac{\kappa\sigma\left(  x\right)
}{\sqrt{6}}\right)  , \label{Egauge}%
\end{equation}
we find that the conformally gauge invariant action (\ref{conformal action})
yields the familiar action in Eq.~(\ref{action gravity}) with an
Einstein-Hilbert term $\frac{1}{2\kappa^{2}}R\left(  g_{E}\right)  $ and a
minimally coupled scalar field $\sigma,$ just as in Eq.~(\ref{action gravity}%
). The gauge invariant $\chi$ in the $E$-gauge becomes%
\begin{equation}
\chi=(-g_{E})^{\frac{1}{4}}.
\end{equation}
In a cosmological solution with the metric
\begin{equation}
ds_{E}^{2}=a_{E}^{2}\left(  \tau\right)  \left(  -d\tau^{2}+ds_{3}^{2}\right)
\end{equation}
where the 3-dimensional metric $ds_{3}^{2}$ will be discussed later,
$(-g_{E})^{\frac{1}{4}}$ is just the scale factor of the universe, $\chi
=a_{E}^{2}\left(  \tau\right)  ,$ so $\chi$ must vanish at the cosmological
singularity at the time of the big bang $a_{E}\left(  \tau_{B}\right)  =0$.
This shows that the big bang corresponds to $\chi\left(  \tau_{B}\right)  =0,$
and since $\chi$ is gauge invariant, the cosmological singularity is at
$\chi\left(  \tau_{B}\right)  =0$ in all gauges.

Note that the geometry completely collapses at the singularity at time
$\tau_{B}$ in the $E$-gauge since $(-g_{E}\left(  \tau_{B}\right)  )=0$, and
this is why geodesics are incomplete. However, we will see that the geometry
does not collapse at all at $\chi\left(  \tau_{B}\right)  =0$ in the other
gauges, and this is how we are able to complete the geodesics and the geometry.

\noindent\textit{$\gamma$-gauge:} In the $\gamma$-gauge we choose
$(-g_{\gamma})^{\frac{1}{4}}$=constant for all $\tau$. Since the cosmological
FRW metric in the $E$-gauge is conformally flat (even when the curvature is
non-zero), $ds_{E}^{2}=a_{E}^{2}\left(  \tau\right)  \left(  -e^{2}d\tau
^{2}-ds_{3FRW}^{2}\right)  $, we can at first discuss the \textit{general
conformally flat metric} which can always be put into the form $ds^{2}%
=a^{2}\left(  x^{\mu}\right)  \eta_{\mu\nu},$ where $\eta_{\mu\nu}$ is the
Minkowski metric. A conformally flat metric becomes fully flat in the $\gamma
$-gauge $ds^{2}=\eta_{\mu\nu}$, namely the gauge choice $a_{\gamma}\left(
x\right)  =1,$ leading to $R\left(  g_{\gamma}\right)  =0.$ Hence the degrees
of freedom in the $\gamma$-gauge are $\phi_{\gamma},s_{\gamma}$ in a flat
Minkowski geometry. The gauge invariant $\chi$ now takes the form%
\begin{equation}
\chi\left(  x^{\mu}\right)  =\frac{\kappa^{2}}{6}(\phi_{\gamma}^{2}-s_{\gamma
}^{2})\left(  x^{\mu}\right)  , \label{chi}%
\end{equation}
and the gauge invariant action in (\ref{conformal action}), taken with any
conformally flat metric, reduces to a gauge fixed action in flat Minkowski
space%
\begin{equation}
S_{\gamma}=\int d^{4}x\left[  \frac{1}{2}\left(  (\partial_{\mu}\phi_{\gamma
})^{2}-(\partial_{\mu}s_{\gamma})^{2}\right)  -\phi_{\gamma}^{4}f\left(
\frac{s_{\gamma}}{\phi_{\gamma}}\right)  \right]  ,
\end{equation}
where $\phi_{\gamma}\left(  x\right)  $ is a ghost since it has the wrong-sign
kinetic energy. However, this ghost is removed as follows: before choosing the
$\gamma$-gauge we recall that there was an Einstein equation for the metric,
$G_{\mu\nu}\left(  g\right)  =T_{\mu\nu},$ which must be imposed for any
metric in the $\gamma$-gauge as well. For the conformally flat metric in the
present case, the curvature vanishes in the $\gamma$-gauge, $R_{\mu\nu
\lambda\sigma}\left(  \eta\right)  =0,$ leading to $G_{\mu\nu}\left(
\eta\right)  =0,$ and therefore $T_{\mu\nu}\left(  \phi_{\gamma},s_{\gamma
}\right)  =0.$ The vanishing of the stress tensor for the action $S_{\gamma}$
is a constraint that removes the ghost $\phi_{\gamma}.$

To include the effect of both anisotropy and curvature, the 3-dimensional part
of the FRW metric, $\left(  ds_{3}^{2}\right)  _{FRW}$, must be replaced by
the corresponding 3-dimensional parts of the Kasner metric, Bianchi IX metric,
Bianchi VIII metric, when $K=0,$ $K>0,~K<0$ respectively. We parameterize the
Kasner metric $\left(  K=0\right)  $ as follows
\[
\left(  ds_{3}^{2}\right)  _{Kasner}=e^{-2\sqrt{2/3}\kappa\alpha_{1}}\left(
dz\right)  ^{2}+e^{\sqrt{2/3}\kappa\alpha_{1}}\left(  e^{\sqrt{2}\kappa
\alpha_{2}}\left(  dx\right)  ^{2}+e^{-\sqrt{2}\kappa\alpha_{2}}\left(
dy\right)  ^{2}\right)  ,
\]
where $\alpha_{1}\left(  x^{\mu}\right)  $ and $\alpha_{2}\left(  x^{\mu
}\right)  $ are the anisotropy fields that are taken to depend only on time in
the homogeneous case considered in this paper. The Bianchi IX and VIII metrics
contain the same $\alpha_{1,2}\left(  x^{\mu}\right)  ,$ as well as the
non-zero curvature parameter $K$ that is included by generalizing $\left(
dx,dy,dz\right)  $ to $\left(  d\sigma_{x},d\sigma_{y},d\sigma_{z}\right)  $.
Both of these Bianchi metrics reduce to the Kasner metric when the curvature
$K$ vanishes.

In the homogeneous limit, with an anisotropic cosmological metric of the form
given above, $ds^{2}=a^{2}\left(  \tau\right)  \left(  -e^{2}d\tau^{2}%
-ds_{3}^{2}\right)  ,$ the Weyl invariant action reduces to the following
effective action for the homogeneous cosmological degrees of freedom%
\begin{equation}
S_{\text{eff}}=\int d\tau\left\{
\begin{array}
[c]{c}%
-\frac{1}{2e}\left(  \partial_{\tau}\left(  a\phi\right)  \right)  ^{2}%
+\frac{1}{2e}\left(  \partial_{\tau}\left(  as\right)  \right)  ^{2}%
+\frac{\kappa^{2}}{12e}\left(  \phi^{2}-s^{2}\right)  a^{2}\left(  \dot
{\alpha}_{1}^{2}+\dot{\alpha}_{2}^{2}\right)  \\
-e\left[  a^{4}\phi^{4}f\left(  \frac{s}{\phi}\right)  +\rho_{r}-\frac{1}%
{2}\left(  \phi^{2}-s^{2}\right)  a^{2}v\left(  \alpha_{1},\alpha_{2}\right)
\right]  ,
\end{array}
\right\}  \label{SeffConf}%
\end{equation}
where $\rho_r$ is the radiation density when the scale factor $a=1$.
The effective action is invariant under time dependent Weyl transformations
\begin{equation}
a\left(  \tau\right)  \rightarrow\Omega^{-1}\left(  \tau\right)  a\left(
\tau\right)  ,\;\left(  \phi\left(  \tau\right)  ,s\left(  \tau\right)
\right)  \rightarrow\Omega\left(  \tau\right)  \left(  \phi\left(
\tau\right)  ,s\left(  \tau\right)  \right)
\end{equation}
while $\alpha_{1,2}$ and $e$ are Weyl invariant. Here $v\left(  \alpha
_{1},\alpha_{2}\right)  $ is the anisotropy potential which emerges from the
curvature term $\left(  \phi^{2}-s^{2}\right)  R\left(  g\right)  $
\begin{equation}
v\left(  \alpha_{1},\alpha_{2}\right)  =\frac{K}{1-4~\text{sign}\left(
K\right)  }\left[
\begin{array}
[c]{c}%
e^{-4\sqrt{2/3}\kappa\alpha_{1}}+4e^{2\sqrt{2/3}\kappa\alpha_{1}}\sinh
^{2}\left(  \sqrt{2}\kappa\alpha_{2}\right)  \\
-4~\text{sign}\left(  K\right)  ~e^{-\sqrt{2/3}\kappa\alpha_{1}}\cosh\left(
\sqrt{2}\kappa\alpha_{2}\right)
\end{array}
\right]  .\label{Va}%
\end{equation}
In the isotropic limit the anisotropy potential reduces to a constant
$\lim_{\alpha_{1,2}\rightarrow0}v\left(  \alpha_{1},\alpha_{2}\right)  =K.$
For the Kasner metric the potential energy term is absent even if anisotropy
is present since $K=0.$

For the cosmological solutions discussed in this paper we will concentrate on
the isotropic case $\alpha_{1,2}\rightarrow0,$ and therefore $v\left(
\alpha_{1},\alpha_{2}\right)  \rightarrow K.$ The generalization of our
discussion to homogeneous and anisotropic universes is given in
\cite{BCSTletter} and \cite{BCST}.

In this action $e\left(  \tau\right)  $ is related to the lapse function which
is a part of the metric $g_{\mu\nu}\left(  x\right)  .$ Its presence insures
$\tau$-reparameterization symmetry. The equation of motion with respect to $e$
imposes a constraint on the degrees of freedom. This constraint is equivalent
to the $G_{00}=T_{00}$ Einstein equation. One may choose a gauge for $\tau,$
such that $e\left(  \tau\right)  =1,$ in which case the action simplifies.

In the Einstein gauge of Eq.(\ref{Egauge}) the cosmological action is
\begin{equation}
S_{\text{eff}}^{E}=\int d\tau\left\{
\begin{array}
[c]{c}%
\frac{1}{e}\left[  -\frac{6}{2\kappa^{2}}\dot{a}_{E}^{2}+\frac{1}{2}a_{E}%
^{2}\dot{\sigma}^{2}+\frac{1}{2}a_{E}^{2}\dot{\alpha}_{1}^{2}+\frac{1}{2}%
a_{E}^{2}\dot{\alpha}_{2}^{2}\right] \\
-e\left[  a_{E}^{4}V\left(  \sigma\right)  +\rho_{r}-\frac{6}{2\kappa^{2}%
}a_{E}^{2}v\left(  \alpha_{1},\alpha_{2}\right)  \right]
\end{array}
\right\}  . \label{ssuper1}%
\end{equation}
In this action note that the $a_{E}\left(  \tau\right)  $ degree of freedom
has the wrong sign kinetic energy term, and therefore it is potentially a
ghost. However, the Hamiltonian derived from this action in the $e\left(
\tau\right)  =1$ gauge is required to vanish as a result of the constraint,
and this is just sufficient to compensate for the ghost. This constraint
equation, which is called the zero energy condition, is the same as the first
Friedmann equation.

The $\gamma$-gauge is defined by fixing
\begin{equation}
a_{\gamma}\left(  \tau\right)  =1
\end{equation}
for all $\tau.$ In the $\gamma$-gauge, the cosmological action is
\begin{equation}
S_{\text{eff}}^{\gamma}=\int d\tau\left\{
\begin{array}
[c]{c}%
\frac{1}{e}\left[  -\frac{1}{2}\dot{\phi}_{\gamma}^{2}+\frac{1}{2}\dot
{s}_{\gamma}^{2}+\frac{\kappa^{2}}{12}\left(  \phi_{\gamma}^{2}-s_{\gamma}%
^{2}\right)  \left(  \dot{\alpha}_{1}^{2}+\dot{\alpha}_{2}^{2}\right)  \right]
\\
-e\left[  \phi_{\gamma}^{4}f\left(  s_{\gamma}/\phi_{\gamma}\right)  +\rho
_{r}-\frac{1}{2}\left(  \phi_{\gamma}^{2}-s_{\gamma}^{2}\right)  v\left(
\alpha_{1},\alpha_{2}\right)  \right]
\end{array}
\right\}  . \label{SeffG}%
\end{equation}
When $V\left(  \sigma\right)  $ is given as in Eq.~(\ref{V}), then
$\phi_{\gamma}^{4}f\left(  s_{\gamma}/\phi_{\gamma}\right)  =b\phi_{\gamma
}^{4}+cs_{\gamma}^{4}$ is purely quartic. In this action the $\phi_{\gamma}$
degree of freedom has the wrong sign kinetic energy term. However, as in the
case of $a_{E}$ in the previous paragraph, due to the $\tau$%
-reparameterization symmetry, and the corresponding zero energy constraint
that follows from the equation for $e$, this potential ghost is eliminated.

We can relate the $E\,$-gauge and $\gamma$-gauge degrees of freedom to each
other by a gauge transformation. More easily, we identify the gauge invariants
$\chi$ and $s/\phi$ in both gauges, and from them we extract
\begin{equation}
\text{ }a_{E}^{2}={\frac{\kappa^{2}}{6}}\left\vert \phi_{\gamma}^{2}%
-s_{\gamma}^{2}\right\vert ,\;\sigma=\frac{\sqrt{6}}{2\kappa}\ln\left(
\left\vert \frac{\phi_{\gamma}+s_{\gamma}}{\phi_{\gamma}-s_{\gamma}%
}\right\vert \right)  . \label{link1}%
\end{equation}
From this, we see that, what appeared as a cosmological singularity in the
$E$-gauge at $a_{E}^{2}\left(  \tau_{B}\right)  =0$ does not show up at all as
a geometrical singularity in the $\gamma$-gauge since $R_{\mu\nu\lambda\sigma
}\left(  \eta\right)  =0$. Of course, the big bang of the $E$-gauge must
reflect itself again as the gauge invariant $\chi\left(  \tau_{B}\right)  =0,$
which becomes $(\phi_{\gamma}^{2}-s_{\gamma}^{2})\left(  \tau_{B}\right)
\rightarrow0,$ however this is not a singularity of the equations within the
$\gamma$-gauge.

The absence of a geometrical singularity in the $\gamma$-gauge, and the
simplicity of the equations of motions for $\phi_{\gamma},s_{\gamma}$ is the
key for being able to solve all the cosmological equations and finding the
complete set of solutions analytically. This is also what permits us to
understand geodesic completeness. The geodesically complete geometry includes
space-time regions of \textit{antigravity}. In the antigravity regions the
gravitational parameter $\frac{1}{12}\left(  \phi^{2}-s^{2}\right)  $ that
appears in the general action becomes negative. The switching of the sign in
generic solutions of the theory occurs precisely at $\chi\left(  \tau
_{B}\right)  =0,$ which appears as the cosmological singularity in the
$E$-gauge, but this is a completely smooth point for the geometry in the
$\gamma$-gauge, as well as all other gauges, except the $E$-gauge.

\noindent\textit{Supergravity gauge ($c$-gauge):} In the $c$-gauge
\cite{2Tgravity}, we fix $\phi_{c}=\phi_{0}$ where $\phi_{0}$ is a constant
for all $x^{\mu}$. Then in (\ref{conformal action}) or (\ref{SeffConf}) there
is only one scalar $s_{c}\left(  x\right)  $ while the curvature term takes
the form $\frac{1}{12}\left(  \phi_{0}^{2}-s_{c}^{2}\left(  x\right)  \right)
R\left(  g_{c}\left(  x\right)  \right)  .$ We see that the $\phi_{0}^{2}$
term plays the role of the Hilbert-Einstein term while the overall structure
$\frac{1}{12}\left(  \phi_{0}^{2}-s_{c}^{2}\right)  R\left(  g_{c}\right)  $
is similar to that found in supergravity, including the K\"{a}hler potential.
In fact, this model can be regarded as a toy model for the scalar sector of a
full supergravity model. Then the term $\left(  -s_{c}^{2}R\left(
g_{c}\right)  \right)  $ allows us to identify $s_{c}^{2}\left(  x\right)  $
as the analog of the K\"{a}hler potential in supergravity. The gauge invariant
becomes $\chi\equiv\frac{\kappa^{2}}{6}(-g_{c})^{\frac{1}{4}}(\phi_{0}%
^{2}-s_{c}^{2}\left(  x\right)  ),$ and for a cosmological solution it takes
the form%
\begin{equation}
\chi=\frac{\kappa^{2}}{6}(\phi_{0}^{2}-s_{c}^{2}\left(  \tau\right)
)a_{c}^{2}\left(  \tau\right)  .
\end{equation}
Since we have all the solutions, we can verify that the cosmological
singularity $\chi\left(  \tau_{B}\right)  =0,$ which in the Einstein frame is
at $a_{E}\left(  \tau_{B}\right)  =0,$ occurs when $s_{c}^{2}\left(  \tau
_{B}\right)  =\phi_{0}^{2},$ rather than $a_{c}^{2}\left(  \tau_{B}\right)
=0.$ Hence the metric $g_{c}^{\mu\nu}$ in this gauge is not singular at the
Big Bang. This is because the quantity $s_{c}/\phi_{0}=s_{\gamma}/\phi
_{\gamma}=s_{E}/\phi_{E}$ is gauge invariant and takes the value $\left(
s/\phi\right)  \left(  \tau_{B}\right)  \rightarrow1$ at the singularity in
any gauge. (The reason that the $E$-gauge $\frac{1}{12}\left(  \phi_{E}%
^{2}\left(  x\right)  -s_{E}^{2}\left(  x\right)  \right)  =\left(
2\kappa^{2}\right)  ^{-1}$ can remain finite even at the singularity
$\chi\left(  \tau_{B}\right)  =0,$ is because in the expression $\left(
\phi_{E}^{2}-s_{E}^{2}\right)  \left(  \tau_{B}\right)  =\phi_{E}^{2}\left(
\tau_{B}\right)  \left(  1-s_{E}^{2}/\phi_{E}^{2}\right)  \left(  \tau
_{B}\right)  ,$ when the second factor vanishes, the factor $\phi_{E}%
^{2}\left(  \tau_{B}\right)  $ blows up so that $\left(  \phi_{E}^{2}%
-s_{E}^{2}\right)  $ remains unchanged at all $x^{\mu}$, including at the
singularity at time $\tau_{B}$.) This shows that, similar to $\frac{1}%
{12}\left(  \phi_{0}^{2}-s_{c}^{2}\right)  R\left(  g_{c}\right)  ,$ the
effective gravitational constant in the curvature term in the action of a
supergravity theory is typically expected to switch sign at the big crunch or
big bang. So the phenomena we discuss here, including antigravity regimes
related to geodesic completeness, are expected generically in typical
supergravity theories.

\noindent\textit{String gauge (}$s$\textit{-gauge)}{:} We also discuss the
string gauge ($s$-gauge) to connect to the typical structures in string
theory. The string frame in $d$ dimensions is defined by the following form of
Lagrangian
\begin{equation}
L_{\text{string gauge}}=\frac{1}{2\kappa^{2}}e^{-\sqrt{\frac{d-2}{2}}\Phi
}\left(  R\left(  g_{s}\right)  +\frac{d-2}{2}g_{s}^{\mu\nu}\partial_{\mu}%
\Phi\partial_{\nu}\Phi-V_{s}\left(  \Phi\right)  )\right)  , \label{stringL}%
\end{equation}
Note the wrong sign and unusual normalization of the kinetic term for the
\textquotedblleft dilaton\textquotedblright\ $\Phi$ . When the transformation
from the string-frame to the E-frame is performed by the substitution
\begin{equation}
\left(  g_{s}\right)  _{\mu\nu}=e^{\sqrt{\frac{2}{d-2}}\Phi}\left(
g_{E}\right)  _{\mu\nu},\text{ }\Phi=\sqrt{2\kappa^{2}}\sigma, \label{sToE}%
\end{equation}
then the right sign and normalization of the Einstein-frame $\sigma$ field
emerges%
\begin{equation}
L_{\text{Einstein gauge}}=\left(  \frac{1}{2\kappa^{2}}R\left(  g_{E}\right)
-\frac{1}{2}g_{E}^{\mu\nu}\partial_{\mu}\sigma\partial_{\nu}\sigma-V\left(
\sigma\right)  \right)  .
\end{equation}
The string gauge in Eq.(\ref{stringL}), for $d=4,$ is obtained from our Weyl
invariant action (\ref{conformal action}) by choosing the following gauge in
which $\phi_{s},s_{s}$ are expressed in terms of a single scalar $\Phi$ (this
is analogous to $\phi_{E},s_{E}$ given in Eq.(\ref{Egauge}))
\begin{align*}
\phi_{s}\left(  x\right)   &  =\frac{\sqrt{6}}{\kappa}e^{-\frac{1}{2}%
\Phi\left(  x\right)  }\cosh\left(  \sqrt{\frac{1}{12}}\Phi\left(  x\right)
\right)  ,\\
s_{s}\left(  x\right)   &  =\frac{\sqrt{6}}{\kappa}e^{-\frac{1}{2}\Phi\left(
x\right)  }\sinh\left(  \sqrt{\frac{1}{12}}\Phi\left(  x\right)  \right)  ,
\end{align*}
while the metric $g_{s}^{\mu\nu}$ is labelled with the letter $s$ to
distinguish it from the metric in another Weyl gauge. This shows how the
dilaton field $\Phi$ of the non-Weyl invariant string theory can be related to
the fields $\phi,s$ in the Weyl invariant theory.

Then the effective cosmological action in the string gauge takes the form of
Eq.(\ref{SeffConf}) where the gauge fixed form for $\left(  \phi_{s}%
,s_{s}\right)  $ is inserted. The remaining degrees of freedom in the string
gauge are then $\left(  a_{s},\sigma\right)  ,$where $\sigma$ is the same
field as the one in the Einstein gauge except for the overall normalization in
Eq.(\ref{sToE}), while $a_{s}$ is related to $a_{E}$ simply by the
transformation in Eq.(\ref{sToE}) for $d=4$
\begin{equation}
a_{s}^{2}=e^{\sqrt{2\kappa^{2}}\sigma}a_{E}^{2},\;\Phi=\sqrt{2\kappa^{2}%
}\sigma.
\end{equation}
Therefore, by using the relation between the $E\,$-gauge and the $\gamma
$-gauge in (\ref{link1}), we can relate the $s$-gauge degrees of freedom,
$a_{s},\Phi$, to the $\gamma$-gauge degrees of freedom as follows%
\begin{align}
a_{s}^{2} &  =\frac{\kappa^{2}}{6}\left\vert \frac{\phi_{\gamma}+s_{\gamma}%
}{\phi_{\gamma}-s_{\gamma}}\right\vert ^{\sqrt{3}}\left(  \phi_{\gamma}%
^{2}-s_{\gamma}^{2}\right)  \\
\Phi &  =\sqrt{3}\ln\left\vert \frac{\phi_{\gamma}+s_{\gamma}}{\phi_{\gamma
}-s_{\gamma}}\right\vert \nonumber
\end{align}
The expressions given above are consistent with the gauge invariants
$\chi,\left(  s/\phi\right)  ,a\phi,as,$ as expressed in the $s,\gamma
,E,c$-gauges, such as $a_{s}\phi_{s}=a_{E}\phi_{E}=a_{c}\phi_{c}=a_{\gamma
}\phi_{\gamma},$ etc.

In solving the cosmological equations various gauge choices turn out to be
useful. In particular, for the cases we have solved, the $\gamma$-gauge turned
out to be the most useful. By using the relations between gauges displayed
above, the solutions for $\phi\left(  \tau\right)  ,s_{\gamma}\left(
\tau\right)  $ imply the solutions of the degrees of freedom in the other
gauge choices.

\section{Solving the equations with radiation and curvature}

\label{solve}

The crucial step introduced in \cite{inflationBC}, as suggested by 2T-physics,
is to take advantage of the gauge symmetry of the conformally invariant action
(\ref{conformal action}). The Einstein equations derived from the action
(\ref{conformal action}), taken with only time dependent isotropic fields,
yields gauge covariant cosmological equations for three fields $a\left(
\tau\right)  ,\phi\left(  \tau\right)  ,s\left(  \tau\right)  $ in any gauge.
If one chooses the Einstein gauge given in Eq.~(\ref{Egauge}) then the
corresponding gauge fixed fields $a_{E}\left(  \tau\right)  ,\sigma\left(
\tau\right)  $ satisfy the Friedmann equations including radiation $\rho_{r},$
as given below in Eqs.(\ref{friedmann}-\ref{sigma}). Instead, if one chooses
the $\gamma$-gauge defined by $a_{\gamma}\left(  x\right)  =1,$ the remaining
fields $\phi_{\gamma}\left(  \tau\right)  ,s_{\gamma}\left(  \tau\right)  $
turn out to satisfy the decoupled equations (\ref{phi}-\ref{energy constraint}%
) that can be solved exactly. Then, by performing a gauge transformation that
relates the two gauge fixed configurations we obtain the full set of solutions
of the Friedmann equations.

\begin{figure}
[th]
\begin{center}
\includegraphics[
height=2.2969in,
width=2.7994in
]%
{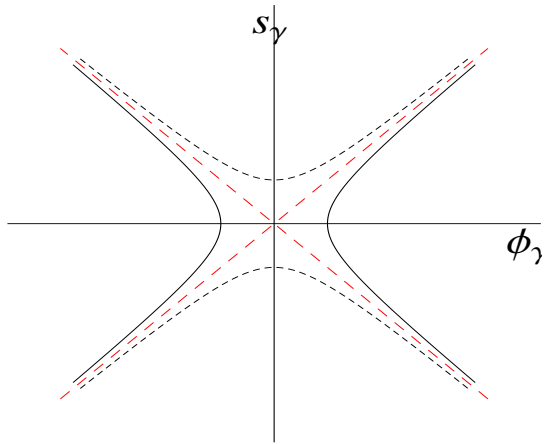}%
\caption{The gravity regime ($\phi_{\gamma}^{2}>s_{\gamma}^{2}$) is in left and right quadrants and the
antigravity regime ($\phi_{\gamma}^{2}<s_{\gamma}^{2}$) is in top and bottom quadrants.}%
\label{CCrFig1}%
\end{center}
\end{figure}

The field re-definition from $\sigma,a_{E}$ to $\phi_{\gamma},s_{\gamma},$ is
derived from the gauge transformation that connects the $E$ and $\gamma$
gauges \cite{inflationBC} and is given in Eq.~(\ref{link1}). Note that
generically $\chi\left(  x^{\mu}\right)  $ can be positive or negative. But
the metric in the Einstein gauge, in either the gravity or antigravity
patches, has always the correct signature metric$,$ hence, the absolute value
in the relation for $a_{E}^{2}$ in Eq.~(\ref{link1}). The inverse of the
transformation (\ref{link1}) involves four quadrants in the $\left(
\phi_{\gamma},s_{\gamma}\right)  $ space depicted in Fig.~\ref{CCrFig1}, and
is given by
\begin{equation}
\phi_{\gamma}=\pm\left\{
\begin{array}
[c]{c}%
\frac{\sqrt{6}}{\kappa}\sqrt{\left\vert \chi\right\vert }\cosh\left(
\frac{\kappa\sigma}{\sqrt{6}}\right)  ,\text{ if }\chi>0\\
\frac{\sqrt{6}}{\kappa}\sqrt{\left\vert \chi\right\vert }\sinh\left(
\frac{\kappa\sigma}{\sqrt{6}}\right)  ,\text{ if }\chi<0
\end{array}
\right.  ,\;s_{\gamma}=\pm\left\{
\begin{array}
[c]{c}%
\frac{\sqrt{6}}{\kappa}\sqrt{\left\vert \chi\right\vert }\sinh\left(
\frac{\kappa\sigma}{\sqrt{6}}\right)  ,\text{ if }\chi>0\\
\frac{\sqrt{6}}{\kappa}\sqrt{\left\vert \chi\right\vert }\cosh\left(
\frac{\kappa\sigma}{\sqrt{6}}\right)  ,\text{ if }\chi<0
\end{array}
\right.  . \label{link2}%
\end{equation}
As indicated in Fig.~\ref{CCrFig1}, in the gravity sector (left/right
quadrants) $\chi\left(  x^{\mu}\right)  $ is positive, while it is negative in
the antigravity sector (top/bottom quadrants).%

The transformations (\ref{link1}) and (\ref{link2}) can be used by starting
directly from the original action (\ref{action gravity}) without ever
mentioning the Weyl symmetry or the Einstein gauge. Furthermore, it can be
used for any spacetime dependence of the fields $\sigma\left(  x\right)
,a_{E}\left(  x\right)  $, $\phi_{\gamma}\left(  x\right)  ,s_{\gamma}\left(
x\right)  $ to rewrite all equations in terms of $\phi_{\gamma}\left(
x\right)  ,s_{\gamma}\left(  x\right)  $ rather than $\sigma\left(  x\right)
,a_{E}\left(  x\right)  .$ If used in that sense then it can be considered to
be an analog of what the Kruskal-Szekeres coordinates are to the Schwarzchild
coordinates in the description of the black hole spacetime. Namely, they
demonstrate that the spacetime regions across the horizon are smoothly
geodesically connected. The analog here is that the field space regions
$\left(  \phi_{\gamma}^{2}-s_{\gamma}^{2}\right)  >0$ and $\left(
\phi_{\gamma}^{2}-s_{\gamma}^{2}\right)  <$ $0$ are naturally connected at
$\phi_{\gamma}^{2}=s_{\gamma}^{2}$; hence, the patches of spacetime $x^{\mu}$
in which each inequality is satisfied are connected geodesically, as seen in
our purely time dependent solutions. Note that, in the $\phi_{\gamma}\left(
\tau\right)  ,s_{\gamma}\left(  \tau\right)  $ form, the Friedmann equations
(\ref{phi}-\ref{energy constraint}) do not display any singularities. Further
thought suggests that the connection $\phi_{\gamma}^{2}\left(  x^{\mu}\right)
=s_{\gamma}^{2}\left(  x\right)  $ (which corresponds to the dashed 45 degree
lines in field space in Fig.~\ref{CCrFig1}) can occur only at spacetimes
$x^{\mu}$ at which the curvature in the Einstein frame diverges $R\left(
g_{E}\left(  x\right)  \right)  =\infty$.

Inserting the above field redefinition into the standard Friedmann equations
and equation of motion for $\sigma$ in the isotropic limit (derived from the
action in (\ref{action gravity}) or (\ref{ssuper1}) ),
\begin{gather}
\frac{\dot{a}_{E}^{2}}{a_{E}^{4}}=\frac{\kappa^{2}}{3}\left[  \frac
{\dot{\sigma}^{2}}{2a_{E}^{2}}+V\left(  \sigma\right)  +\frac{\rho_{r}}%
{a_{E}^{4}}\right]  -\frac{K}{a_{E}^{2}}\label{friedmann}\\
\frac{\ddot{a}_{E}}{a_{E}^{3}}-\frac{\dot{a}_{E}^{2}}{a_{E}^{4}}=-\frac
{\kappa^{2}}{3}\left[  \frac{\dot{\sigma}^{2}}{a_{E}^{2}}-V\left(
\sigma\right)  +\frac{\rho_{r}}{a_{E}^{4}}\right]  ,\label{f2}\\
\frac{\ddot{\sigma}}{a_{E}^{2}}+2\frac{\dot{a}_{E}\dot{\sigma}}{a_{E}^{3}%
}+V^{\prime}\left(  \sigma\right)  =0, \label{sigma}%
\end{gather}
where dot denotes derivative with respect to the conformal time $\tau$, we
obtain
\begin{align}
0  &  =\ddot{\phi}_{\gamma}-4b\phi_{\gamma}^{3}+K\phi_{\gamma},\label{phi}\\
0  &  =\ddot{s}_{\gamma}+4cs_{\gamma}^{3}+Ks_{\gamma},\label{s}\\
0  &  =-\left(  \frac{1}{2}\dot{\phi}_{\gamma}^{2}-b\phi_{\gamma}^{4}+\frac
{1}{2}K\phi_{\gamma}^{2}\right)  +\left(  \frac{1}{2}\dot{s}_{\gamma}%
^{2}+cs_{\gamma}^{4}+\frac{1}{2}Ks_{\gamma}^{2}\right)  +\rho_{r}.
\label{energy constraint}%
\end{align}
These are precisely the equations of motion derived from the isotropic limit
of the action (\ref{conformal action}) in the $a_{\gamma}=1$ gauge
\cite{inflationBC}. The fields $\phi_{\gamma},s_{\gamma}$ are decoupled except
for the zero energy condition in Eq.~(\ref{energy constraint}). These
equations are valid not only when $\chi>0$ but also when $\chi<0,$ hence, they
provide a smooth continuation from the gravity sector to the antigravity
sector. The question is whether this continuation is required by the dynamics
as part of the evolution of the universe in a geodesically complete geometry.
We demonstrate that the generic geodesically complete solutions do require
this continuation.

Integrating the first two equations, and using the energy constraint, leads to
the following decoupled first order differential equations%
\begin{equation}
\frac{1}{2}\dot{\phi}_{\gamma}^{2}-b\phi_{\gamma}^{4}+\frac{K}{2}\phi_{\gamma
}^{2}=E_{\phi},\;~~\frac{1}{2}\dot{s}_{\gamma}^{2}+cs_{\gamma}^{4}+\frac{K}%
{2}s_{\gamma}^{2}=E_{s}, \label{Erelation}%
\end{equation}
with a relation between the two integration constants $\left(  E_{\phi}%
,E_{s}\right)  $ that reduces the unknowns to the single energy parameter $E$
\begin{equation}
E_{s}\equiv E,\;E_{\phi}=E+\rho_{r}. \label{E}%
\end{equation}

Equations (\ref{Erelation}) are analogous to the equations satisfied by two
non-relativistic particles with \textquotedblleft position\textquotedblright%
\ coordinates, $\phi_{\gamma}$ and $s_{\gamma},$ moving independently from
each other as controlled by the potentials $V\left(  \phi_{\gamma}\right)
=-b\phi_{\gamma}^{4}+\frac{K}{2}\phi_{\gamma}^{2}$ and $V\left(  s_{\gamma
}\right)  =cs_{\gamma}^{4}+\frac{K}{2}s_{\gamma}^{2},$ at energy levels
$E_{\phi}=E+\rho_{r}$ and $E_{s}=E$ respectively. So, the nature of the
solution and the corresponding physics is easily obtained through this
analogy. It is sufficient to draw a picture of the potentials $V\left(
\phi_{\gamma}\right)  ,V\left(  s_{\gamma}\right)  $ and indicate the energy
levels $E_{\phi}=E+\rho_{r}$ and $E_{s}=E$ on these pictures, and then let
each particle begin its motion at some arbitrary points. The reader is invited
to examine the figures in the Appendix to follow our arguments below.

We can choose to begin the motion at $\tau_{0}$ with initial values
$\phi_{\gamma}\left(  \tau_{0}\right)  ,s\left(  \tau_{0}\right)  $ that
insure the gauge invariant factor $\left(  1-s^{2}/\phi^{2}\right)  $ is
positive in all gauges at the initial time $\tau_{0}$ . Due to the time
translation symmetry one of these initial values may be fixed once and for all
without losing generality. For example, we may begin the motion somewhere on
the horizontal line in Fig.~\ref{CCrFig1} which is in the gravity sector in
any gauge
\begin{equation}
\left(  1-s^{2}\left(  \tau_{0}\right)  /\phi^{2}\left(  \tau_{0}\right)
\right)  =\left(  1-s_{\gamma}^{2}\left(  \tau_{0}\right)  /\phi_{\gamma}%
^{2}\left(  \tau_{0}\right)  \right)  =1>0.
\end{equation}
This motion begins in the right or left quadrants and can be described
initially by choosing the Einstein gauge (\ref{Egauge}). The ensuing motion
gives the time dependence of the generic solution $\phi_{\gamma}\left(
\tau\right)  $ and $s_{\gamma}\left(  \tau\right)  .$ The solution is
controlled by 6 parameters, namely the four parameters that define the model
$\left(  b,c,K,\rho_{r}\right)  ,$ one initial value energy parameter $E$ and
one initial value $\phi_{\gamma}\left(  \tau_{0}\right)  .$ We need to analyze
various regions of the 6-parameter space because the motion can be
qualitatively different in different ranges of the parameters. This is easily
seen by staring at the pictures of the potentials in the Appendix (see
\cite{cyclic BCT} for the discussion).

The generic motion is oscillatory with each particle moving independently with
independent oscillation periods. Each particle may pass through zero at
various times independently from each other. Hence $\left(  \phi_{\gamma}%
^{2}\left(  \tau\right)  -s_{\gamma}^{2}\left(  \tau\right)  \right)  $ keeps
changing sign in the $\gamma$-gauge for the generic solution. This shows that
generically the \textit{gauge invariant factor} $\left(  1-s_{\gamma}%
^{2}\left(  \tau\right)  /\phi_{\gamma}^{2}\left(  \tau\right)  \right)  ,$
which is the same in every gauge $\left(  1-s_{\gamma}^{2}\left(  \tau\right)
/\phi_{\gamma}^{2}\left(  \tau\right)  \right)  =\left(  1-s^{2}\left(
\tau\right)  /\phi^{2}\left(  \tau\right)  \right)  $, changes sign
periodically at times $\tau=\tau_{n}$ in \textit{every gauge}, where $\tau
_{n}$ is defined by the zeros of the gauge invariant factor computed in the
$\gamma$-gauge $\left(  1-s_{\gamma}^{2}\left(  \tau_{n}\right)  /\phi
_{\gamma}^{2}\left(  \tau_{n}\right)  \right)  =0.$ At precisely these times
the scale factor in the Einstein gauge vanishes as seen from Eq.~(\ref{link1}%
), $a_{E}^{2}\left(  \tau_{n}\right)  =\frac{\kappa^{2}}{6}\left(
\phi_{\gamma}^{2}\left(  \tau_{n}\right)  -s_{\gamma}^{2}\left(  \tau
_{n}\right)  \right)  =0,$ and hence, this is when there is a big bang or a
big crunch. In the Einstein gauge the generic motion must be terminated
artificially at these instants of time since $a_{E}^{2}$ is positive by
definition. However, in the $\phi_{\gamma},s_{\gamma}$ space the motion
continues smoothly to the antigravity regime where $\phi^{2}\left(
\tau\right)  <s^{2}\left(  \tau\right)  ,$ which shows that the Einstein frame
is geodesically incomplete for the generic motion. There exists a special
subset of solutions that is geodesically complete in the Einstein frame
without wandering into the antigravity sector, but for now let us continue
with the generic motion.

\begin{figure}
[th]
\begin{center}
\includegraphics[
height=2.2969in,
width=2.7994in
]%
{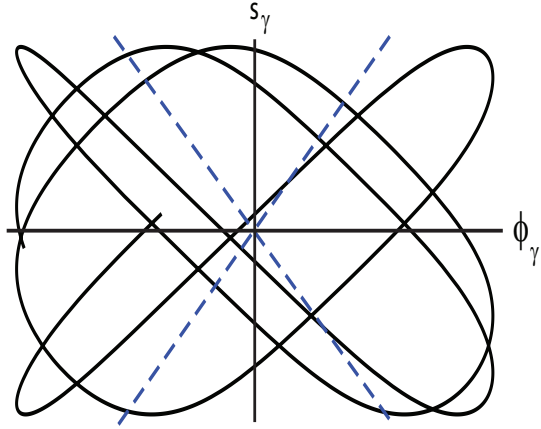}%
\caption{The generic isotropic solution crosses back and forth
through  the gravity (left and right quadrants) and antigravity
 (top and bottom) quadrants.}%
\label{figGeneric}%
\end{center}
\end{figure}

We have seen that even though the motion began in the gravity sector in
Fig.~\ref{CCrFig1} the particle reaches some point on the \textquotedblleft
light cone\textquotedblright\ in $\left(  \phi,s\right)  $ space where
$\phi^{2}\left(  \tau_{1}\right)  =s^{2}\left(  \tau_{1}\right)  $ at
$\tau=\tau_{1}$ in any gauge; it then moves smoothly into the antigravity
region where $\phi^{2}\left(  \tau\right)  <s^{2}\left(  \tau\right)  $ for
some period of time $\tau_{1}<\tau<\tau_{2}$. It then turns around and passes
through some other point on the lightcone at $\phi^{2}\left(  \tau_{2}\right)
=s^{2}\left(  \tau_{2}\right)  $ at $\tau=\tau_{2}$, thus reaching the gravity
region again. The generic motion continues periodically in this way,
oscillating back and forth between the gravity and the antigravity regions.
This generic motion is represented by a curve in the $\left(  \phi_{\gamma
},s_{\gamma}\right)  $ plane. Since we have the analytic solutions, we can
construct the curve explicitly as a parametric plot $\phi_{\gamma}\left(
\tau\right)  ,s_{\gamma}\left(  \tau\right)  $ as shown in
Fig.~\ref{figGeneric}. The precise curve of the parametric plot is determined
by the values of the 6 parameters $\left(  b,c,K,\rho_{r},E,\phi\left(
\tau_{0}\right)  \right)  $. Generically the curve winds around the $\left(
\phi_{\gamma},s_{\gamma}\right)  $ plane since $\phi_{\gamma}\left(
\tau\right)  ,s_{\gamma}\left(  \tau\right)  $ are each periodic, although
their periods are generically incommensurate. The curve becomes a closed curve
if the ratio of the periods is a rational number. This indicates that the
generic solution has repeated big bangs and crunches and can be cyclic when
the periods are commensurate. Within each cycle there is \textit{a period of
antigravity sandwiched between every crunch and the following big bang}.

The mathematical expressions of the solutions are given in the Appendix. The
methods we used here follow those of \cite{inflationBC,cyclic BCT}. The
solutions are parametrized in terms of Jacobi elliptic functions $sn(z|m)$,
$cn\left(  z|m\right)  $, $dn\left(  z|m\right)  $ \cite{Abramowitz} in
combinations chosen appropriately in all the relevant regions of the parameter
space. For example, in a given region of parameter space the solution looks
like
\begin{equation}
\phi_{\gamma}\left(  \tau\right)  =A\frac{sn(\frac{\tau+\tau_{0}}{T}%
|m)}{dn(\frac{\tau+\tau_{0}}{T}|m)},
\end{equation}
with a similar expression for $s_{\gamma}\left(  \tau\right)  ,$ where the
factors $A,T,m$ are determined in terms of the 6 parameters $\left(
b,c,K,\rho_{r},E,\phi\left(  \tau_{0}\right)  \right)  $. The interested
reader will find these expressions in our recent paper \cite{cyclic BCT} for
the case of $\rho_{r}=0$. The only modification to generalize to $\rho_{r}%
\neq0$ is that previously we had used $E_{\phi}=E_{s}=E,$ while presently we
have $E_{s}=E$ and $E_{\phi}=E+\rho_{r}$. Since $E_{\phi}\geq E_{s}$ due to
$\rho_{r}\geq0,$ there are more cases to investigate depending on the regions
of the parameter space.%

\section{Geodesically complete bounces without antigravity}

\label{geoComplete}

Are there solutions that avoid the antigravity period in the cycle? Yes, there
is a subset of the parameter space for which the universe completely avoids
antigravity. Although this behavior is not generic, such solutions can be
characterized as the only ones that are geodesically complete in only the
gravity regime of the Einstein frame, which means they are fully described by
the action (\ref{action gravity}) with the additional condition of geodesic
completeness. This special subset is obtained by demanding that $\left\vert
\phi_{\gamma}\left(  \tau\right)  \right\vert \geq\left\vert s_{\gamma}\left(
\tau\right)  \right\vert $ at all times. So for the solutions of $\phi
_{\gamma}\left(  \tau\right)  $ that oscillate between positive and negative
values, at the points in time when $\left\vert \phi_{\gamma}\left(
\tau\right)  \right\vert $ vanishes $\left\vert s_{\gamma}\left(  \tau\right)
\right\vert $ must also vanish. This is possible only if the period of $\phi$
is an integer multiple of the period of $s.$ Since there is a time translation
symmetry in the differential equations, without losing generality we can
choose that the first instance they both vanish is at $\tau=0.$ Hence, for
$\phi_{\gamma}\left(  \tau\right)  ,s_{\gamma}\left(  \tau\right)  $ we must
require $\phi_{\gamma}\left(  0\right)  =s_{\gamma}\left(  0\right)  =0$ which
synchronizes their initial values to be both zero. This point in time is the
big bang since we compute from Eq.~(\ref{link1}) that at this time the scale
factor vanishes $a_{E}\left(  0\right)  =0.$ There are regions of parameter
space that yield such solutions, but as compared to the full 6-parameter space
it may be considered a set of measure zero. In the case of no radiation,
$\rho_{r}=0,$ this geodesically complete subset of solutions is given
analytically in \cite{cyclic BCT}.

\begin{figure}
\begin{center}
\includegraphics[width=4.0in]{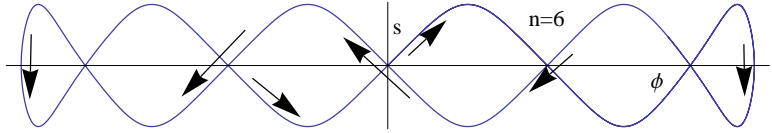}
\end{center}
\caption{Non-generic, zero-size bounce cyclic solution that never crosses into
the antigravity region. This figure is for $b<0$ and $c>0$. If $b,c>0$ the
figure extends to $\infty$ in the $\phi$ direction (see \cite{cyclic BCT}).}%
\label{CCfig2}
\end{figure}

A special example of such a solution is given in the parametric plot in
Fig.~\ref{CCfig2} borrowed from \cite{cyclic BCT}. This is a solution in a
region of the parameter space where there is no radiation $\rho_{r}=0;$ no
curvature $K=0;$ special initial conditions $\phi_{\gamma}\left(  0\right)
=s_{\gamma}\left(  0\right)  =0;$ and a quantized relation $b=-c/n^{4}$ for
integer $n.$ This quantization arises from asking the relative quantization of
the periods for $\phi,s.$ Besides these restrictions the parameters are free
to be in the regions $c>0$ , $E>0$ and $n\geq2$.

In Fig.~\ref{CCfig2}, with $n=6,$ the fields $\phi_{\gamma},s_{\gamma}$ start
out initially both vanishing $\phi_{\gamma}\left(  0\right)  =s_{\gamma
}\left(  0\right)  =0$ at the big bang (the arrow at the origin of the
figure); then while $\phi_{\gamma}\left(  \tau\right)  $ keeps growing, the
field $s_{\gamma}\left(  \tau\right)  $ oscillates several times until
$\phi_{\gamma}\left(  \tau\right)  $ reaches its maximum and turns around;
then $\phi_{\gamma}\left(  \tau\right)  $ decreases to zero while $s_{\gamma
}\left(  \tau\right)  $ oscillates several times and vanishes at the same time
as $\phi_{\gamma}$. This point in time represents a big crunch. Then the
motion continues smoothly to negative values of $\phi_{\gamma}$ and repeats
the same behavior of big bang then turnaround and big crunch. The full cycle
is repeated again and again periodically which is described by Fig.~5 in the
Appendix. Note the 5 nodes in this figure are determined by the choice of the
integer $n=6.$%

When $\rho_{r}>0$, the quantization requirement for the periods puts a less
severe restriction on the parameters $\left(  b,c,K,\rho_{r},E,\phi\left(
\tau_{0}\right)  \right)  .$ Although the synchronization of initial
conditions $\phi\left(  0\right)  =s\left(  0\right)  =0$ and the relative
quantization of the periods are still necessary, these conditions no longer
require that $b/c$ is quantized by itself because the additional parameter
$\rho_{r}$ also enters in the quantization of the periods. Instead, the
parameters $\left(  b,c,K,\rho_{r},E\right)  $ collectively are subject to one
quantization condition; \textit{e.g.\.{,}} the integration parameter $E$ may
be quantized in terms of the other four parameters plus an integer. An example
of such a solution with radiation (but with $K=0$ for illustration) is given
in the first line of Table I, by the parameters that satisfy $\frac{b\left(
E+\rho_{r}\right)  }{cE}=\frac{-1}{n^{4}}$, with $n=\operatorname{integer}.$
This is solved by a quantized integration parameter $E_{n}=-\frac{b\rho
_{r}n^{4}}{bn^{4}+c}.$ On the other hand when $\rho_{r}$ vanishes we have,
$\frac{b\left(  E+0\right)  }{cE}=\frac{-1}{n^{4}},$ where the parameter $E$
drops out and there is a solution like the one in Fig.~\ref{CCfig2}, only if
the parameters of the model are quantized $b=-c/n^{4}$. The inclusion of
radiation changes the parametric plot above in a simple way: the trajectory
extends further out in the $\phi_{\gamma}$ direction as $\rho_{r}$ increases
due to the higher energy in the $\phi_{\gamma}$ field.

We list below all the cases of parameter subspaces that permit purely gravity
(i.e. no antigravity regime) geodesically complete solutions and point out the
corresponding figures and formulas shown in the Appendix. All of these
describe a universe that always remains in the gravity regime of the Einstein
frame, and either: (i) bounces at zero size for $K=0$; (ii) bounces at zero
size for $K\neq0$; or (iii) bounces at finite size for $K>0$. These are found
by setting
\begin{equation}
\phi\left(  0\right)  =s\left(  0\right)  =0,\label{initial}%
\end{equation}
(which implies $\delta=0$) and then replacing $E_{\phi}=E+\rho_{r}$ and
$E_{s}=E,$ instead of $E_{\phi}=E_{s}=E,$ in the quantization of the periods.
These necessary conditions cannot be satisfied for all the solutions given in
the figures in the Appendix; the cases that are compatible with these
conditions are indicated on the right side of Tables I,IIa,IIb,III.

\begin{itemize}
\item If $K=0$, there are two regimes of parameter space in which there can be
a singular bounce without violating the null energy condition:
\[%
\begin{tabular}
[c]{|l|l|l|l|l|l|l|}\hline
& $b$ & $c$ & $E$ & $\rho_{r}$ & Table I:$\;\text{conditions when }K=0$ & FIG
\#\\\hline
$1.$ & $<0$ & $>0$ & $>0$ & $\geq0$ & $\;\frac{b\left(  E+\rho_{r}\right)
}{cE}=\frac{-1}{n^{4}},\text{ }\left\{
\begin{array}
[c]{c}%
n=1,2,3\ldots\text{if }\rho_{r}>0\\
n=2,3,4\ldots\text{ if }\rho_{r}=0
\end{array}
\right.  $ & Fig.5\\\hline
$2.$ & $>0$ & $>0$ & $>0$ & $\geq0$ & $\;\frac{b}{4c}\frac{E+\rho_{r}}%
{E}=\frac{1}{n^{4}},\text{ }n=1,2,3\ldots$ & Fig.6\\\hline
\end{tabular}
\ \ \ \
\]

\newpage
\item If $K>0,$ there exist two categories of cyclic solutions, the ones that
bounce at zero size without violating the null energy condition, and the ones
that bounce at finite size.

The conditions on the parameters for \textit{bouncing at finite size} are%

\[
\bigskip
\begin{tabular}
[c]{|l|l|l|l|l|l|l|}\hline
& $b$ & $c$ & $E+\rho_{r}$ & $\rho_{r}$ & Table IIa:$\;\text{conditions when
}K>0$ & FIG \#\\\hline
$1.$ & $>0$ & $>0$ & $\geq0,\leq\frac{K^{2}}{16b}$ & $<\frac{K^{2}}{16b}$ &
$\left\vert \phi\left(  \tau_{0}\right)  \right\vert \geq\sqrt{\frac{K}{4b}}$
& Fig.12\\\hline
$2.$ & $>0$ & $<0$ & $>0,\leq\frac{K^{2}}{16b}$ & $<\frac{K^{2}}{16b}$ &
$\left\vert \phi\left(  \tau_{0}\right)  \right\vert \geq\sqrt{\frac{K}{4b}%
},\left\vert s\left(  \tau_{0}\right)  \right\vert \leq\left\vert \phi\left(
\tau_{0}\right)  \right\vert $ & Fig.18\\\hline
\end{tabular}
\
\]

The conditions on the parameters for \textit{bouncing at zero size} are given
below. In these expressions $\mathcal{K}\left(  m\right)  \equiv
EllipticK\left(  m\right)  $ is a well known special function that corresponds
to the quarter period of the Jacobi elliptic functions, such as $sn(z|m),$
with label $m$ \cite{Abramowitz}.
\[%
\begin{tabular}
[c]{|l|l|l|l|l|l|l|}\hline
& $b$ & $c$ & $E+\rho_{r}$ & $\rho_{r}$ & Table IIb:$\;\text{conditions when
}K>0$ & FIG \#\\\hline
$1.$ & $\leq0$ & $>0$ & $>0$ & $\geq0$ & $%
\begin{array}
[c]{c}%
\frac{\left(  1+\frac{16c}{K^{2}}E\right)  ^{1/4}\times\mathcal{K}\left(
\frac{1}{2}-\frac{1}{2}\left(  1-\frac{16b}{K^{2}}\left(  E+\rho_{r}\right)
\right)  ^{-1/2}\right)  }{\left(  1-\frac{16b}{K^{2}}\left(  E+\rho
_{r}\right)  \right)  ^{1/4}\times\mathcal{K}\left(  \frac{1}{2}-\frac{1}%
{2}\left(  1+\frac{16c}{K^{2}}E\right)  ^{-1/2}\right)  }=n\\
\left\{
\begin{array}
[c]{c}%
n=1,2,3\ldots\text{if }\rho_{r}>0\\
n=2,3,4\ldots\text{ if }\rho_{r}=0
\end{array}
\right.  ,E>0
\end{array}
$ & Fig.10\\\hline
$2.$ & $>0$ & $\geq0$ & $>\frac{K^{2}}{16b}$ & $\geq0$ & $%
\begin{array}
[c]{c}%
\frac{\sqrt{2}\left(  1+\frac{16c}{K^{2}}E\right)  ^{1/4}\times\mathcal{K}%
\left(  \frac{1}{2}+\frac{1}{2}\left(  \frac{16b}{K^{2}}\left(  E+\rho
_{r}\right)  \right)  ^{-1/2}\right)  }{\left(  \frac{16b}{K^{2}}\left(
E+\rho_{r}\right)  \right)  ^{1/4}\times\mathcal{K}\left(  \frac{1}{2}%
-\frac{1}{2}\left(  1+\frac{16c}{K^{2}}E\right)  ^{-1/2}\right)  }=n\\
n=1,2,3\ldots
\end{array}
$ & Fig.13\\\hline
$3.$ & $>0$ & $\geq0$ & $\geq0$ & $\geq0$ & $%
\begin{array}
[c]{c}%
\frac{\left(  1+\frac{16Ec}{K^{2}}\right)  ^{1/4}\times\mathcal{K}\left(
\left(  \frac{1}{2}+\frac{1}{2}\sqrt{1-\frac{16b}{K^{2}}\left(  E+\rho
_{r}\right)  }\right)  ^{-1}-1\right)  }{\left(  \frac{1}{2}+\frac{1}{2}%
\sqrt{1-\frac{16b}{K^{2}}\left(  E+\rho_{r}\right)  }\right)  ^{1/2}%
\times\mathcal{K}\left(  \frac{1}{2}-\frac{1}{2}\left(  1+\frac{16Ec}{K^{2}%
}\right)  ^{-1/2}\right)  }=n\\
n=1,2,3\ldots,E+\rho_{r}\leq\frac{K^{2}}{16b},\left\vert \phi_{\gamma}\left(
\tau_{0}\right)  \right\vert <\sqrt{\frac{K}{4b}}%
\end{array}
$ & Fig.11\\\hline
$4.$ & $>0$ & $<0$ & $>\frac{K^{2}}{16b}$ & $\geq0$ & $%
\begin{array}
[c]{c}%
\frac{\sqrt{2}\left(  \frac{1}{2}+\frac{1}{2}\sqrt{1+\frac{16cE}{K^{2}}%
}\right)  ^{1/2}\times\mathcal{K}\left(  \frac{1}{2}+\frac{1}{2}\sqrt
{\frac{16b}{K^{2}}\left(  E+\rho_{r}\right)  }\right)  }{\left(  \frac
{16b}{K^{2}}\left(  E+\rho_{r}\right)  \right)  ^{1/4}\times\mathcal{K}\left(
\left(  \frac{1}{2}+\frac{1}{2}\sqrt{1+\frac{16cE}{K^{2}}}\right)
^{-1}-1\right)  }=n\\
n=1,2,3\ldots,\frac{K^{2}}{16\left(  -c\right)  }>E>0,~\left\vert s_{\gamma
}\left(  \tau_{0}\right)  \right\vert <\sqrt{\frac{K}{4\left(  -c\right)  }}%
\end{array}
$ & Fig.17\\\hline
$5.$ & $>0$ & $c<0$ & $>\frac{K^{2}}{16b}$ & $\geq0$ & $\frac{\left(
-\frac{16c}{K^{2}}E\right)  ^{1/4}\ast\mathcal{K}\left(  \frac{1}{2}+\frac
{1}{2}\left(  \frac{16b}{K^{2}}\left(  E+\rho_{r}\right)  \right)
^{-1/2}\right)  }{\left(  \frac{16b}{K^{2}}\left(  E+\rho_{r}\right)  \right)
^{1/4}\ast\mathcal{K}\left(  \frac{1}{2}+\left(  -\frac{16c}{K^{2}}E\right)
^{-1/2}\right)  }=1;\;\frac{K^{2}}{16\left(  -c\right)  }<E$ & Fig.16\\\hline
\end{tabular}
\ \ \ \
\]

\bigskip

\item If $K<0,$ the conditions on the parameters for \textit{bouncing at zero
size} are
\[%
\begin{tabular}
[c]{|l|l|l|l|l|l|l|}\hline
& $b$ & $c$ & $E+\rho_{r}$ & $\rho_{r}$ & Table III:$\;\text{condition when
}K<0$ & FIG \#\\\hline
$1.$ & $<0$ & $>0$ & $>0$ & $\geq0$ & $%
\begin{array}
[c]{c}%
\frac{\left(  1+\frac{16c}{K^{2}}E\right)  ^{1/4}\times\mathcal{K}\left(
\frac{1}{2}-\frac{1}{2}\left(  1-\frac{16b}{K^{2}}\left(  E+\rho_{r}\right)
\right)  ^{-1/2}\right)  }{\left(  1-\frac{16b}{K^{2}}\left(  E+\rho
_{r}\right)  \right)  ^{1/4}\times\mathcal{K}\left(  \left(  \frac{1}{2}%
-\frac{1}{2}\left(  1+\frac{16c}{K^{2}}E\right)  ^{-1/2}\right)  \right)
}=n\\
\left\{
\begin{array}
[c]{c}%
n=1,2,3\ldots\text{if }\rho_{r}>0\\
n=2,3,4\ldots\text{ if }\rho_{r}=0
\end{array}
\right.  ,E>0
\end{array}
$ & Fig.21\\\hline
$2.$ & $>0$ & $>0$ & $>0$ & $\geq0$ & $%
\begin{array}
[c]{c}%
\frac{\left(  1+\frac{16c}{K^{2}}E\right)  ^{1/4}\times\mathcal{K}\left(
\frac{1}{2}+\frac{1}{2}\left(  \frac{16b}{K^{2}}\left(  E+\rho_{r}\right)
\right)  ^{-1/2}\right)  }{\sqrt{2}\left(  \frac{16b}{K^{2}}\left(  E+\rho
_{r}\right)  \right)  ^{1/4}\times\mathcal{K}\left(  \frac{1}{2}-\frac{1}%
{2}\left(  1+\frac{16c}{K^{2}}E\right)  ^{-1/2}\right)  }=n\\
n=1,2,3\ldots,E>0
\end{array}
$ & Fig.24\\\hline
\end{tabular}
\ \ \ \ \
\]

\item In addition, for any values of $b,c,$ there is the special solution in
which $s_{\gamma}\left(  \tau\right)  =0$ for all $\tau$ (sitting at the $s=0$
extremum of $V_{s}=\frac{1}{2}Ks^{2}+cs^{4}$) while $\phi\left(  \tau\right)
$ performs any motion at energy $E_{\phi}=\rho_{r}$.
\end{itemize}

\section{Geodesically complete bounces with antigravity}

\label{antig}

In addition to the solutions described in the previous section, there are ones
that are geodesically complete provided an antigravity regime is included (see
Fig.~2). In Refs.~\cite{BCSTletter,BCST}, we show that, when anisotropy is
added to the curvature and radiation, there is a strong attractor behavior
such that almost all solutions pass through the origin and undergo a period of antigravity (a loop) between each
big crunch and big bang. This is illustrated in Fig.~4. As discussed in
\cite{BCSTletter,BCST}, the zero-size bounce solutions that evolved from
crunch to bang in the purely gravity region in the absence of anisotropy (as
listed in Tables I,IIb,III) and illustrated in Fig. 3, as well as the other
generic solutions in the Appendix illustrated in Fig.~2, are strongly
modified near the singularity by the anisotropy, such that the trajectory
cannot avoid the antigravity region. Furthermore, given some initial
conditions, the global behavior of a trajectory, far away from the
singularity, can also be altered even by a small amount of anisotropy
\cite{BCST}. The finite-size bounce solutions could avoid the antigravity
region despite anisotropy, but this may occur only in a very narrow region of
parameter space.

\begin{figure}
\begin{center}
\includegraphics[width=3.0in]{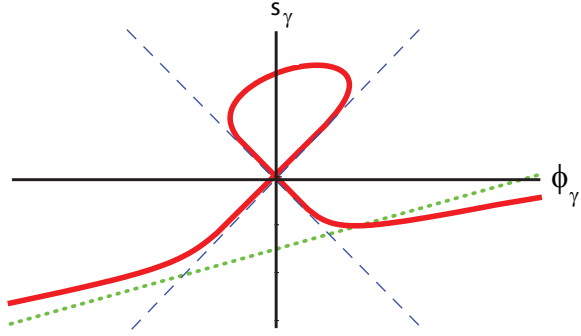}
\end{center}
\caption{Comparison of a solution without anisotropy (green dotted path) and with anisotropy added (red thick solid curve).  An attractor mechanism caused by the anisotropy distorts the path so that it passes through the origin at the crunch and undergoes a loop  in the antigravity region, through the origin again, and then re-emerging in the gravity regime.  }
\label{comppaths}
\end{figure}

We are, therefore, faced with trying to understand physical phenomena in the
antigravity regime. Since physical intuition for gravity is developed mainly
in the Einstein frame, we begin with the Einstein gauge. When $\left(
\phi^{2}-s^{2}\right)  $ is negative, it is again possible to use the Weyl
symmetry to choose an Einstein gauge, $\phi_{E}^{2}-s_{E}^{2}=-1/2\kappa^{2},$
but this is in a new domain of field space, namely in the top and bottom
quadrants of the $\left(  \phi,s\right)  $ space, as shown in
Fig.~\ref{CCrFig1}. The new $\phi_{E},s_{E}$ are given by interchanging the
$\sinh$ and $\cosh$ in Eq.~(\ref{Egauge}), namely $\phi_{E}\left(  x\right)
=\pm\frac{\sqrt{6}}{\kappa}\sinh(\kappa\sigma\left(  x\right)  /\sqrt
{6}),\;s_{E}\left(  x\right)  =\pm\frac{\sqrt{6}}{\kappa}\cosh(\kappa
\sigma\left(  x\right)  /\sqrt{6}),$ such that $\phi_{E}^{2}-s_{E}%
^{2}=-1/2\kappa^{2}.$ Then the gauge fixed form of the action
(\ref{conformal action}) looks like the action in Eq.~(\ref{action gravity}),
except that the first two terms change sign. The potential $V\left(
\sigma\right)  $ does not change sign, but it is a new function $\bar
{V}\left(  \sigma\right)  ,$ which is related to Eq.~(\ref{V}) by
interchanging $\sinh$ and $\cosh$. The metric $g_{E}^{\mu\nu}$ in this gauge
has no signature change. Hence, for matter fields, including radiation, the
signs of their kinetic terms remain the same in the gravity and antigravity
sectors. The gauge-fixed action in the antigravity regime looks as follows
\begin{equation}
S_{antigravity}=\int d^{4}x\sqrt{-g}\left\{
\begin{array}
[c]{c}%
-\frac{1}{2\kappa^{2}}R\left(  g_{E}\right)  +\frac{1}{2}g^{\mu\nu}%
\partial_{\mu}\sigma\partial_{\nu}\sigma-\bar{V}\left(  \sigma\right)  \\
+radiation+matter
\end{array}
\right\}  .\label{antiaction}%
\end{equation}
Because of the sign change in the first two terms, $\sigma$ now looks like a
ghost while the $a_{E}$ degree of freedom is no longer a ghost. The zero total
energy constraint ($G_{00}=T_{00}$ Einstein equation) compensates for one
ghost, as it did in the usual gravity regime, so there are no unitarity.
concerns regarding the $\sigma$ degree of freedom. However, the other
fluctuations of the metric, namely the spin-2 gravitons, now have the wrong
sign kinetic terms. Note that some of the spin-2 degrees of freedom are in the
form of the anisotropy fields; including them does not seem to show any
particular instability or other unusual behavior \cite{BCST}.

The discussion of the relativistic harmonic oscillator, as treated in
\cite{bars harmonic oscillator}, is a good toy model to understand how to
correctly quantize the theory while maintaining unitarity when some degrees of
freedom have the wrong sign kinetic energy. The basic technique is to
interchange the roles of creation-annihilation operators when the kinetic
energy has the wrong sign; then the resulting Fock space has only positive
norms. As seen in \cite{bars harmonic oscillator} a similar approach also
occurs in the construction of unitary representations of non-compact groups by
using oscillators. Similarly, in the antigravity regime, the theory should be
quantized without negative norm ghosts by interchanging the roles of
creation-annihilation operators for gravitons. The price is that the energy of
the spin-2 gravitons is unbounded below, so potentially there is an
instability. At the linearized level, which defines a perturbative Hilbert
space, there is no consequence. But when interactions are included, due to the
availability of negative energy states, it may be possible to emit abundantly
spin-2 gravitons with negative energy; however this must be accompanied with
the emission of positive energy matter to maintain the zero energy constraint.
So the theory must react in some interesting ways through the interactions as
soon as the antigravity regime is reached.

In the cases where there is anisotropy \cite{BCSTletter}\cite{BCST}, we find
that the trajectory of the antigravity depends on the radiation density such
that, if $\rho_{r}$ increases due to the spontaneous production of negative
energy gravitons (as noted above), the effect is to decrease the duration of
the antigravity period and return the universe more rapidly to the big bang
and a period of pure gravity expansion. As noted in \cite{BCSTletter}%
\cite{bars DPF} this is an indication that the dynamics tries to minimize the
effects of antigravity, but the details of how this works is currently cloudy.

Of course, quantum gravity effects need to be also included. Therefore, it
would be very interesting to study similar circumstances in the framework of
string theory. To formulate the antigravity aspects in string theory we could
use the field transformations given in Eqs.(\ref{link1},\ref{link2}), but even
better would be the inclusion of the analog of the Weyl symmetry in the
framework of string theory.

It is worth mentioning that it seems possible to connect the state of the
universe before the crunch to the state of the universe after the crunch by
solving our classical equations analytically along a path in the complex
$\tau$ plane, such that the path completely avoids the antigravity regime, and
also stays sufficiently far away from the singularities, so that quantum
corrections become negligible. Such an approach is very desirable for the
cyclic universe scenario. We will report on this type of solution in a
separate paper.

\section{Summary}

\label{discussion}

In this paper we have used analytic solutions of cosmological equations to
discuss geodesic completeness through the big bang singularity. In the context
of the path integral, our complete set of classical solutions provide a
semi-classical approximation to the quantum theory.

The computations presented in this paper mostly ignored anisotropy and used a
special potential energy $V\left(  \sigma\right)  $ to obtain \textit{all} the
analytic solutions of the Friedmann equations, in a model that includes
radiation and spatial curvature. The solutions are characterized by six
parameters that include initial values and model parameters. We learned that
the generic solution, in which none of the six parameters are restricted,
shows that the trajectory of the universe goes smoothly through the
crunch/bang singularities while traversing from gravity to antigravity
spacetime patches, and doing this repeatedly in a periodic manner. The generic
trajectory can cross the \textquotedblleft lightcone\textquotedblright\ in
field space, shown in Fig.~\ref{CCrFig1}, at any place. The crossing points
on the \textquotedblleft lightcone\textquotedblright\ depend on the values of
the six parameters. Although our general exact results are obtained in a
specific model, the presence of antigravity is likely to occur generically in
any model that is geodesically complete. Therefore, the phenomenon of
antigravity should be considered seriously in discussing cosmology.

We found that it is possible to avoid antigravity and still have a
geodesically complete geometry within a smaller (but still infinite) subset of
solutions (Tables I,IIb,III). These are the only geodesically complete
solutions contained totally within the traditional Einstein frame. One group
of trajectories passes through the center of the \textquotedblleft
lightcone\textquotedblright repeatedly, resulting in a cyclic universe. These
solutions, which do not violate the null energy condition, provide a set of
examples that bouncing at zero size is possible classically in cosmological
scenarios with or without spatial curvature.

It should be emphasized that our new results transcend the specific
simple model above. The phenomena we have found should also be
expected generically in supergravity theories coupled to matter
whose formulation include a similar factor that multiplies $R\left(
g\right)  .$ In supergravity, that factor is related to the Kahler
potential, and this factor, combined with the usual Einstein-Hilbert
term, is not generally positive definite \cite{weinberg}. In fact,
in a gauge that we call the supergravity gauge, or $c$-gauge, in
which $\phi(x)$ is set to a constant $\phi_0$
\cite{BCST,2Tgravity}, our term $(\phi_0^2-s^2)R(g)$ reduces
precisely to the familiar form in supergravity including a
K\"ahler-like potential.
 In the past, it was assumed that the overall factor is positive and
investigations of supergravity proceeded only in the positive regime. A
discussion of the field space in the positive sector for general $\mathcal{N}%
$=2 supergravity can be found in \cite{deWit}. However, our results
suggest that generically the overall factor can and will change sign
dynamically, in every gauge, and therefore antigravity sectors
similar to our discussion in this paper should be expected in
typical supergravity theories. This is illustrated with an example
in \cite{BCST}.

Until better understood in the context of quantum gravity, or string theory,
our results should be considered to be a first pass for the types of new
questions they raise and the answers they provide.

Much remains to be understood, including quantum gravity and string theory
effects, but it is clear that previously unsuspected phenomena, including
antigravity, come into play classically close to the cosmological singularity.
The technical tools to study such issues in the context of a full quantum
theory of gravity are yet to be developed. This is an important challenge to
the theory community, since the results have profound implications for both
fundamental physics and our understanding of the origin, evolution and future
of the universe.

\begin{acknowledgments}
IB and PJS thank the Perimeter Institute for its generous hospitality and
support. Research at the Perimeter Institute is support by the Government of
Canada and the Province of Ontario through the Ministry of Research and
Innovation.  This research was also partially supported by the U.S. Department of
Energy under grant number DE-FG03-84ER40168 (IB) and under grant number
DE-FG02-91ER40671 (PJS).
\end{acknowledgments}

\appendix

\section{The Analytic Solutions}

\label{analytic}

The intuitive approach for solving the Friedmann equations in the separable
form Eqs.$\left(  \ref{phi}-\ref{energy constraint}\right)  $ was described in
Sec.~\ref{solve}. In Eqs.(\ref{energy constraint},\ref{E}) we showed that a
first integral of these second order equations takes the form of the analog
problem of a particle in a potential, separately for the \textquotedblleft
particles\textquotedblright\ $\phi_{\gamma},s_{\gamma}.$ Namely, $\frac{1}%
{2}\dot{\phi}_{\gamma}^{2}+V_{\phi}=E_{\phi},\;$and$~\frac{1}{2}\dot
{s}_{\gamma}^{2}+V_{s}=E_{s},$ with the potentials $V_{\phi}=-b\phi_{\gamma
}^{4}+\frac{K}{2}\phi_{\gamma}^{2}$ and $V_{s}=cs_{\gamma}^{4}+\frac{K}%
{2}s_{\gamma}^{2},$ and also with the energy constraint generally solved by
$E_{s}=E$ and $E_{\phi}=E+\rho_{r}$ for any $E.$ In this appendix we list the
complete set of analytic solutions to these equations in all possible regions
of the six parameter space $\left(  b,c,K,\rho_{r},E,\phi\left(  \tau
_{0}\right)  \right)  .$ In this Appendix, we restrict ourselves to the case
$b+c>0$, which corresponds in the Einstein frame to a potential $V\left(
\sigma\right)  $ in Eq.~(\ref{V}) that is bounded from below. The
generalization to unstable potentials is straightforward.

Figs.~5-29 below represent the potentials $V_{\phi}$ (solid line) and $V_{s}$
(dashed line) and the energy levels $E_{\phi},E_{s}$; they are drawn in the
various regions of parameter space. The $\phi$-level $E_{\phi}=E+\rho_{r}$ is
higher than the $s$-level $E_{s}=E$ because the radiation energy density
$\rho_{r}$ is positive. The parameter $E$ is allowed to slide vertically
within the physical range permitted for the $s$-particle's motion in the
potential $V_{s}.$ From the figures alone one can obtain the intuitive
solution by invoking the analogy of a particle moving in a potential for
either $s_{\gamma}$ or $\phi_{\gamma}$. Next to each figure we give the
corresponding analytic solution to Eq.$\left(  \ref{phi}%
-\ref{energy constraint}\right)  .$ We did not include separately some trivial
cases such as the case when the fields sit still at the top or bottom of the
potential extrema, or in the cases of $b=0$ or $c=0$ where Eqs.$\left(
\ref{phi}-\ref{energy constraint}\right)  $ become linear differential
equations with simple trigonometric solutions. These cases are recovered in
the appropriate limit of the expressions below.

In the analytic expressions below instead of the free parameter $\phi\left(
\tau_{0}\right)  $ for an arbitrary initial value, we have inserted the
arbitrary phase shift parameter $\delta.$ Note also that we have used the time
translation symmetry of the equations to choose a specific value for the
initial value for $s_{\gamma}\left(  \tau\right)  $ at $\tau=0.$ In this way
we give here the generic solutions that describe all possible geodesically
complete trajectories of the universe, including those that move between the
gravity and antigravity regions.

A subset of solutions stay only in the gravity sector with geodesically
complete trajectories. They are obtained by putting constraints on the
parameters. These include setting $\delta=0$ (synchronized initial conditions
for $\phi,s$) and requiring a quantization of the periods of $\phi$ relative
to the period of $s.$ The corresponding parameter space is given in section
(\ref{geoComplete}).

The reader can verify that the following expressions solve the differential
equations and that the plots of $\phi\left(  \tau\right)  ,s\left(
\tau\right)  $ as functions of $\tau$ correspond to the motion intuitively
expected for a particle in the corresponding potential at the given energy
level. To verify the solution the following properties of the Jacobi elliptic
functions $sn\left(  z|m\right)  ,$~$cn\left(  z|m\right)  $ and $dn\left(
z|m\right)  $ are useful \cite{Abramowitz}. The derivative of Jacobi elliptic
functions are given in terms of expressions somewhat similar to those for
trigonometric functions
\begin{align}
\frac{d}{dz}sn\left(  z|m\right)   &  =cn\left(  z|m\right)  \times dn\left(
z|m\right)  ,\\
\frac{d}{dz}cn\left(  z|m\right)   &  =-sn\left(  z|m\right)  \times dn\left(
z|m\right)  ,\\
\frac{d}{dz}dn\left(  z|m\right)   &  =-m\times sn\left(  z|m\right)  \times
cn\left(  z|m\right)  .
\end{align}
They also satisfy quadratic relations, such as%
\begin{equation}
\left(  sn\left(  z|m\right)  \right)  ^{2}+\left(  cn\left(  z|m\right)
\right)  ^{2}=1;\;\ m\left(  sn\left(  z|m\right)  \right)  ^{2}+\left(
dn\left(  z|m\right)  \right)  ^{2}=1.
\end{equation}

When $m$ is a real number in the range $-\infty<m<1$, the function $sn\left(
z|m\right)  $ oscillates between the values $-1$ to $+1,$ similar to the
trigonometric function $\sin\left(  z\right)  $, vanishing at $z\rightarrow0,$
and reaching a maximum at the quarter period $z=K\left(  m\right)  $, where
$K\left(  m\right)  $ is the elliptic integral as a function of $m.$ From the
quadratic relations above, it is deduced that the behavior of $cn\left(
z|m\right)  $ is that it oscillates similar to a cosine $\cos\left(  z\right)
,$ while $dn\left(  z|m\right)  $ oscillates between the positive values $+1$
and and $\left(  1-m\right)  ^{1/2}.$ When $m>1$ the behavior is still
oscillatory but quite different than $\sin\left(  z\right)  ,$ $\cos\left(
z\right)  ,$etc.. However, it is possible to use identities to rewrite the
solution in terms of $sn\left(  z|m^{\prime}\right)  ,cn\left(  z|m^{\prime
}\right)  ,dn\left(  z|m^{\prime}\right)  $ where $m^{\prime}=1-m$ is again in
the range $-\infty<m^{\prime}<1.$ We have used such identities so that all the
$m$ values that appear in our solutions below are in the range $-\infty<m<1.$
Then the reader can get a feeling of the behavior of the solutions by the
analogy to trigonometric $\sin\left(  z\right)  ,$ $\cos\left(  z\right)  ,$etc..

\bigskip

\newpage
We begin with the $K=0$ cases; there are five different regions for the
remaining parameters as listed in Figs.(5-9)

\bigskip%

\begin{tabular}
[c]{||c||l||}\hline\hline
$%
\begin{array}
[c]{c}%
{\includegraphics[
height=0.7567in,
width=1.9268in
]%
{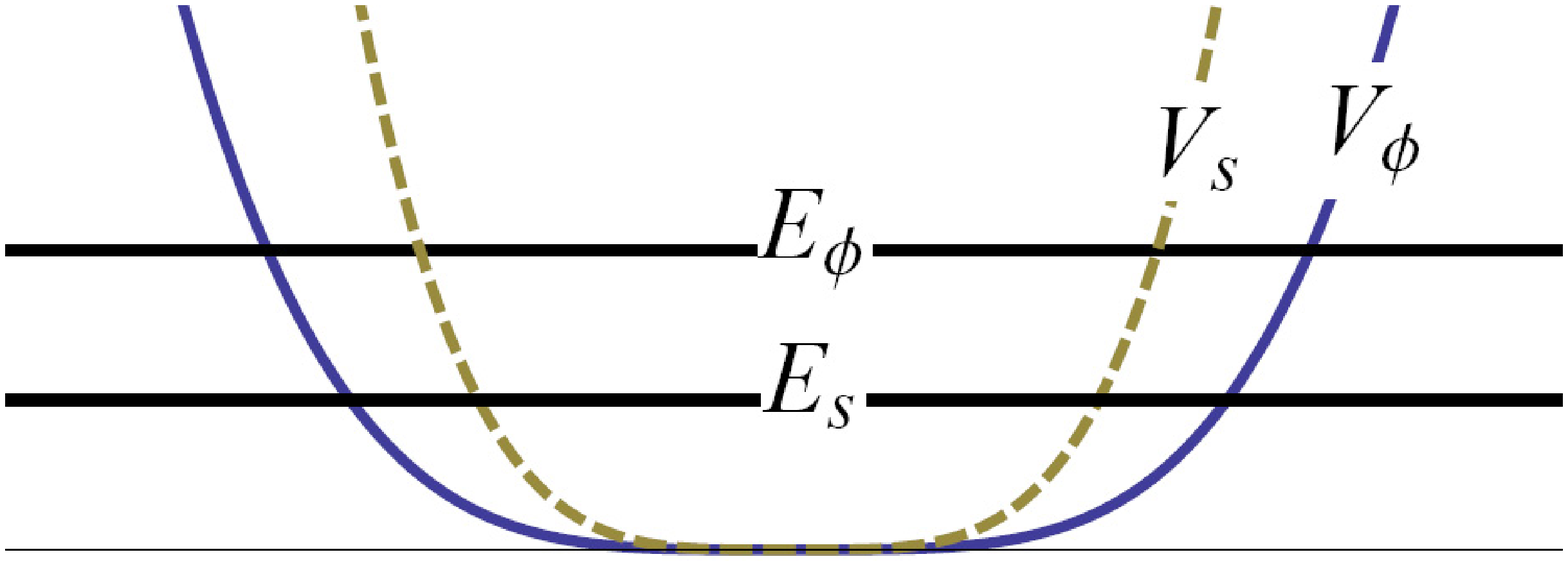}%
}%
\\
\text{FIG. 5}\\
b<0,\text{ }c>0,\text{ }E_{\phi}\geq E_{s}>0
\end{array}
$ & $%
\begin{array}
[c]{l}%
\phi=\left(  \frac{E+\rho_{r}}{-4b}\right)  ^{1/4}\frac{sn(\frac{\tau+\delta
}{T_{\phi}}|\frac{1}{2})}{dn(\frac{\tau+\delta}{T_{\phi}}|\frac{1}{2})},\text{
}T_{\phi}=\left(  -16b\left(  E+\rho_{r}\right)  \right)  ^{-1/4}\\
s=\left(  \frac{E}{4c}\right)  ^{1/4}\frac{sn(\frac{\tau}{T_{s}}|\frac{1}{2}%
)}{dn(\frac{\tau}{T_{s}}|\frac{1}{2})},\text{ }T_{s}=\left(  16Ec\right)
^{-1/4}%
\end{array}
$\\\hline\hline
\end{tabular}
\medskip%

\begin{tabular}
[c]{||c||l||}\hline\hline
$%
\begin{array}
[c]{c}%
{\includegraphics[
height=0.8873in,
width=1.6622in
]%
{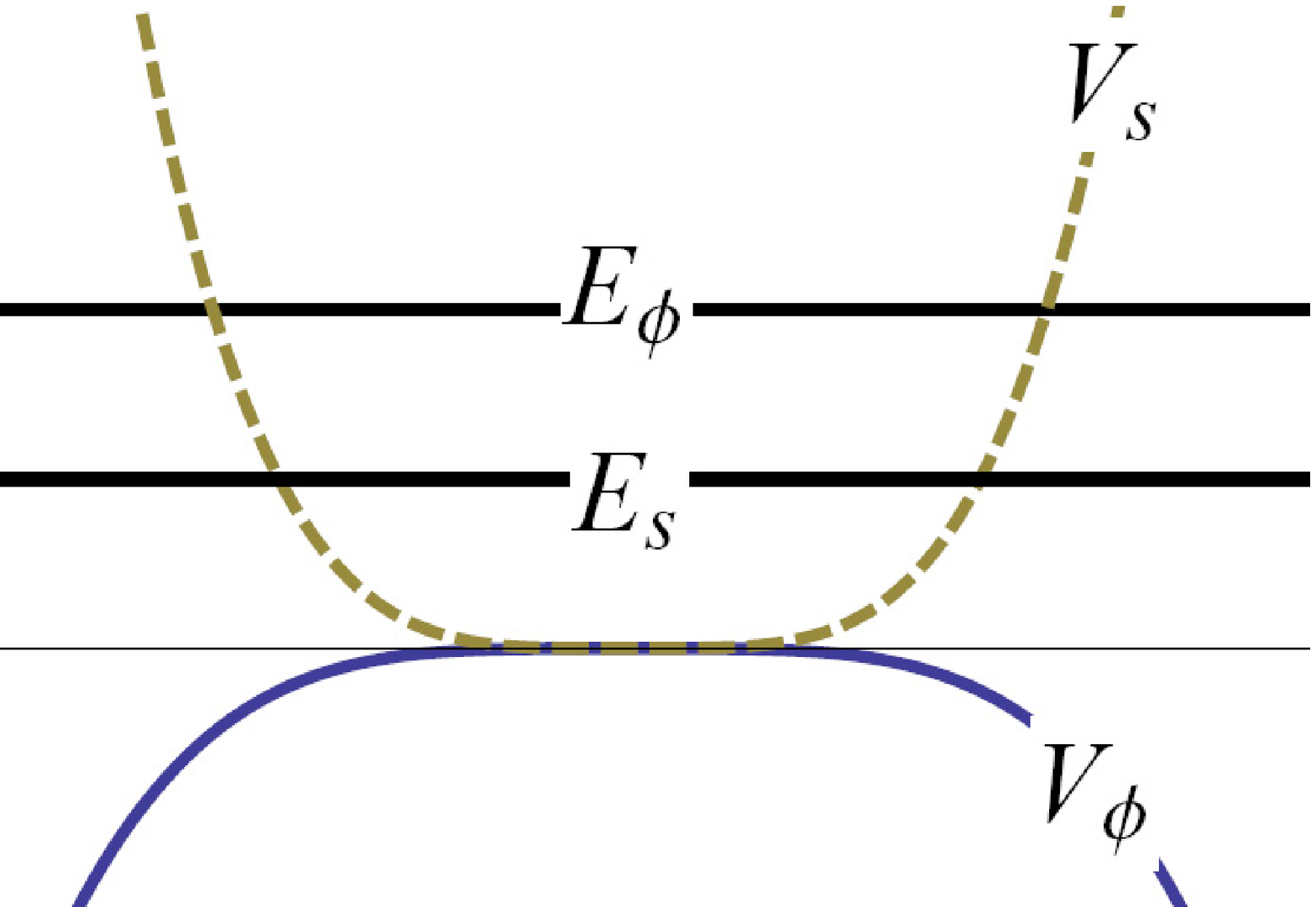}%
}%
\\
\text{FIG. 6}\\
b>0,\text{ }c>0,\text{ }E_{\phi}\geq E_{s}>0
\end{array}
$ & $%
\begin{array}
[c]{l}%
\phi=\left(  \frac{E+\rho_{r}}{b}\right)  ^{1/4}\frac{sn\left(  \frac
{\tau+\delta}{T_{\phi}}|\frac{1}{2}\right)  }{1+cn\left(  \frac{\tau+\delta
}{T_{\phi}}|\frac{1}{2}\right)  },\text{ }T_{\phi}=\frac{1}{\sqrt{2}}\left(
16b\left(  E+\rho_{r}\right)  \right)  ^{-1/4}\\
s=\left(  \frac{E}{4c}\right)  ^{1/4}\frac{sn(\frac{\tau}{T_{s}}|\frac{1}{2}%
)}{dn(\frac{\tau}{T_{s}}|\frac{1}{2})},\text{ }T_{s}=\left(  16cE\right)
^{-1/4}%
\end{array}
$\\\hline\hline
\end{tabular}
\medskip%

\begin{tabular}
[c]{||c||l||}\hline\hline
$%
\begin{array}
[c]{c}%
{\includegraphics[
height=0.7991in,
width=1.8325in
]%
{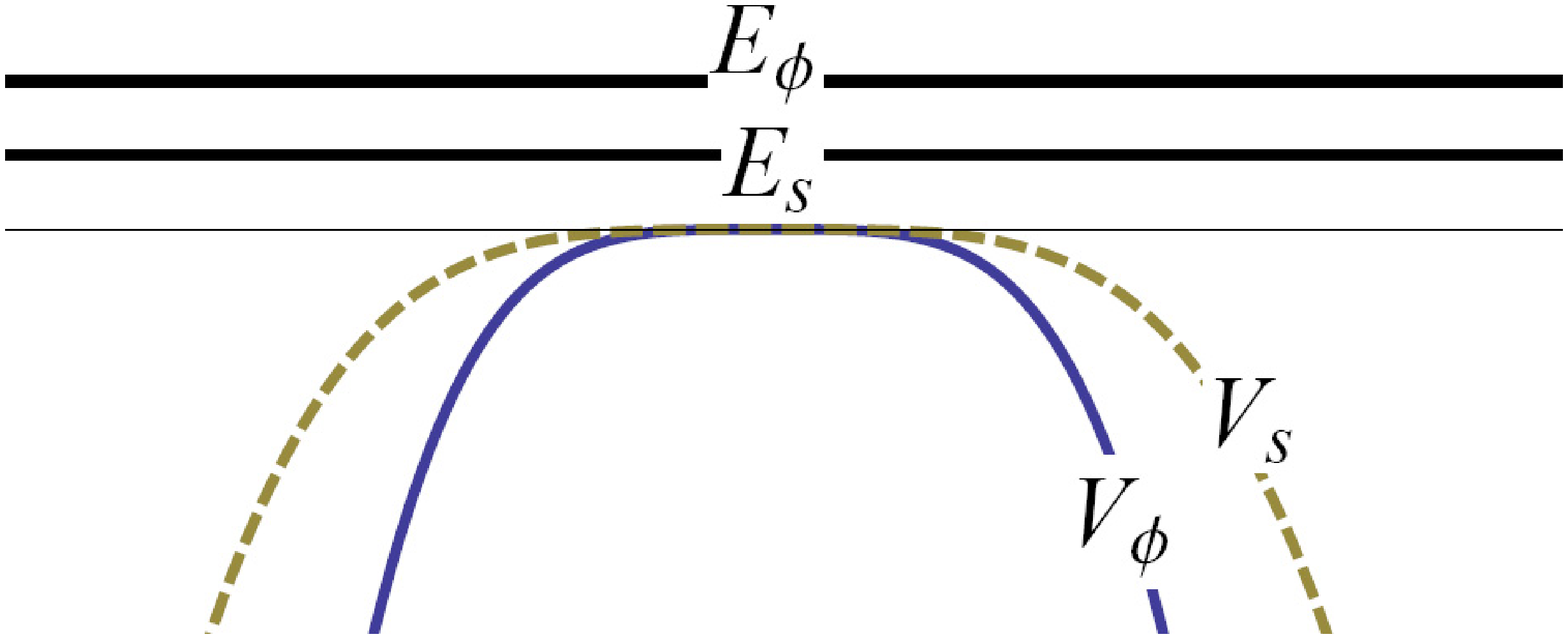}%
}%
\\
\text{FIG. 7}\\
b>0,\text{ }c<0,\text{ }E_{\phi}\geq E_{s}>0
\end{array}
$ & $%
\begin{tabular}
[c]{l}%
$\phi=\left(  \frac{E+\rho_{r}}{b}\right)  ^{1/4}\frac{sn\left(  \frac
{\tau+\delta}{T_{\phi}}|\frac{1}{2}\right)  }{1+cn\left(  \frac{\tau+\delta
}{T_{\phi}}|\frac{1}{2}\right)  },$ $T_{\phi}=\frac{1}{\sqrt{2}}\left(
16b\left(  E+\rho_{r}\right)  \right)  ^{-1/4}$\\
$s=\left(  \frac{E}{-c}\right)  ^{1/4}\frac{sn\left(  \frac{\tau}{T_{s}}%
|\frac{1}{2}\right)  }{1+cn\left(  \frac{\tau}{T_{s}}|\frac{1}{2}\right)  },$
$T_{s}=\frac{1}{\sqrt{2}}\left(  -16cE\right)  ^{-1/4}$%
\end{tabular}
\ \ \ \ $\\\hline\hline
\end{tabular}
\medskip%

\begin{tabular}
[c]{||c||l||}\hline\hline
$%
\begin{array}
[c]{c}%
{\includegraphics[
height=0.7887in,
width=1.9337in
]%
{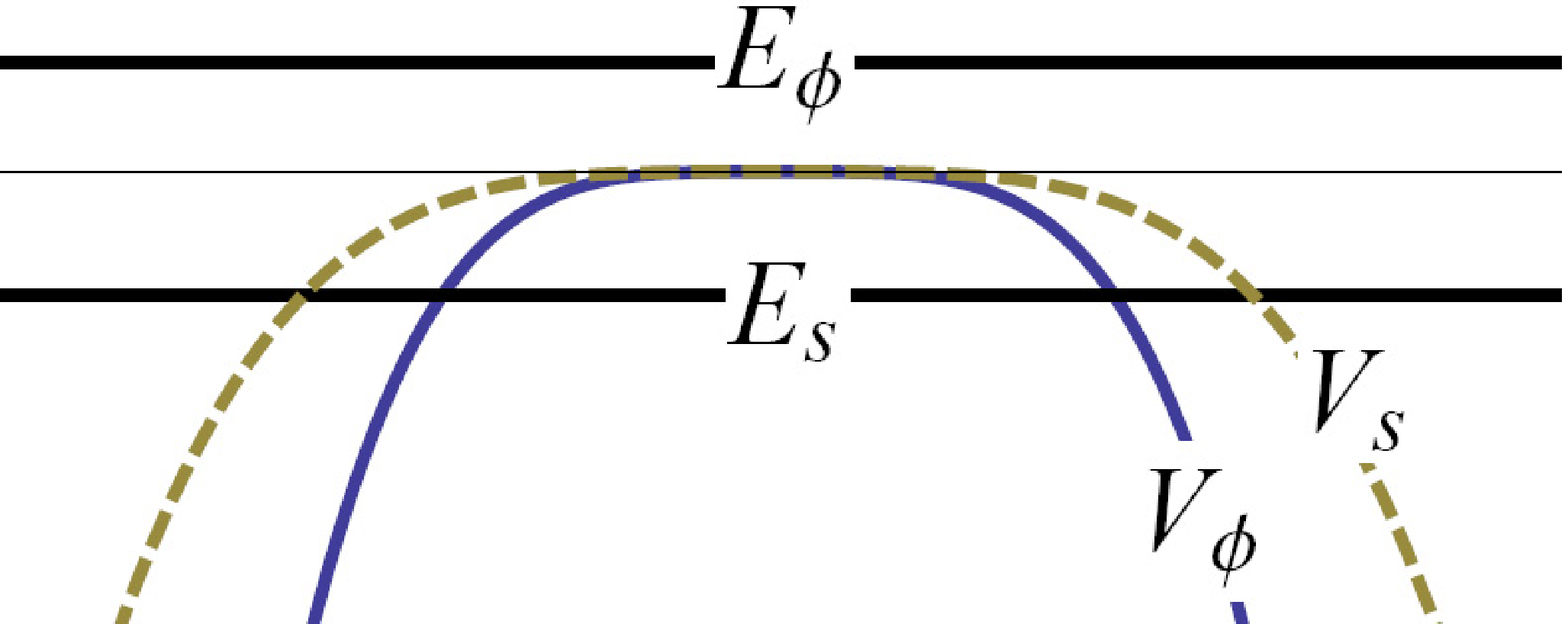}%
}%
\\
\text{FIG. 8}\\
b>0,\text{ }c<0,\text{ }E_{\phi}\geq E_{s}\\
E_{\phi}>0,\text{ }E_{s}<0
\end{array}
$ & $%
\begin{tabular}
[c]{l}%
$\phi=\left(  \frac{E+\rho_{r}}{b}\right)  ^{1/4}\frac{sn\left(  \frac
{\tau+\delta}{T_{\phi}}|\frac{1}{2}\right)  }{1+cn\left(  \frac{\tau+\delta
}{T_{\phi}}|\frac{1}{2}\right)  },$ $T_{\phi}=\frac{1}{\sqrt{2}}\left(
16b\left(  E+\rho_{r}\right)  \right)  ^{-1/4}$\\
$s=\left(  \frac{E}{c}\right)  ^{1/4}\frac{1}{cn\left(  \frac{\tau}{T_{s}%
}|\frac{1}{2}\right)  },$ $T_{s}=\left(  16cE\right)  ^{-1/4}$%
\end{tabular}
\ \ \ \ \ $\\\hline\hline
\end{tabular}
\medskip

$%
\begin{tabular}
[c]{||c||l||}\hline\hline
$%
\begin{array}
[c]{c}%
{\includegraphics[
height=0.7896in,
width=1.9441in
]%
{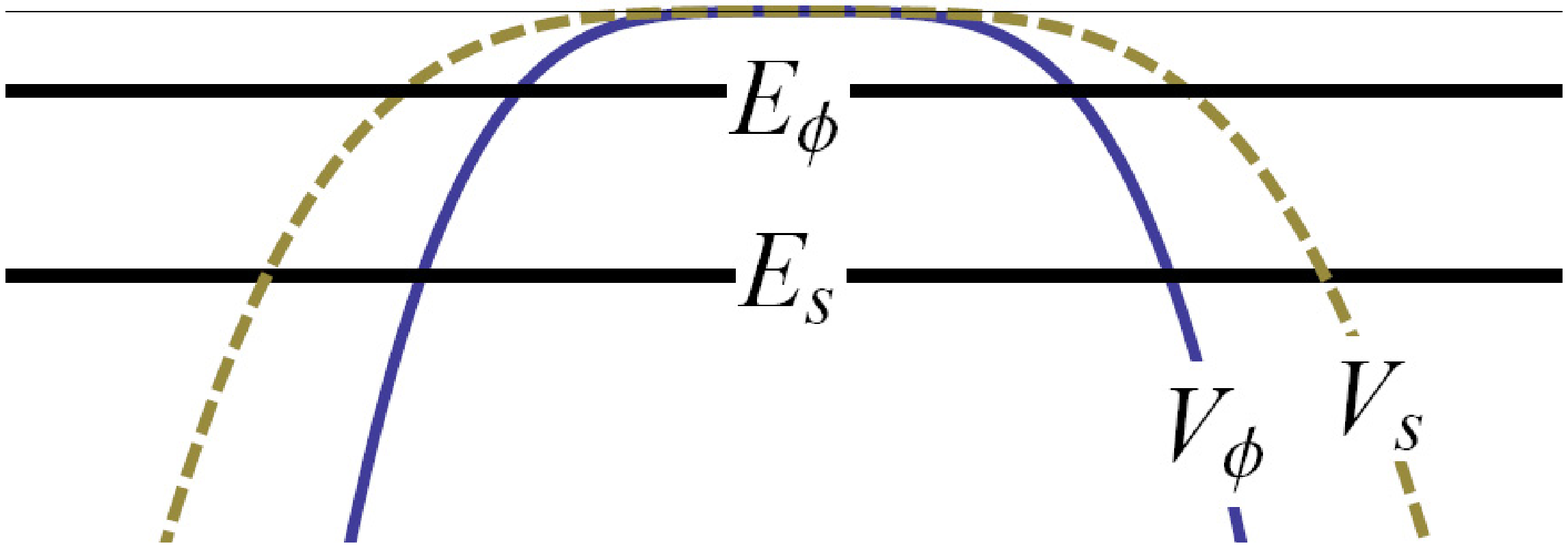}%
}%
\\
\text{FIG. 9}\\
b>0,\text{ }c<0,\text{ }0>E_{\phi}\geq E_{s}%
\end{array}
$ & $%
\begin{tabular}
[c]{l}%
$\phi=\left(  \frac{-\left(  E+\rho_{r}\right)  }{b}\right)  ^{1/4}\frac
{1}{cn\left(  \frac{\tau+\delta}{T_{\phi}}|\frac{1}{2}\right)  },$ \ $T_{\phi
}=\left(  -16b\left(  E+\rho_{r}\right)  \right)  ^{-1/4}$\\
$s=\left(  \frac{E}{c}\right)  ^{1/4}\frac{1}{cn\left(  \frac{\tau}{T_{s}%
}|\frac{1}{2}\right)  },$ \ $T_{s}=\left(  16cE\right)  ^{-1/4}$%
\end{tabular}
\ \ \ \ $\\\hline\hline
\end{tabular}
\ \ \ \ $\medskip

Note that as the parameters $b,c,E,\left(  E+\rho_{r}\right)  $ change signs
the corresponding solutions and physical behaviors change qualitatively.
Nevertheless, the mathematical expressions in Figs.(5-9) are related to each
other by the following rules for analytic continuation, where $x$ is real
\begin{align}
&  \left.
\begin{array}
[c]{l}%
\frac{1}{\sqrt{2}}sn\left(  \frac{e^{\pm i\pi/4}}{\sqrt{2}}x|\frac{1}%
{2}\right)  =sn\left(  x|\frac{1}{2}\right) \\
1+cn\left(  \frac{e^{\pm i\pi/4}}{\sqrt{2}}x|\frac{1}{2}\right)  =dn\left(
x|\frac{1}{2}\right)
\end{array}
\right\}  ,\;\text{relates }\left\{
\begin{array}
[c]{l}%
\text{Figs.(6) to (5),~}b\text{ flips sign}\\
\text{Figs.(7) to (6),~}c\text{ flips sign}%
\end{array}
\right. \\
&  \left.  cn\left(  xe^{\pm i\pi/4}|\frac{1}{2}\right)  =\frac{1+cn\left(
\frac{x}{\sqrt{2}}|\frac{1}{2}\right)  }{sn\left(  \frac{x}{\sqrt{2}}|\frac
{1}{2}\right)  }\right.  ,\;\text{relates }\left\{
\begin{array}
[c]{l}%
\text{Figs.(9) to (7),~}E\text{ flips sign}\\
\text{Figs.(9) to (8),~}(E+\rho)\text{ flips sign}%
\end{array}
\right.
\end{align}
So, it is possible to write a single formula to cover all the solutions (such
as the formulas in Fig.9, modified with appropriate absolute signs)
\begin{equation}%
\begin{tabular}
[c]{l}%
$\phi\left(  \tau\right)  =\left\vert \frac{E+\rho_{r}}{b}\right\vert
^{1/4}\left\{  cn\left[  \left(  -16b\left(  E+\rho_{r}\right)  \right)
^{1/4}\left(  \tau+\delta\right)  ~|~\frac{1}{2}\right]  \right\}  ^{-1},$\\
$s\left(  \tau\right)  =\left\vert \frac{E}{c}\right\vert ^{1/4}\left\{
cn\left[  \left(  16cE\right)  ^{1/4}\tau~|~\frac{1}{2}\right]  \right\}
^{-1},$%
\end{tabular}
\end{equation}
and then analytically continue the argument of the Jacobi elliptic functions
to obtain the other expressions. In this form all signs of the parameters
$b,c,E,\left(  E+\rho_{r}\right)  $ are permitted, thus capturing the physical
behavior of all corresponding regions of parameter space with a single
expression (when $K=0$). Under these flips of signs the functions $\phi,s$
remain real even though the argument of the function is complex. This unified
version is convenient to feed it to a computer to obtain plots of the solutions.

There is a similar analytic continuation for the cases with nonzero curvature
given below, but the formulas for analytic continuation are considerably more
involved, so we will not bother to discuss them.

\newpage

Next we have the $K>0$ cases, with eleven combinations which are listed in Figs.(10-19)

\bigskip%

\begin{tabular}
[c]{||c||l||}\hline\hline%
\begin{tabular}
[c]{c}%
{\includegraphics[
height=0.9046in,
width=1.836in
]%
{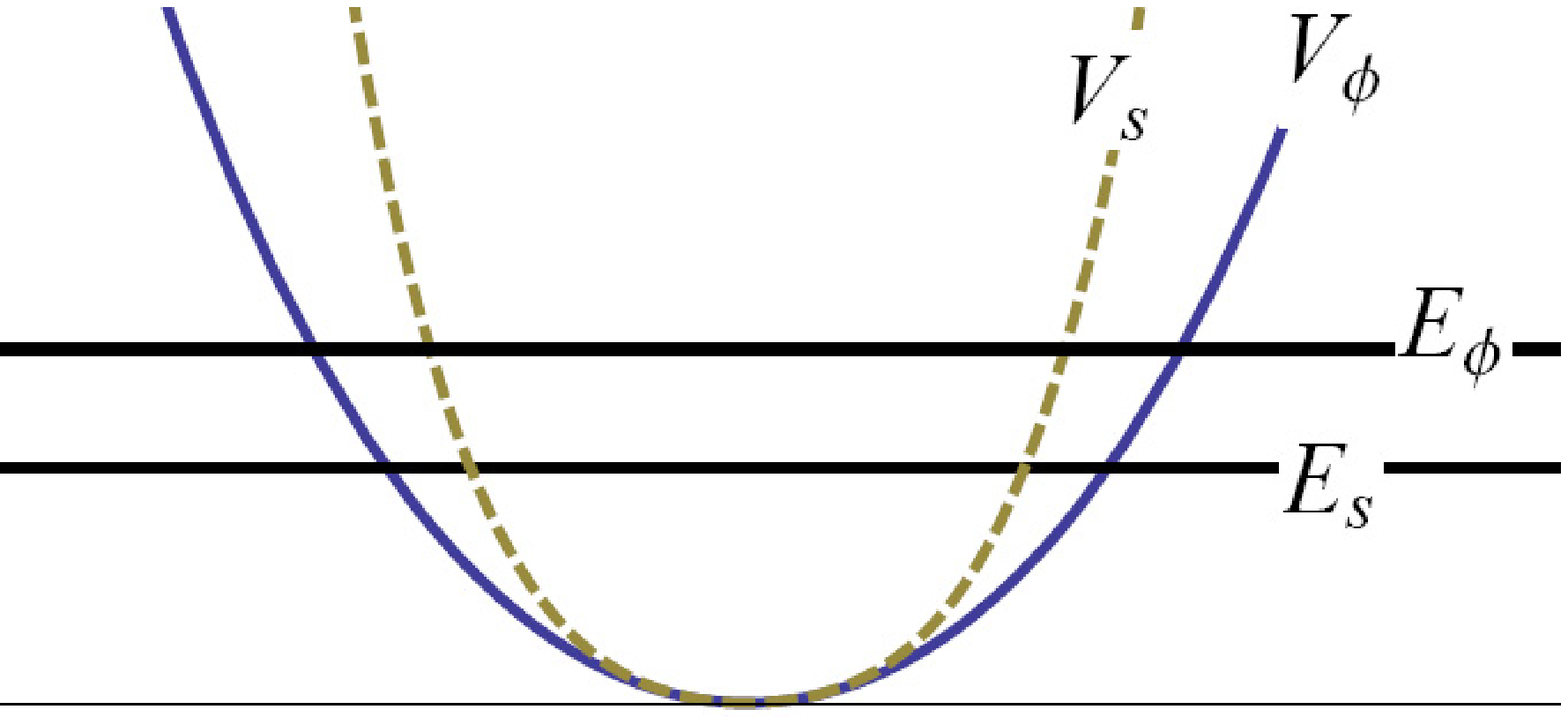}%
}%
\\
FIG. 10\\
$b<0,$ $c>0,E_{\phi}\geq E_{s}>0$%
\end{tabular}
& $%
\begin{array}
[c]{l}%
{\small \phi=}\sqrt{\frac{1-K^{2}T_{\phi}^{4}}{8\left\vert b\right\vert
T_{\phi}^{2}}}\frac{sn(\frac{\tau+\delta}{T_{\phi}}|m_{\phi})}{dn(\frac
{\tau+\delta}{T_{\phi}}|m_{\phi})}{\small ,}%
\begin{array}
[c]{l}%
m_{\phi}=\frac{1}{2}\left(  1-KT_{\phi}^{2}\right)  \leq\frac{1}{2}\\
T_{\phi}=\frac{1}{\sqrt{K}}\left(  1-\frac{16b}{K^{2}}\left(  E+\rho
_{r}\right)  \right)  ^{-1/4}%
\end{array}
\\
{\small s=}\sqrt{\frac{1-K^{2}T_{s}^{4}}{8cT_{s}^{2}}}\frac{sn(\frac{\tau
}{T_{s}}|m_{s})}{dn(\frac{\tau}{T_{s}}|m_{s})}{\small ,}%
\begin{array}
[c]{l}%
m_{s}=\frac{1}{2}\left(  1-KT_{s}^{2}\right)  \leq\frac{1}{2}\\
T_{s}=\frac{1}{\sqrt{K}}\left(  1+\frac{16Ec}{K^{2}}\right)  ^{-1/4}%
\end{array}
\end{array}
$\\\hline\hline
\end{tabular}
\medskip%

\begin{tabular}
[c]{||c||l||}\hline\hline%
\begin{tabular}
[c]{c}%
{\includegraphics[
height=0.9115in,
width=1.8593in
]%
{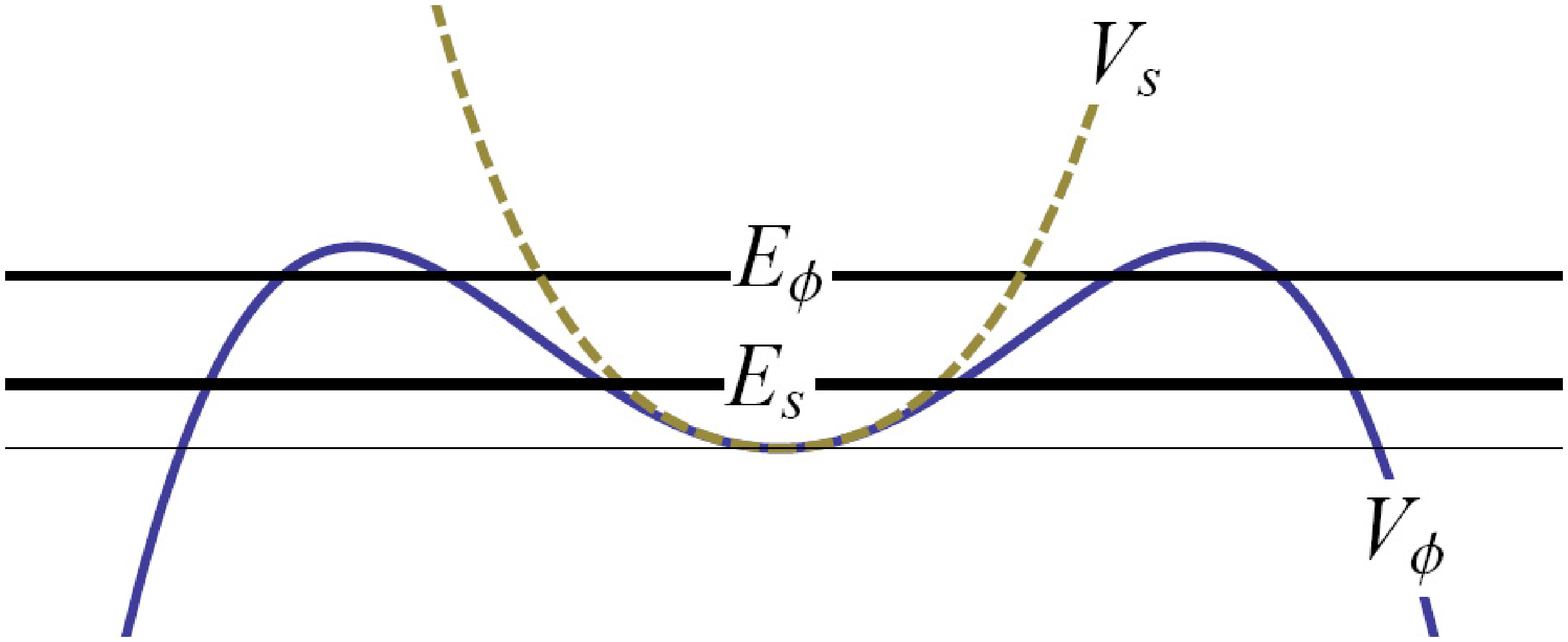}%
}%
\\
FIG. 11\\
$b>0,$ $c>0,E_{\phi}\geq E_{s}>0$\\
$E_{\phi}<\frac{K^{2}}{16b},$ $\left\vert \phi\left(  0\right)  \right\vert
<\sqrt{\frac{K}{4b}}$%
\end{tabular}
& $%
\begin{array}
[c]{l}%
{\small \phi=}\sqrt{\frac{KT_{\phi}^{2}-1}{2bT_{\phi}}}{\small sn}\left(
\frac{\tau+\delta}{T_{\phi}}|m_{\phi}\right)  {\small ,}%
\begin{array}
[c]{l}%
m_{\phi}=KT_{\phi}^{2}-1\leq1\\
T_{\phi}=\frac{\sqrt{2}}{\sqrt{K}}\left(  1+\sqrt{1-\frac{16b\left(
E+\rho_{r}\right)  }{K^{2}}}\right)  ^{-1/2}%
\end{array}
\\
{\small s=}\sqrt{\frac{1-K^{2}T_{s}^{4}}{8cT_{s}^{2}}}\frac{sn(\frac{\tau
}{T_{s}}|m_{s})}{dn(\frac{\tau}{T_{s}}|m_{s})}{\small ,}%
\begin{array}
[c]{l}%
m_{s}=\frac{1}{2}\left(  1-KT_{s}^{2}\right)  \leq\frac{1}{2}\\
T_{s}=\frac{1}{\sqrt{K}}\left(  1+\frac{16Ec}{K^{2}}\right)  ^{-1/4}%
\end{array}
\end{array}
$\\\hline\hline
\end{tabular}
\medskip%

\begin{tabular}
[c]{||c||l||}\hline\hline%
\begin{tabular}
[c]{c}%
{\includegraphics[
height=0.9349in,
width=1.9009in
]%
{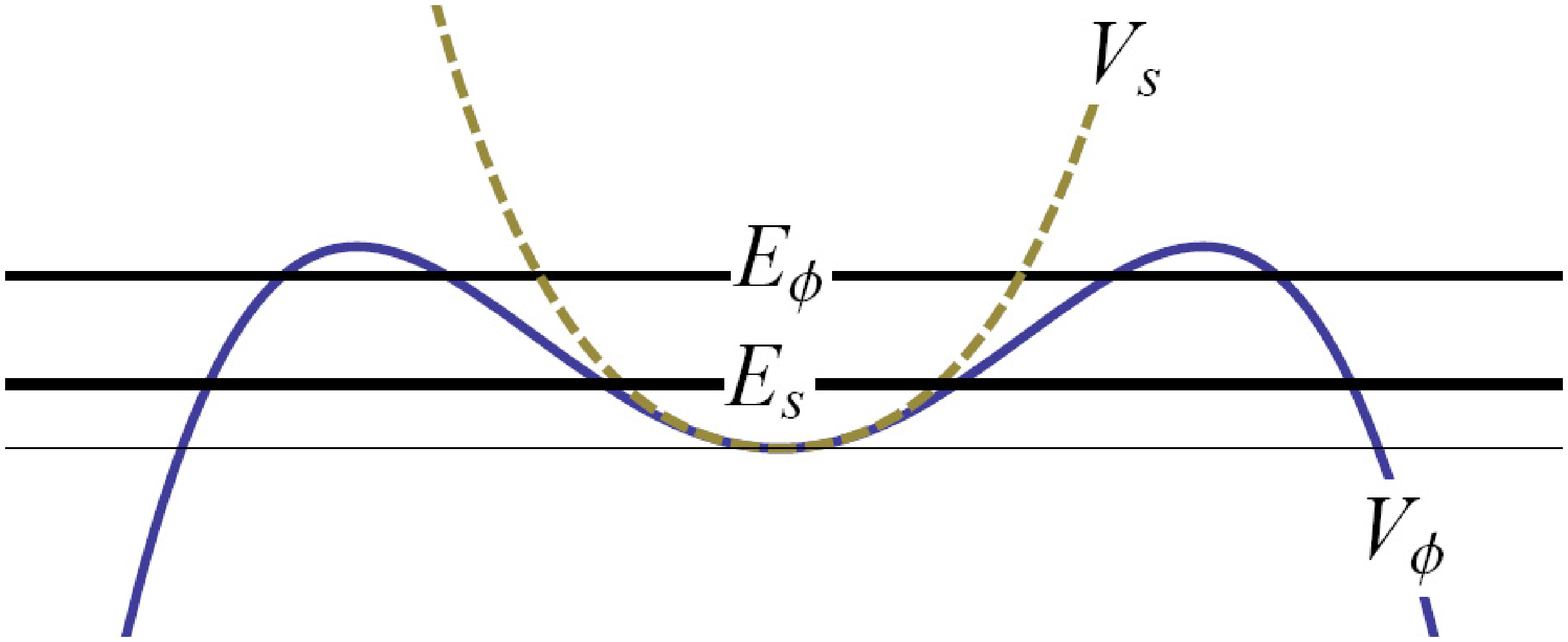}%
}%
\\
FIG. 12\\
$b>0,$ $c>0,E_{\phi}\geq E_{s}>0$\\
$E_{\phi}<\frac{K^{2}}{16b},$ $\left\vert \phi\left(  0\right)  \right\vert
>\sqrt{\frac{K}{4b}}$%
\end{tabular}
& $%
\begin{array}
[c]{l}%
{\small \phi=}\sqrt{\frac{KT_{\phi}^{2}+1}{4bT_{\phi}^{2}}}\frac{1}{cn\left(
\frac{\tau+\delta}{T_{\phi}}|m_{\phi}\right)  }{\small ,}%
\begin{array}
[c]{l}%
m_{\phi}=\frac{1}{2}\left(  1-KT_{\phi}^{2}\right)  \leq0\\
T_{\phi}=\frac{1}{\sqrt{K}}\left(  1-\frac{16b}{K^{2}}\left(  E+\rho
_{r}\right)  \right)  ^{-1/4}%
\end{array}
\\
{\small s=}\sqrt{\frac{1-K^{2}T_{s}^{4}}{8cT_{s}^{2}}}\frac{sn(\frac{\tau
}{T_{s}}|m_{s})}{dn(\frac{\tau}{T_{s}}|m_{s})}{\small ,}%
\begin{array}
[c]{l}%
m_{s}=\frac{1}{2}\left(  1-KT_{s}^{2}\right)  \leq\frac{1}{2}\\
T_{s}=\left(  K^{2}+16cE\right)  ^{-1/4}%
\end{array}
\end{array}
$\\\hline\hline
\end{tabular}
\medskip%

\begin{tabular}
[c]{||c||l||}\hline\hline
$%
\begin{tabular}
[c]{c}%
$%
{\includegraphics[
height=0.7619in,
width=1.8706in
]%
{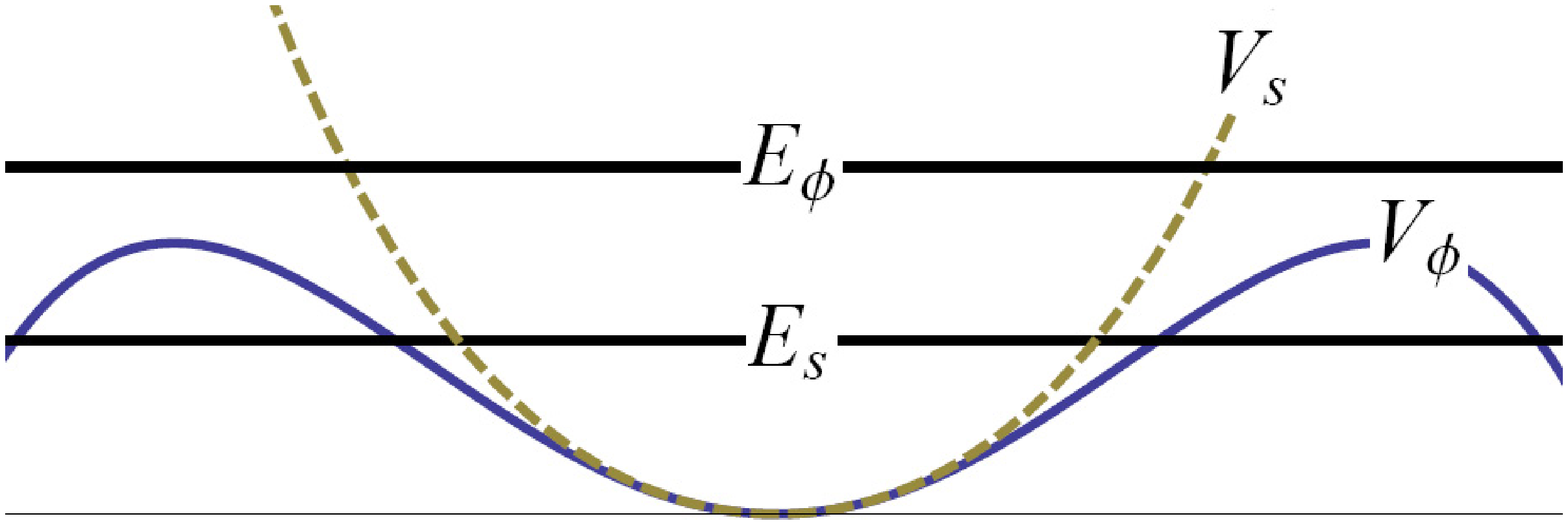}%
}%
$\\
$\text{FIG. }$13\\
$\text{\ }b>0,\text{ }c>0,$\\
$E_{\phi}\geq E_{s}>0,$ \ $E_{\phi}>\frac{K^{2}}{16b}$%
\end{tabular}
\ \ \ \ \ $ & $%
\begin{array}
[c]{l}%
\phi\left(  \tau\right)  =\left(  \frac{E+\rho_{r}}{b}\right)  ^{1/4}%
\frac{sn\left(  \frac{\tau+\delta}{T_{\phi}}|m_{\phi}\right)  }{1+cn\left(
\frac{\tau+\delta}{T_{\phi}}|m_{\phi}\right)  },%
\begin{array}
[c]{l}%
m_{\phi}=\frac{1}{2}+KT_{\phi}^{2}\leq1\\
T_{\phi}=\left(  64b\left(  E+\rho_{r}\right)  \right)  ^{-1/4}%
\end{array}
\\
s\left(  \tau\right)  =\sqrt{\frac{1-K^{2}T_{s}^{4}}{8cT_{s}^{2}}}%
\frac{sn(\frac{\tau}{T_{s}}|m_{s})}{dn(\frac{\tau}{T_{s}}|m_{s})},%
\begin{array}
[c]{l}%
m_{s}=\frac{1}{2}\left(  1-KT_{s}^{2}\right)  \leq\frac{1}{2}\\
T_{s}=\left(  K^{2}+16cE\right)  ^{-1/4}%
\end{array}
\end{array}
$\\\hline\hline
\end{tabular}
\medskip%

\begin{tabular}
[c]{||c||l||}\hline\hline%
\begin{tabular}
[c]{c}%
$%
{\includegraphics[
height=0.5984in,
width=1.7582in
]%
{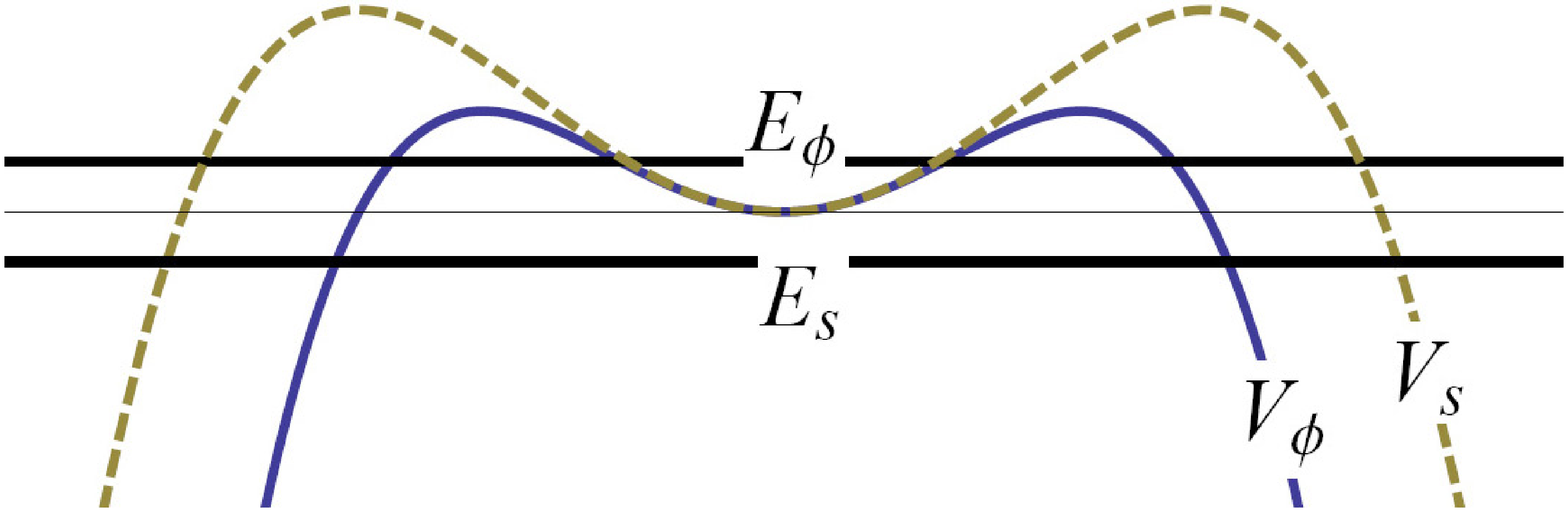}%
}%
$\\
$\text{FIG. }$14\\
$b>0,\text{ }c<0,$\\
$E_{s}\leq0\leq E_{\phi}<\frac{K^{2}}{16b}$\\
$\left\vert \phi\left(  0\right)  \right\vert >\sqrt{\frac{K}{4b}},\left\vert
s\left(  0\right)  \right\vert >\sqrt{\frac{K}{4\left\vert c\right\vert }}$%
\end{tabular}
& $%
\begin{tabular}
[c]{l}%
$\phi=\sqrt{\frac{KT_{\phi}^{2}+1}{4bT_{\phi}^{2}}}\frac{1}{cn\left(
\frac{\tau+\delta}{T_{\phi}}|m_{\phi}\right)  },%
\begin{array}
[c]{l}%
m_{\phi}=\frac{1}{2}\left(  1-KT_{\phi}^{2}\right)  \leq0\\
T_{\phi}=\left(  K^{2}-16b\left(  E+\rho_{r}\right)  \right)  ^{-1/4}%
\end{array}
$\\
$s=\sqrt{\frac{KT_{s}^{2}+1}{4\left\vert c\right\vert T_{s}^{2}}}\frac
{1}{cn\left(  \frac{\tau}{T_{s}}|m_{s}\right)  },%
\begin{array}
[c]{l}%
m_{s}=\frac{1}{2}\left(  1-KT_{s}^{2}\right)  \leq\frac{1}{2}\\
T_{s}=\left(  K^{2}+16\left\vert c\right\vert \left\vert E\right\vert \right)
^{-1/4}%
\end{array}
$%
\end{tabular}
\ \ $\\\hline\hline
\end{tabular}
\medskip%

\begin{tabular}
[c]{||c||l||}\hline\hline%
\begin{tabular}
[c]{c}%
$%
{\includegraphics[
height=0.7074in,
width=1.8343in
]%
{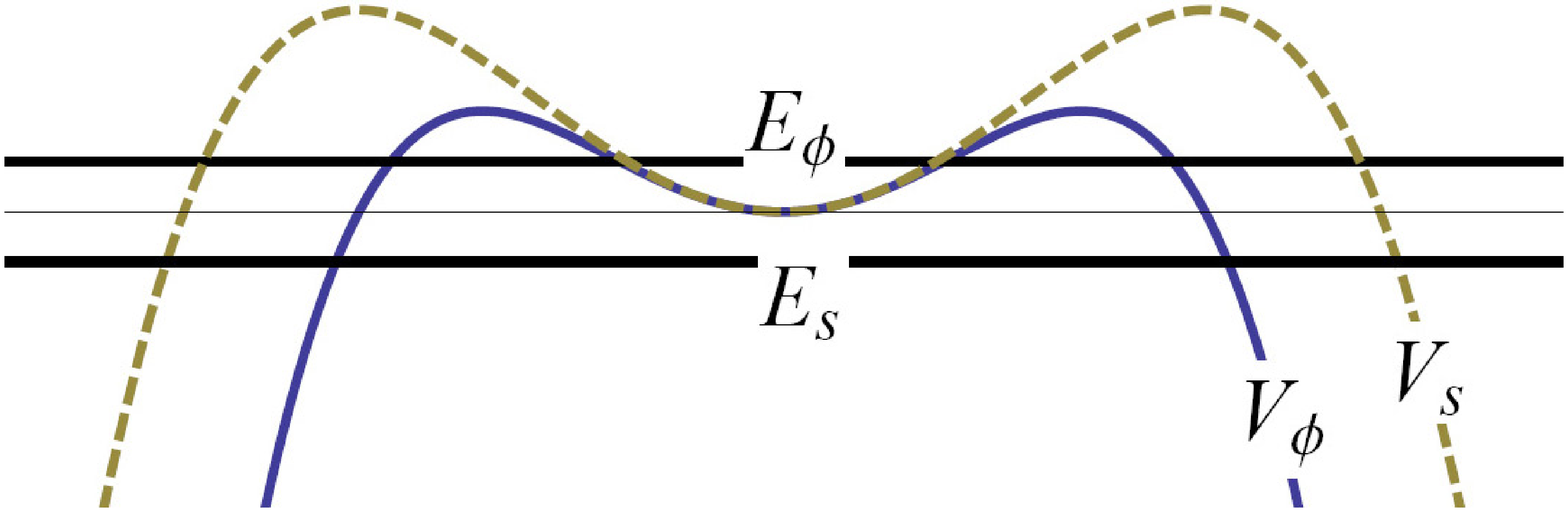}%
}%
$\\
$\text{FIG. 1}$5\\
$b>0,\text{ }c<0,$\\
$E_{\phi}\geq E_{s},$ $E_{\phi}<\frac{K^{2}}{16b},$\\
$\left\vert \phi\left(  0\right)  \right\vert <\sqrt{\frac{K}{4b}},\left\vert
s\left(  0\right)  \right\vert >\sqrt{\frac{K}{4\left\vert c\right\vert }}$%
\end{tabular}
& $%
\begin{tabular}
[c]{l}%
$\phi=\sqrt{\frac{KT_{\phi}^{2}-1}{2bT_{\phi}}}sn\left(  \frac{\tau+\delta
}{T_{\phi}}|m_{\phi}\right)  ,%
\begin{array}
[c]{l}%
m_{\phi}=KT_{\phi}^{2}-1\leq1\\
T_{\phi}=\left(  \frac{K}{2}+\sqrt{\frac{K^{2}}{4}-4b\left(  E+\rho
_{r}\right)  }\right)  ^{-\frac{1}{2}}%
\end{array}
$\\
$s=\sqrt{\frac{KT_{s}^{2}+1}{4\left\vert c\right\vert T_{s}^{2}}}\frac
{1}{cn\left(  \frac{\tau}{T_{s}}|m_{s}\right)  },%
\begin{array}
[c]{l}%
m_{s}=\frac{1}{2}\left(  1-KT_{s}^{2}\right)  \leq\frac{1}{2}\\
T_{s}=\left(  K^{2}-16\left\vert c\right\vert E\right)  ^{-1/4}%
\end{array}
$%
\end{tabular}
\ $\\\hline\hline
\end{tabular}
\medskip%

\begin{tabular}
[c]{||c||l||}\hline\hline%
\begin{tabular}
[c]{c}%
$%
{\includegraphics[
height=0.6685in,
width=1.8645in
]%
{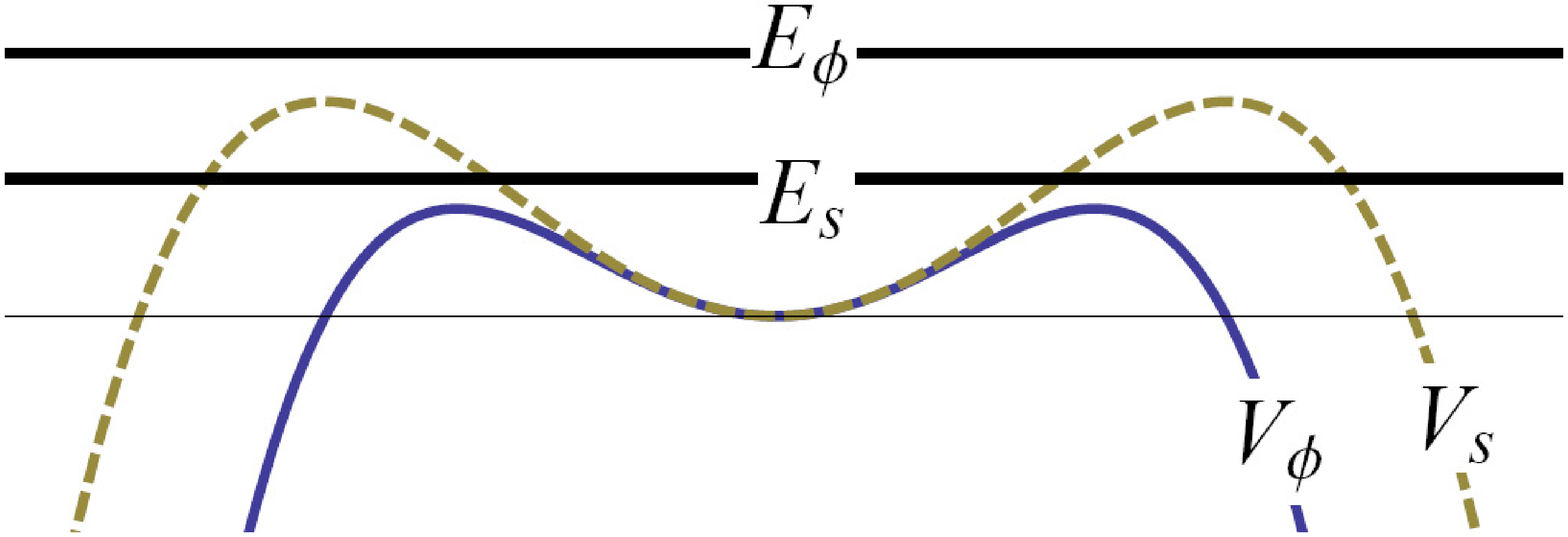}%
}%
$\\
$\text{FIG. 1}$6\\
\ $b>0,\text{ }c<0,\text{ }E_{\phi}\geq E_{s}$ \ \\
$E_{\phi}>\frac{K^{2}}{16b},\text{ }E_{s}<\frac{K^{2}}{16\left\vert
c\right\vert }$\\
$\left\vert s\left(  0\right)  \right\vert >\sqrt{\frac{K}{4\left\vert
c\right\vert }}$%
\end{tabular}
& $%
\begin{tabular}
[c]{l}%
$\phi\left(  \tau\right)  =\left(  \frac{E+\rho_{r}}{b}\right)  ^{1/4}%
\frac{sn\left(  \frac{\tau+\delta}{T_{\phi}}|m_{\phi}\right)  }{1+cn\left(
\frac{\tau+\delta}{T_{\phi}}|m_{\phi}\right)  },%
\begin{array}
[c]{l}%
m_{\phi}=\frac{1}{2}+KT_{\phi}^{2}\leq1\\
T_{\phi}=\left(  64b\left(  E+\rho_{r}\right)  \right)  ^{-1/4}%
\end{array}
$\\
$s\left(  \tau\right)  =\sqrt{\frac{KT_{s}^{2}+1}{4\left\vert c\right\vert
T_{s}^{2}}}\frac{1}{cn\left(  \frac{\tau}{T_{s}}|m_{s}\right)  },%
\begin{array}
[c]{l}%
m_{s}=\frac{1}{2}\left(  1-KT_{s}^{2}\right)  \leq0\\
T_{s}=\left(  K^{2}-16\left\vert c\right\vert E\right)  ^{-1/4}%
\end{array}
$%
\end{tabular}
\ $\\\hline\hline
\end{tabular}
\medskip%

\begin{tabular}
[c]{||c||l||}\hline\hline%
\begin{tabular}
[c]{c}%
$%
{\includegraphics[
height=0.6685in,
width=1.8645in
]%
{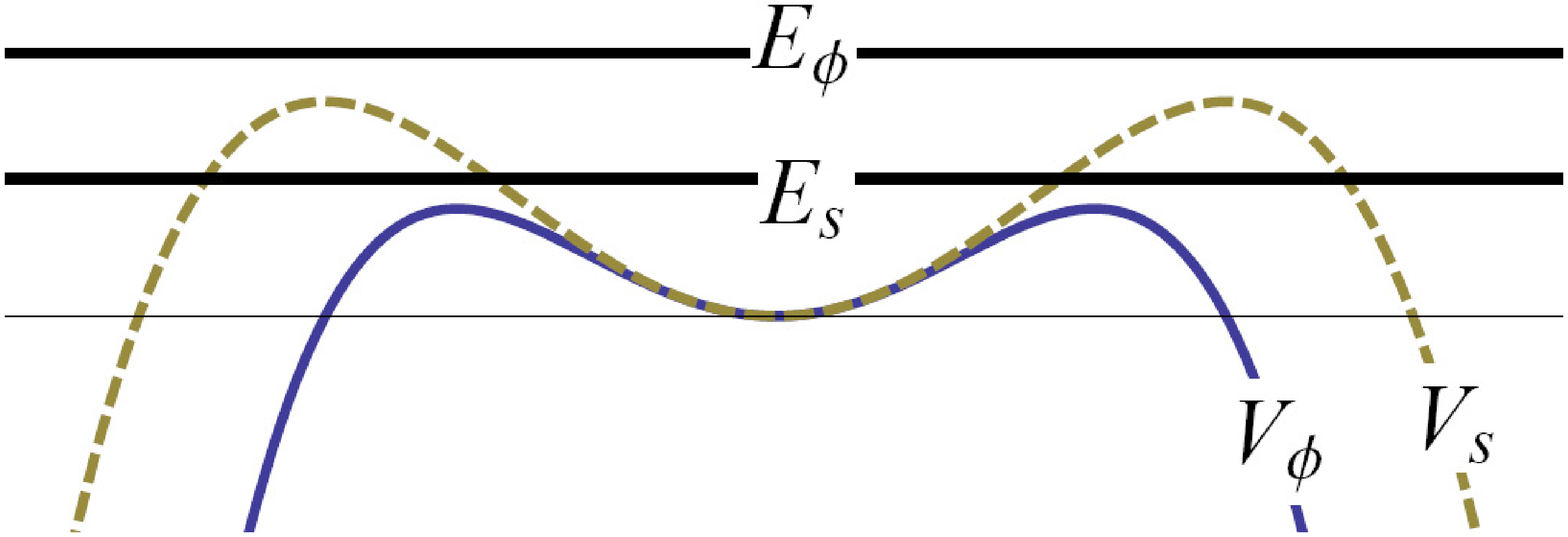}%
}%
$\\
$\text{FIG. 1}$7\\
$\ \ b>0,\text{ }c<0,\text{ }E_{\phi}\geq E_{s}$ \ \\
$E_{\phi}>\frac{K^{2}}{16b},\text{ }0<E_{s}<\frac{K^{2}}{16\left\vert
c\right\vert }$\\
$\left\vert s\left(  0\right)  \right\vert <\sqrt{\frac{K}{4\left\vert
c\right\vert }}$%
\end{tabular}
& $%
\begin{tabular}
[c]{l}%
$\phi=\left(  \frac{E+\rho_{r}}{b}\right)  ^{1/4}\frac{sn\left(  \frac
{\tau+\delta}{T_{\phi}}|m_{\phi}\right)  }{1+cn\left(  \frac{\tau+\delta
}{T_{\phi}}|m_{\phi}\right)  },%
\begin{array}
[c]{l}%
m_{\phi}=\frac{1}{2}+KT_{\phi}^{2}\leq1\\
T_{\phi}=\left(  64b\left(  E+\rho_{r}\right)  \right)  ^{-1/4}%
\end{array}
$\\
$s=\frac{\sqrt{KT_{s}^{2}-1}}{\sqrt{2\left\vert c\right\vert T_{s}}}sn\left(
\frac{\tau}{T_{s}}|m_{s}\right)  ,%
\begin{array}
[c]{l}%
m_{s}=KT_{s}^{2}-1\leq1\\
T_{s}=\left(  \frac{K}{2}+\sqrt{\frac{K^{2}}{4}-4\left\vert c\right\vert
E}\right)  ^{-1/2}%
\end{array}
$%
\end{tabular}
$\\\hline\hline
\end{tabular}
\medskip

$%
\begin{tabular}
[c]{||c||l||}\hline\hline%
\begin{tabular}
[c]{c}%
$%
{\includegraphics[
height=0.665in,
width=1.855in
]%
{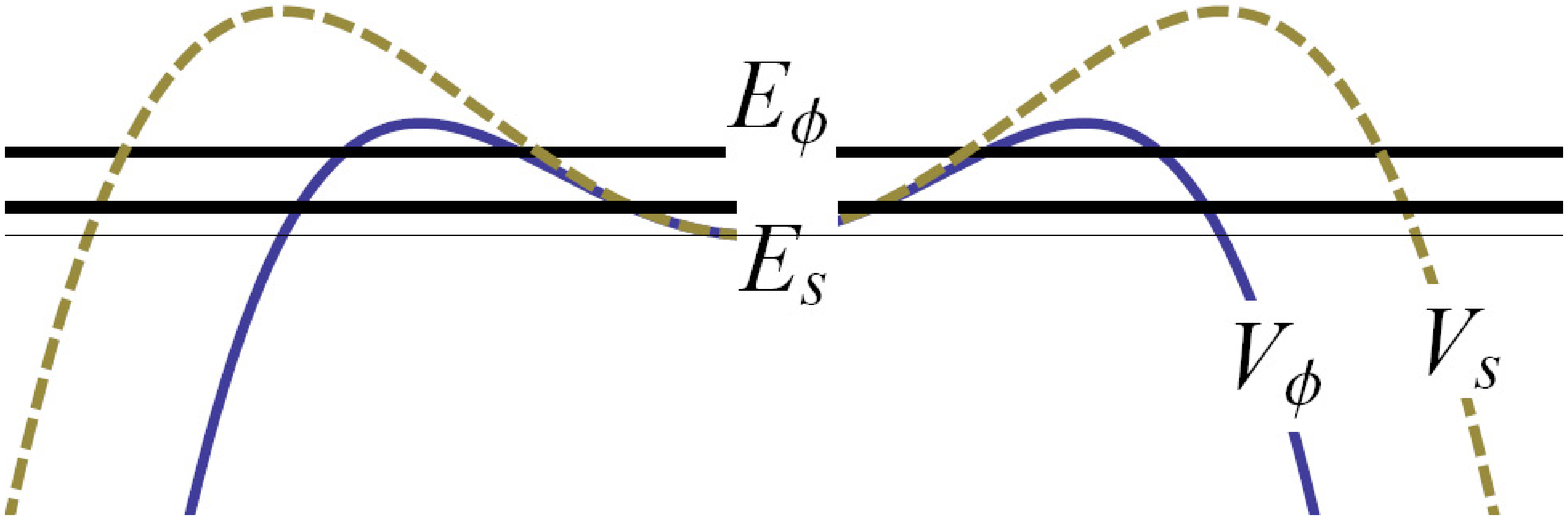}%
}%
$\\
$\text{FIG. 1}$8\\
$b>0,\text{ }c<0,\text{ }$\\
$E_{\phi}\geq E_{s}>0,$ $E_{\phi}<\frac{K^{2}}{16b},$\\
$\left\vert \phi\left(  0\right)  \right\vert >\sqrt{\frac{K}{4b}},\left\vert
s\left(  0\right)  \right\vert <\sqrt{\frac{K}{4b}}$%
\end{tabular}
& $%
\begin{tabular}
[c]{l}%
$\phi=\sqrt{\frac{KT_{\phi}^{2}+1}{4bT_{\phi}^{2}}}\frac{1}{cn\left(
\frac{\tau+\delta}{T_{\phi}}|m_{\phi}\right)  },$ \ $%
\begin{array}
[c]{l}%
m_{\phi}=\frac{1}{2}\left(  1-KT_{\phi}^{2}\right)  \leq0\\
T_{\phi}=\left(  K^{2}-16b\left(  E+\rho_{r}\right)  \right)  ^{-1/4}%
\end{array}
$\\
$s=\frac{\sqrt{KT_{s}^{2}-1}}{\sqrt{2\left\vert c\right\vert T_{s}}}sn\left(
\frac{\tau}{T_{s}}|m_{s}\right)  ,%
\begin{array}
[c]{l}%
m_{s}=KT_{s}^{2}-1\leq1\\
T_{s}=\left(  \frac{K}{2}+\sqrt{\frac{K^{2}}{4}-4\left\vert c\right\vert
E}\right)  ^{-1/2}%
\end{array}
$%
\end{tabular}
$\\\hline\hline
\end{tabular}
\ $%

\begin{tabular}
[c]{||c||l||}\hline\hline%
\begin{tabular}
[c]{c}%
$%
{\includegraphics[
height=0.6529in,
width=1.8204in
]%
{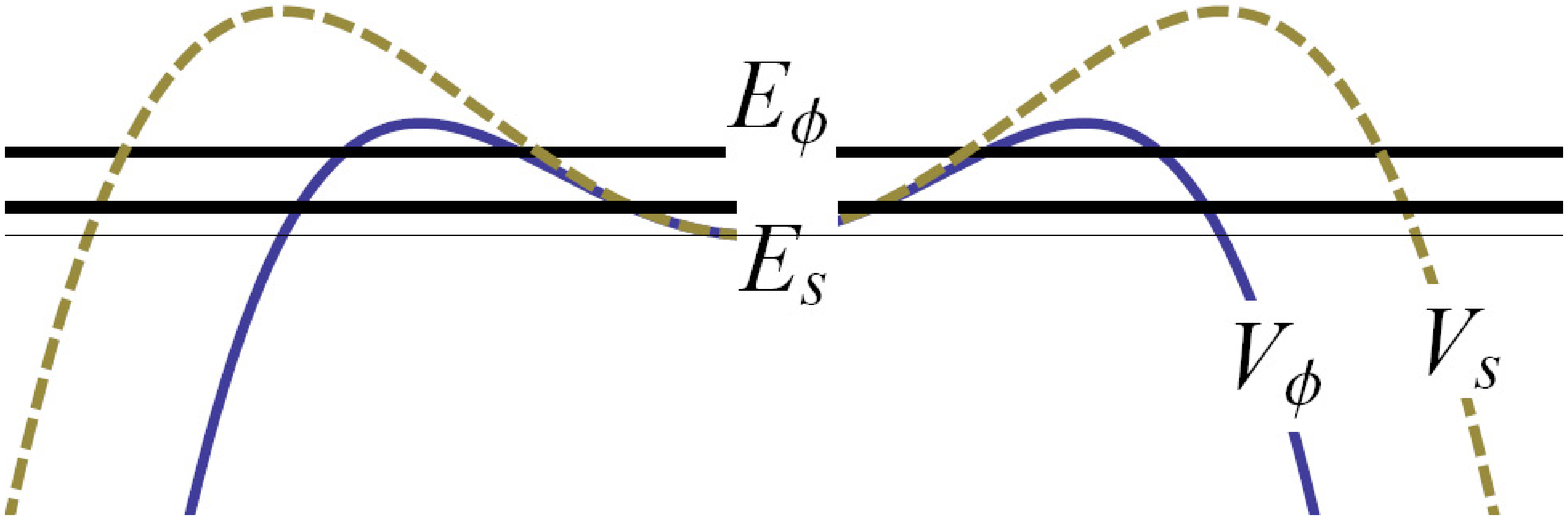}%
}%
$\\
$\text{FIG. 1}$9\\
$b>0,\text{ }c<0,\text{ }$\\
$E_{\phi}\geq E_{s},$ $0<E_{\phi}<\frac{K^{2}}{16b},$\\
$\left\vert \phi\left(  0\right)  \right\vert <\sqrt{\frac{K}{4b}},\left\vert
s\left(  0\right)  \right\vert <\sqrt{\frac{K}{4\left\vert c\right\vert }}$%
\end{tabular}
& $%
\begin{tabular}
[c]{l}%
$\phi=\sqrt{\frac{KT_{\phi}^{2}-1}{2bT_{\phi}}}sn\left(  \frac{\tau+\delta
}{T_{\phi}}|m_{\phi}\right)  ,%
\begin{array}
[c]{l}%
m_{\phi}=KT_{\phi}^{2}-1\leq1\\
T_{\phi}=\left(  \frac{K}{2}+\sqrt{\frac{K^{2}}{4}-4b\left(  E+\rho
_{r}\right)  }\right)  ^{-\frac{1}{2}}%
\end{array}
$\\
$s=\sqrt{\frac{KT_{s}^{2}-1}{2\left\vert c\right\vert T_{s}}}sn\left(
\frac{\tau}{T_{s}}|m_{s}\right)  ,%
\begin{array}
[c]{l}%
m_{s}=KT_{s}^{2}-1\leq1\\
T_{s}=\left(  \frac{K}{2}+\sqrt{\frac{K^{2}}{4}-4\left\vert c\right\vert
E}\right)  ^{-1/2}%
\end{array}
$%
\end{tabular}
$\\\hline\hline
\end{tabular}
\medskip%

\begin{tabular}
[c]{||c||l||}\hline\hline%
\begin{tabular}
[c]{c}%
$%
{\includegraphics[
height=0.8207in,
width=1.8593in
]%
{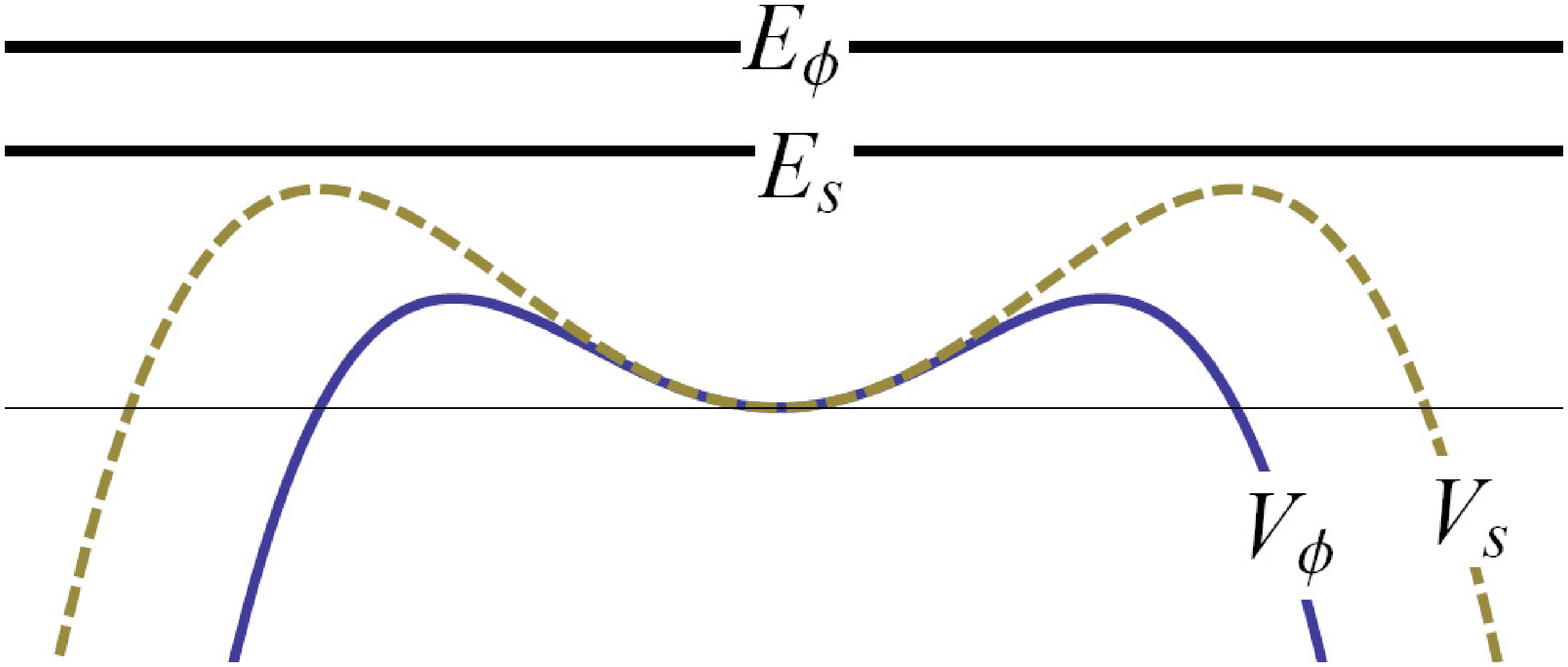}%
}%
$\\
$\text{FIG. }$20\\
$b>0,\text{ }c<0,$\\
$E_{\phi}\geq E_{s},\text{ }E_{s}>\frac{K^{2}}{16\left\vert c\right\vert }$%
\end{tabular}
& $%
\begin{tabular}
[c]{l}%
$\phi=\left(  \frac{E+\rho_{r}}{b}\right)  ^{1/4}\frac{sn\left(  \frac
{\tau+\delta}{T_{\phi}}|m_{\phi}\right)  }{1+cn\left(  \frac{\tau+\delta
}{T_{\phi}}|m_{\phi}\right)  },%
\begin{array}
[c]{l}%
m_{\phi}=\frac{1}{2}+KT_{\phi}^{2}\leq1\\
T_{\phi}=\left(  64b\left(  E+\rho_{r}\right)  \right)  ^{-1/4}%
\end{array}
$\\
$s=\left(  \frac{E}{\left\vert c\right\vert }\right)  ^{1/4}\frac{sn\left(
\frac{\tau}{T_{s}}|m_{s}\right)  }{1+cn\left(  \frac{\tau}{T_{s}}%
|m_{s}\right)  },%
\begin{array}
[c]{l}%
m_{s}=\frac{1}{2}+KT_{s}^{2}\leq1\\
T_{s}=\left(  64\left\vert c\right\vert E\right)  ^{-1/4}%
\end{array}
$%
\end{tabular}
$\\\hline\hline
\end{tabular}
\medskip

\newpage Finally, for $K<0$ there are nine combinations which are listed in Figs.(21-29)

\bigskip%
\begin{tabular}
[c]{||c||l||}\hline\hline
$%
\begin{array}
[c]{c}%
{\includegraphics[
height=0.7524in,
width=1.7469in
]%
{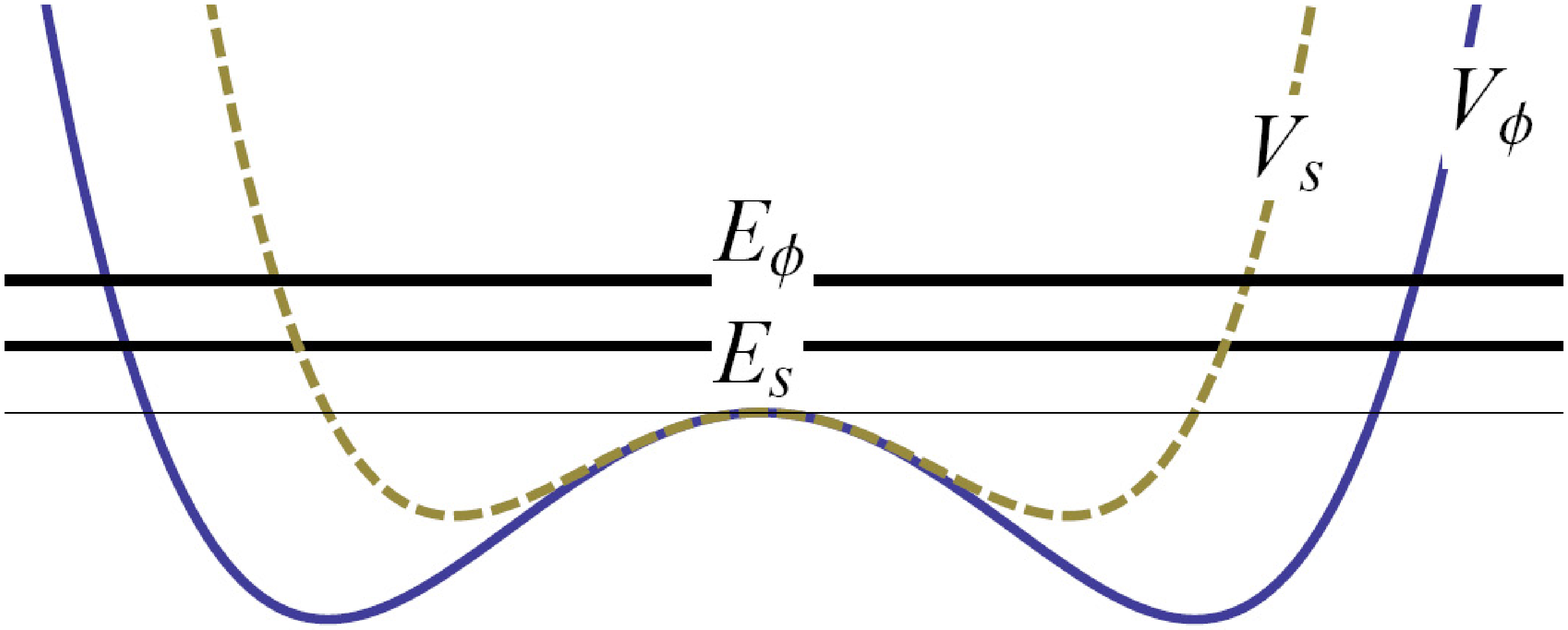}%
}%
\\
\text{FIG. 21}\\
b<0,\text{ }c>0,\text{ }E_{\phi}\geq E_{s}>0
\end{array}
$ & $%
\begin{array}
[c]{l}%
\phi=\sqrt{\frac{1-K^{2}T_{\phi}^{4}}{8\left\vert b\right\vert T_{\phi}^{2}}%
}\frac{sn\left(  \frac{\tau+\delta}{T_{\phi}}|m_{\phi}\right)  }{dn\left(
\frac{\tau+\delta}{T_{\phi}}|m_{\phi}\right)  },%
\begin{array}
[c]{l}%
m_{\phi}=\frac{1}{2}\left(  1+\left\vert K\right\vert T_{\phi}^{2}\right)
\leq1\\
T_{\phi}=\left(  K^{2}+16\left\vert b\right\vert \left(  E+\rho_{r}\right)
\right)  ^{-1/4}%
\end{array}
\\
s=\sqrt{\frac{1-K^{2}T_{s}^{4}}{8cT_{s}^{2}}}\frac{sn\left(  \frac{\tau}%
{T_{s}}|m_{s}\right)  }{dn\left(  \frac{\tau}{T_{s}}|m_{s}\right)  },%
\begin{array}
[c]{l}%
m_{s}=\frac{1}{2}\left(  1+\left\vert K\right\vert T_{s}^{2}\right)  \leq1\\
T_{s}=\left(  K^{2}+16cE\right)  ^{-1/4}%
\end{array}
\end{array}
$\\\hline\hline
\end{tabular}
\medskip%

\begin{tabular}
[c]{||c||l||}\hline\hline
$%
\begin{array}
[c]{c}%
{\includegraphics[
height=0.7135in,
width=1.8939in
]%
{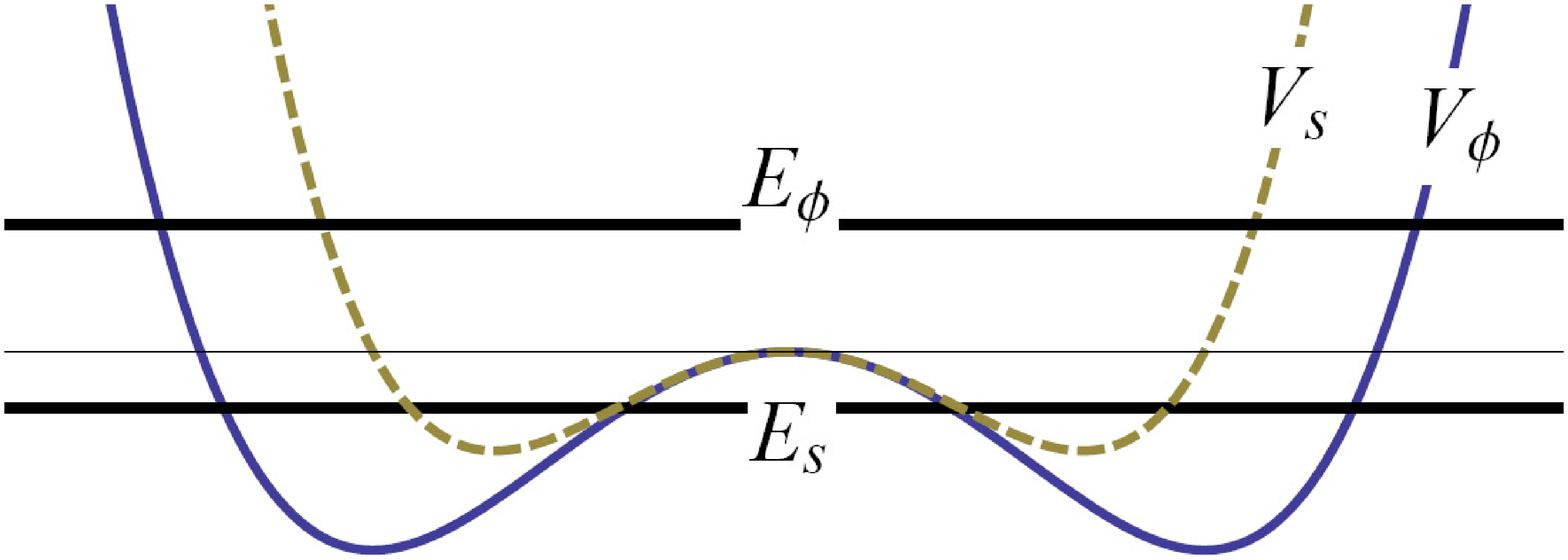}%
}%
\\
\text{FIG. 22}\\
b<0,\text{ }c>0,\\
\text{ }\frac{-K^{2}}{16c}\leq E_{s}\leq0\leq E_{\phi}%
\end{array}
$ & $%
\begin{array}
[c]{l}%
\phi=\sqrt{\frac{1-K^{2}T_{\phi}^{4}}{8\left\vert b\right\vert T_{\phi}^{2}}%
}\frac{sn\left(  \frac{\tau+\delta}{T_{\phi}}|m_{\phi}\right)  }{dn\left(
\frac{\tau+\delta}{T_{\phi}}|m_{\phi}\right)  },%
\begin{array}
[c]{l}%
m_{\phi}=\frac{1}{2}\left(  1+\left\vert K\right\vert T_{\phi}^{2}\right)
\leq1\\
T_{\phi}=\left(  K^{2}+16\left\vert b\right\vert \left(  E+\rho_{r}\right)
\right)  ^{-1/4}%
\end{array}
\\
s=\pm\sqrt{\frac{\left\vert K\right\vert T_{s}^{2}-1}{4cT_{s}^{2}}}dn\left(
\frac{\tau}{T_{s}}|m_{s}\right)  ,\
\begin{array}
[c]{l}%
m_{s}=(2-\frac{1}{2}\left\vert K\right\vert T_{s}^{2})\leq1\\
T_{s}=2\left(  \left\vert K\right\vert +\left(  K^{2}-16c\left\vert
E\right\vert \right)  ^{1/2}\right)  ^{-1/2}%
\end{array}
\end{array}
$\\\hline\hline
\end{tabular}
\medskip%

\begin{tabular}
[c]{||c||l||}\hline\hline
$%
\begin{array}
[c]{c}%
{\includegraphics[
height=0.6754in,
width=1.8049in
]%
{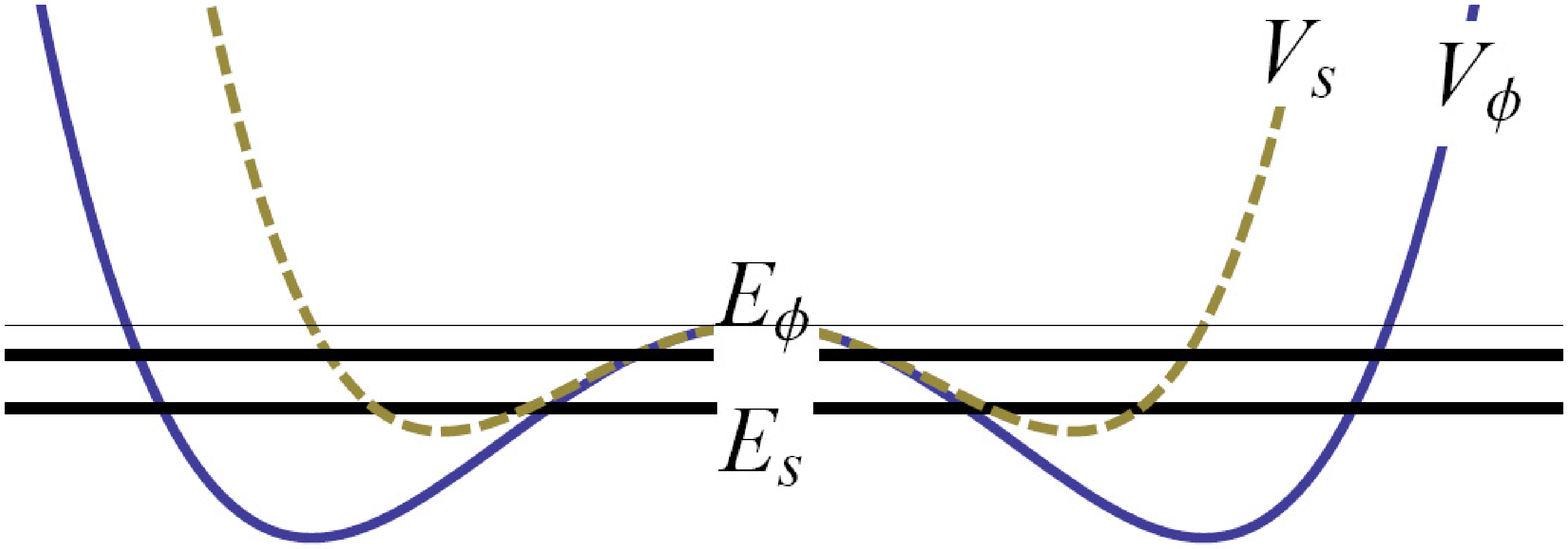}%
}%
\\
\text{FIG. 23}\\
b<0,\text{ }c>0,\text{ }\\
\frac{-K^{2}}{16c}\leq E_{s}\leq E_{\phi}\leq0
\end{array}
$ & $%
\begin{array}
[c]{l}%
\phi=\pm\sqrt{\frac{\left\vert K\right\vert T_{\phi}^{2}-1}{4\left\vert
b\right\vert T_{\phi}^{2}}}dn\left(  \frac{\tau}{T_{\phi}}|m_{\phi}\right)  ,%
\begin{array}
[c]{l}%
m_{\phi}=(2-\frac{1}{2}\left\vert K\right\vert T_{\phi}^{2})\leq1\\
T_{\phi}=2\left(  \left\vert K\right\vert +\sqrt{K^{2}-16\left\vert
b\right\vert \left\vert E+\rho_{r}\right\vert }\right)  ^{-1/2}%
\end{array}
\\
s=\pm\sqrt{\frac{\left\vert K\right\vert T_{s}^{2}-1}{4cT_{s}^{2}}}dn\left(
\frac{\tau}{T_{s}}|m_{s}\right)  ,\
\begin{array}
[c]{l}%
m_{s}=(2-\frac{1}{2}\left\vert K\right\vert T_{s}^{2})\leq1\\
T_{s}=2\left(  \left\vert K\right\vert +\sqrt{K^{2}-16c\left\vert E\right\vert
}\right)  ^{-1/2}%
\end{array}
\end{array}
$\\\hline\hline
\end{tabular}
\medskip%

\begin{tabular}
[c]{||c||l||}\hline\hline
$%
\begin{array}
[c]{c}%
{\includegraphics[
height=0.8389in,
width=1.7789in
]%
{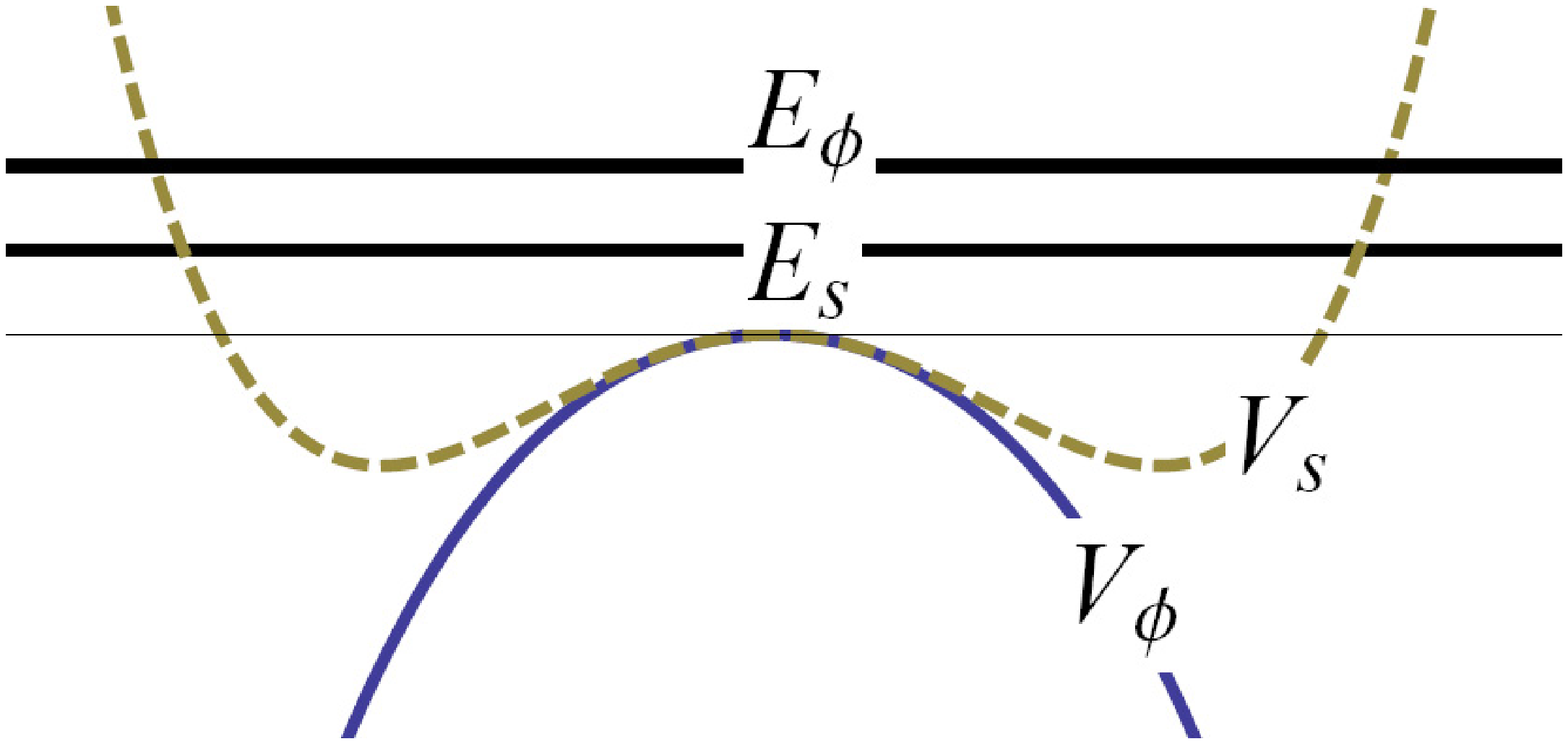}%
}%
\\
\text{FIG. 24}\\
b>0,\text{ }c>0,\text{ }E_{\phi}\geq E_{s}>0
\end{array}
$ & $%
\begin{array}
[c]{l}%
\phi=\sqrt{\frac{1}{8bT_{\phi}^{2}}}\frac{sn\left(  \frac{\tau+\delta}%
{T_{\phi}}|m_{\phi}\right)  }{1+cn\left(  \frac{\tau+\delta}{T_{\phi}}%
|m_{\phi}\right)  },%
\begin{array}
[c]{l}%
m_{\phi}=\frac{1}{2}-\left\vert K\right\vert T_{\phi}^{2}\leq\frac{1}{2}\\
T_{\phi}=\left(  64b\left(  E+\rho_{r}\right)  \right)  ^{-1/4}%
\end{array}
\\
s=\sqrt{\frac{1-K^{2}T_{s}^{4}}{8cT_{s}^{2}}}\frac{sn\left(  \frac{\tau}%
{T_{s}}|m_{s}\right)  }{dn\left(  \frac{\tau}{T_{s}}|m_{s}\right)  },%
\begin{array}
[c]{l}%
m_{s}=\frac{1}{2}\left(  1+\left\vert K\right\vert T_{s}^{2}\right)  \leq1\\
T_{s}=\left(  K^{2}+16cE\right)  ^{-1/4}%
\end{array}
\end{array}
$\\\hline\hline
\end{tabular}
\medskip%

\begin{tabular}
[c]{||c||l||}\hline\hline
$%
\begin{array}
[c]{c}%
{\includegraphics[
height=0.8086in,
width=1.8732in
]%
{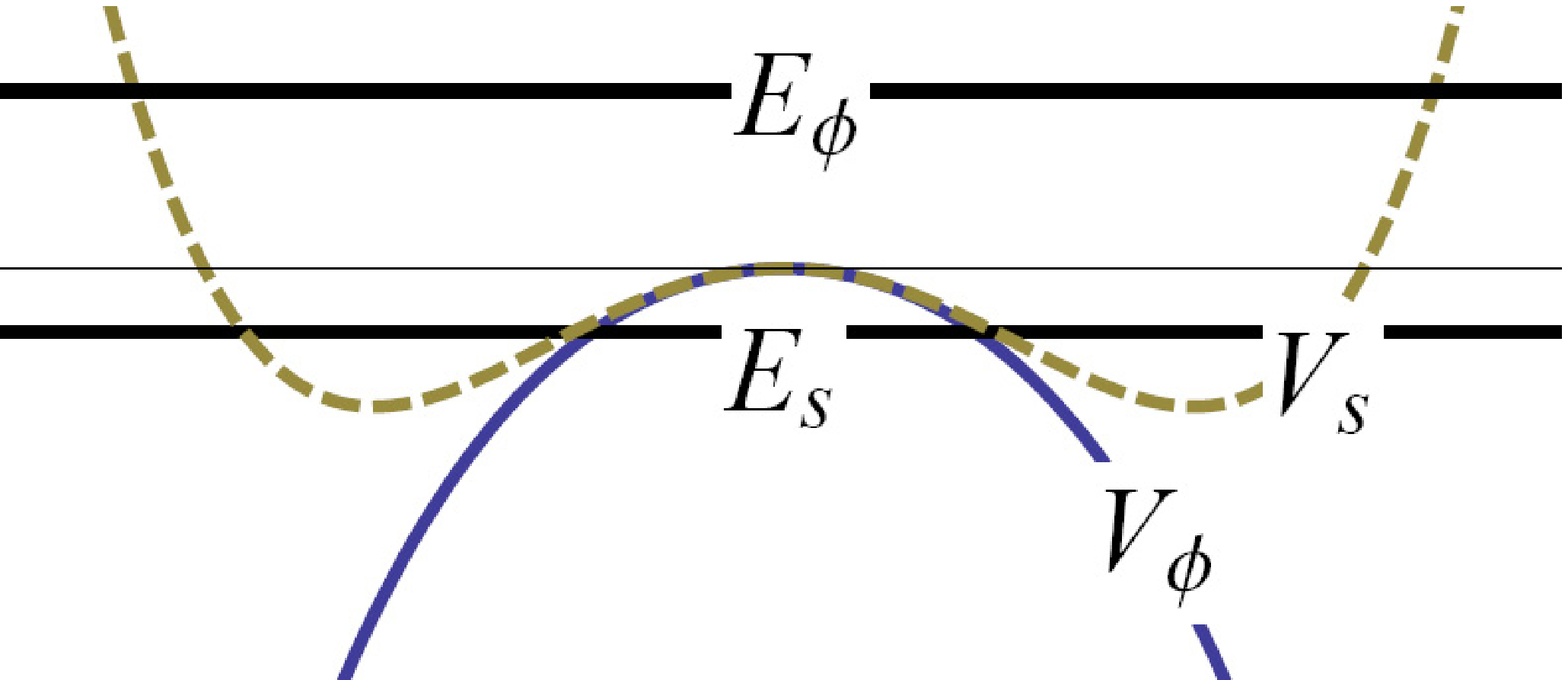}%
}%
\\
\text{FIG. 25}\\
b>0,\text{ }c>0,\text{ }E_{\phi}>E_{s}\\
E_{\phi}>0,\text{ }0>E_{s}>\frac{-K^{2}}{16c}%
\end{array}
$ & $%
\begin{array}
[c]{l}%
\phi=\left(  \frac{-\left\vert E\right\vert +\rho_{r}}{b}\right)  ^{1/4}%
\frac{sn\left(  \frac{\tau+\delta}{T_{\phi}}|m_{\phi}\right)  }{1+cn\left(
\frac{\tau+\delta}{T_{\phi}}|m_{\phi}\right)  },%
\begin{array}
[c]{l}%
m_{\phi}=\frac{1}{2}-\left\vert K\right\vert T_{\phi}^{2}\leq\frac{1}{2}\\
T_{\phi}=\left(  64b\left(  E+\rho_{r}\right)  \right)  ^{-1/4}%
\end{array}
\\
s=\pm\sqrt{\frac{\left\vert K\right\vert T_{s}^{2}-1}{4cT_{s}^{2}}}dn\left(
\frac{\tau}{T_{s}}|m_{s}\right)  ,\
\begin{array}
[c]{l}%
m_{s}=(2-\frac{1}{2}\left\vert K\right\vert T_{s}^{2})\leq1\\
T_{s}=2\left(  \left\vert K\right\vert +\sqrt{K^{2}-16c\left\vert E\right\vert
}\right)  ^{-1/2}%
\end{array}
\end{array}
$\\\hline\hline
\end{tabular}
\medskip%

\begin{tabular}
[c]{||c||l||}\hline\hline
$%
\begin{array}
[c]{c}%
{\includegraphics[
height=0.7152in,
width=1.836in
]%
{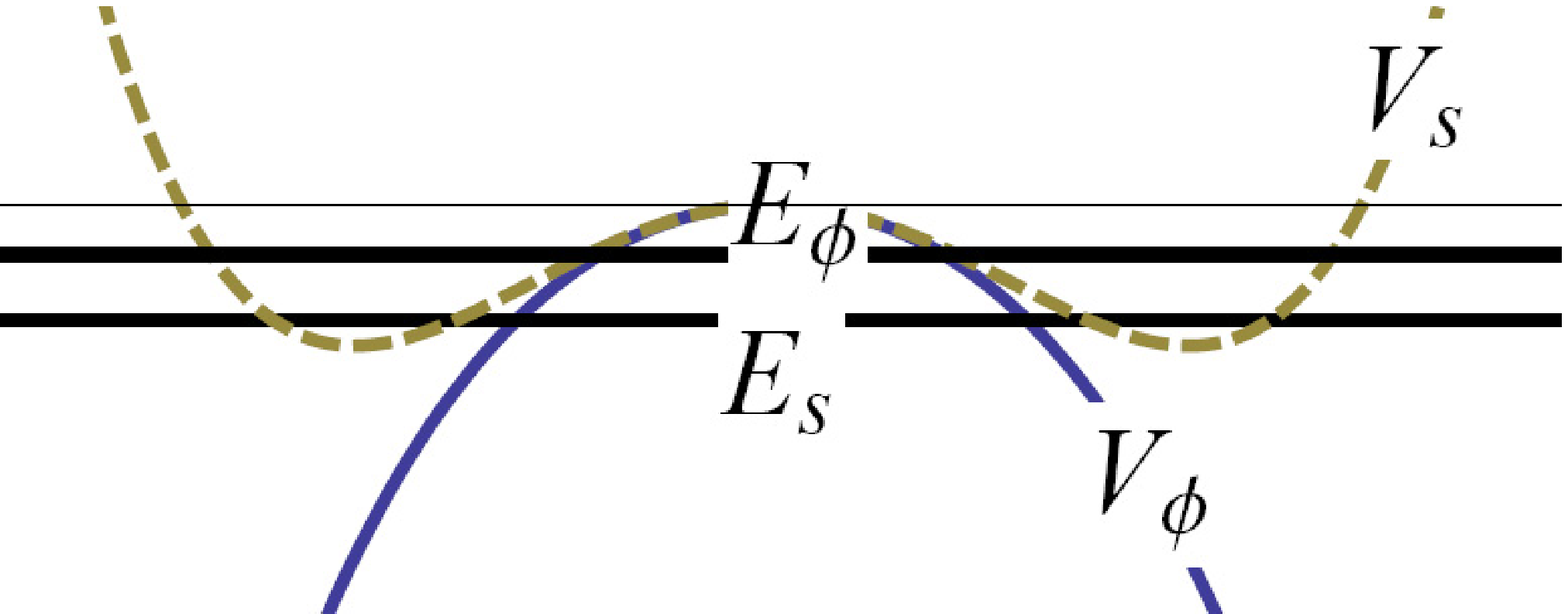}%
}%
\\
\text{FIG. 26}\\
b>0,\text{ }c>0,\text{ }\\
0>E_{\phi}\geq E_{s}>\frac{-K^{2}}{16c}%
\end{array}
$ & $%
\begin{array}
[c]{l}%
\phi=\sqrt{\frac{1-\left\vert K\right\vert T_{\phi}^{2}}{4bT_{\phi}^{2}}}%
\frac{1}{cn\left(  \frac{\tau+\delta}{T_{\phi}}|m_{\phi}\right)  },\
\begin{array}
[c]{l}%
m_{\phi}=\frac{1}{2}\left(  1+\left\vert K\right\vert T_{\phi}^{2}\right)
\leq1\\
T_{\phi}=\left(  K^{2}+16b\left\vert E+\rho_{r}\right\vert \right)  ^{-1/4}%
\end{array}
\\
s=\pm\sqrt{\frac{\left\vert K\right\vert T_{s}^{2}-1}{4cT_{s}^{2}}}dn\left(
\frac{\tau}{T_{s}}|m_{s}\right)  ,\
\begin{array}
[c]{l}%
m_{s}=(2-\frac{1}{2}\left\vert K\right\vert T_{s}^{2})\leq1\\
T_{s}=2\left(  \left\vert K\right\vert +\sqrt{K^{2}-16c\left\vert E\right\vert
}\right)  ^{-1/2}%
\end{array}
\end{array}
$\\\hline\hline
\end{tabular}
\medskip%

\begin{tabular}
[c]{||c||l||}\hline\hline
$%
\begin{array}
[c]{c}%
{\includegraphics[
height=0.9366in,
width=1.8386in
]%
{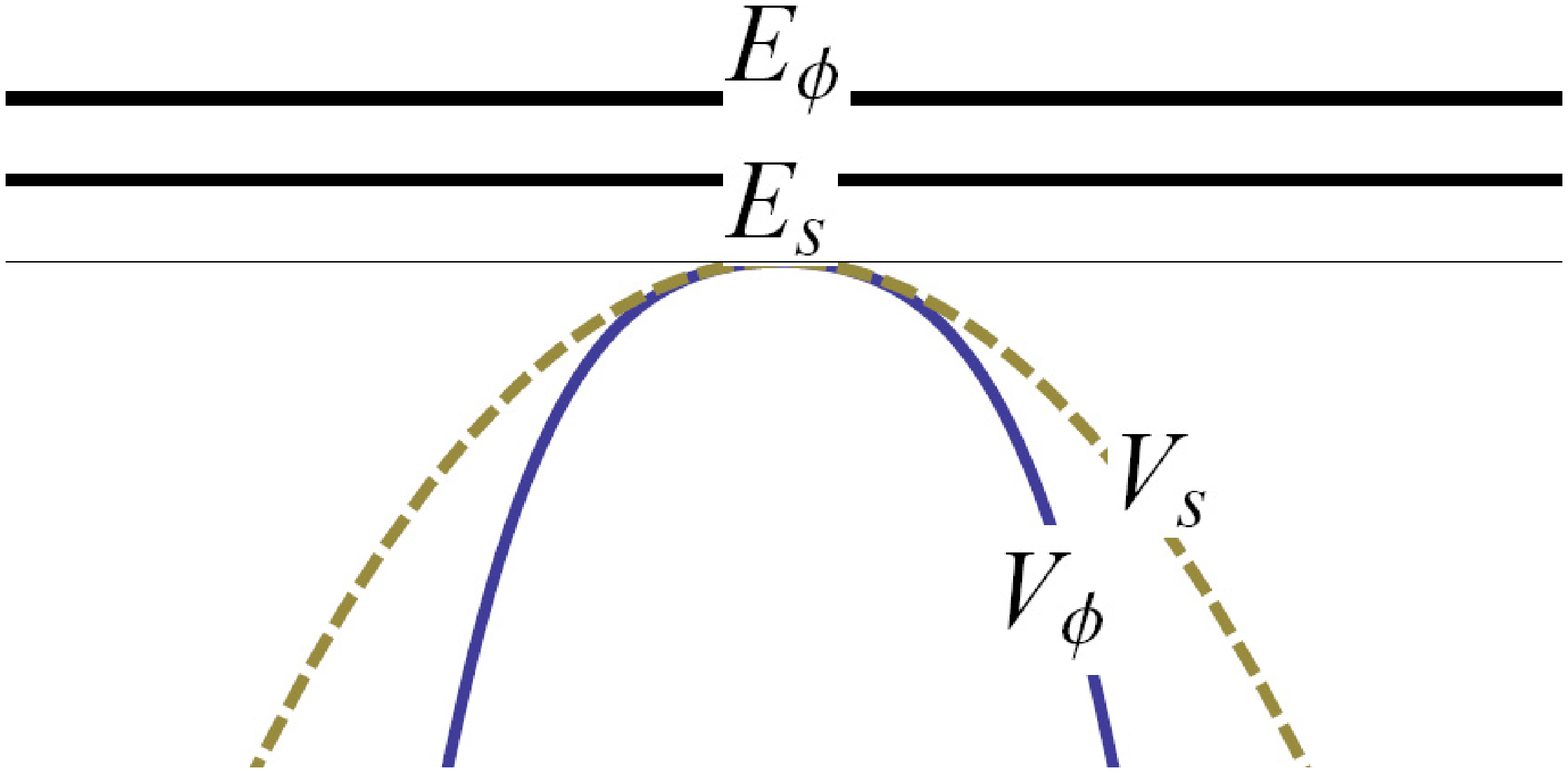}%
}%
\\
\text{FIG. 27}\\
b>0,\text{ }c<0,\text{ }E_{\phi}\geq E_{s}>0
\end{array}
$ & $%
\begin{array}
[c]{l}%
\phi=\left(  \frac{E+\rho_{r}}{b}\right)  ^{1/4}\frac{sn\left(  \frac
{\tau+\delta}{T_{\phi}}|m_{\phi}\right)  }{1+cn\left(  \frac{\tau+\delta
}{T_{\phi}}|m_{\phi}\right)  },%
\begin{array}
[c]{l}%
m_{\phi}=\frac{1}{2}-\left\vert K\right\vert T_{\phi}^{2}\leq\frac{1}{2}\\
T_{\phi}=\left(  64b\left(  E+\rho_{r}\right)  \right)  ^{-1/4}%
\end{array}
\\
s=\left(  \frac{E}{\left\vert c\right\vert }\right)  ^{1/4}\frac{sn\left(
\frac{\tau}{T_{s}}|m_{s}\right)  }{1+cn\left(  \frac{\tau}{T_{s}}%
|m_{s}\right)  },%
\begin{array}
[c]{l}%
m_{s}=\frac{1}{2}-\left\vert K\right\vert T_{s}^{2}\leq\frac{1}{2}\\
T_{s}=\left(  64\left\vert c\right\vert E\right)  ^{-1/4}%
\end{array}
\end{array}
$\\\hline\hline
\end{tabular}
\medskip%

\begin{tabular}
[c]{||c||l||}\hline\hline
$%
\begin{array}
[c]{c}%
{\includegraphics[
height=0.9539in,
width=1.7504in
]%
{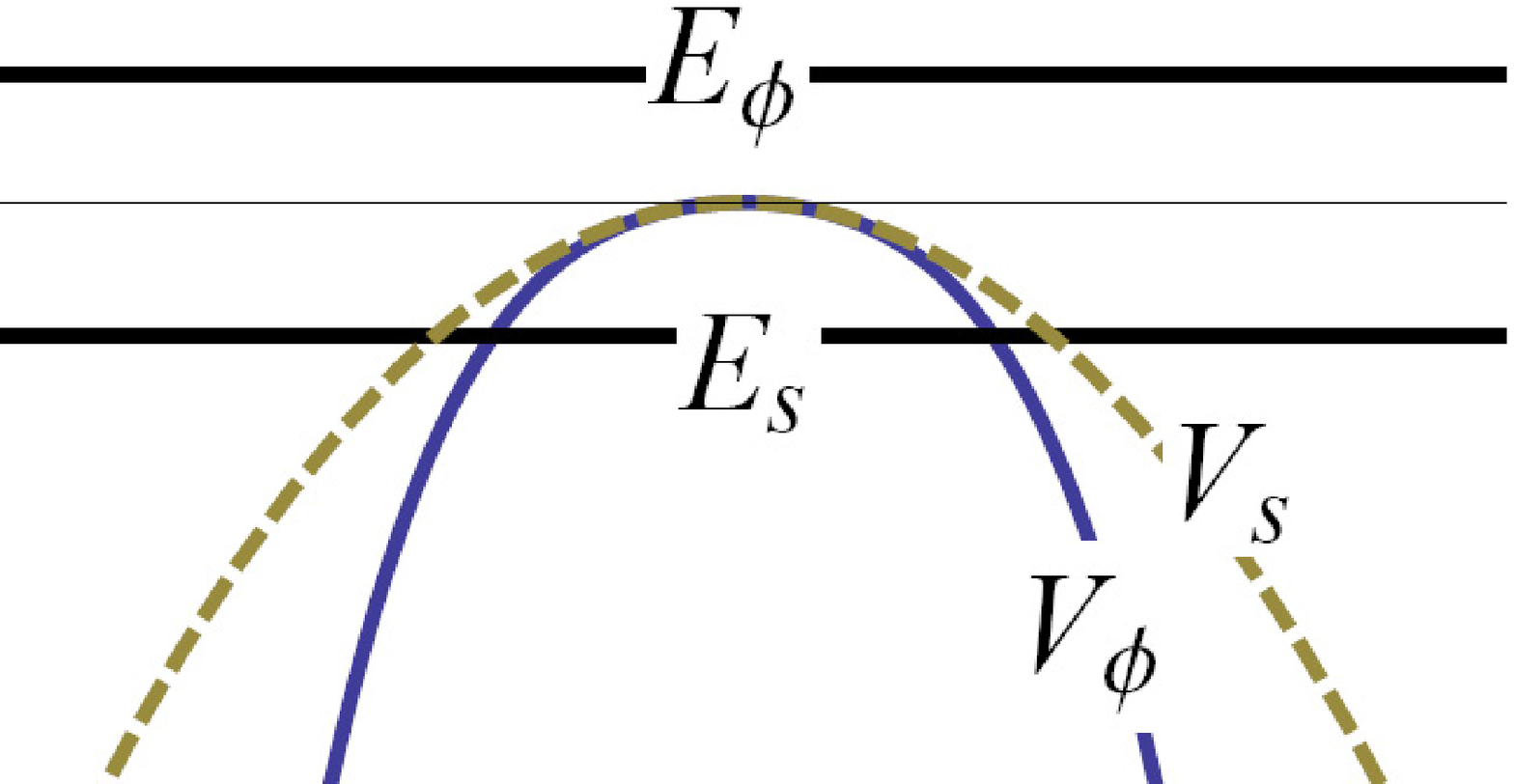}%
}%
\\
\text{FIG. 28}\\
b>0,\text{ }c<0,\text{ }E_{\phi}>E_{s}\\
E_{\phi}>0,\text{ }E_{s}<0
\end{array}
$ & $%
\begin{array}
[c]{l}%
\phi=\sqrt{\frac{1}{8bT_{\phi}^{2}}}\frac{sn\left(  \frac{\tau+\delta}%
{T_{\phi}}|m_{\phi}\right)  }{1+cn\left(  \frac{\tau+\delta}{T_{\phi}}%
|m_{\phi}\right)  },%
\begin{array}
[c]{l}%
m_{\phi}=\frac{1}{2}-\left\vert K\right\vert T_{\phi}^{2}\leq\frac{1}{2}\\
T_{\phi}=\left(  64b\left(  E+\rho_{r}\right)  \right)  ^{-1/4}%
\end{array}
\\
s=\sqrt{\frac{1-\left\vert K\right\vert T_{s}^{2}}{4\left\vert c\right\vert
T_{s}^{2}}}\frac{1}{cn\left(  \frac{\tau}{T_{s}}|m_{s}\right)  },\
\begin{array}
[c]{l}%
m_{s}=\frac{1}{2}\left(  1+\left\vert K\right\vert T_{s}^{2}\right)  \leq1\\
T_{s}=\left(  K^{2}+16\left\vert c\right\vert \left\vert E\right\vert \right)
^{-1/4}%
\end{array}
\end{array}
$\\\hline\hline
\end{tabular}
\medskip%

\begin{tabular}
[c]{||c||l||}\hline\hline
$%
\begin{array}
[c]{c}%
{\includegraphics[
height=0.9444in,
width=1.7435in
]%
{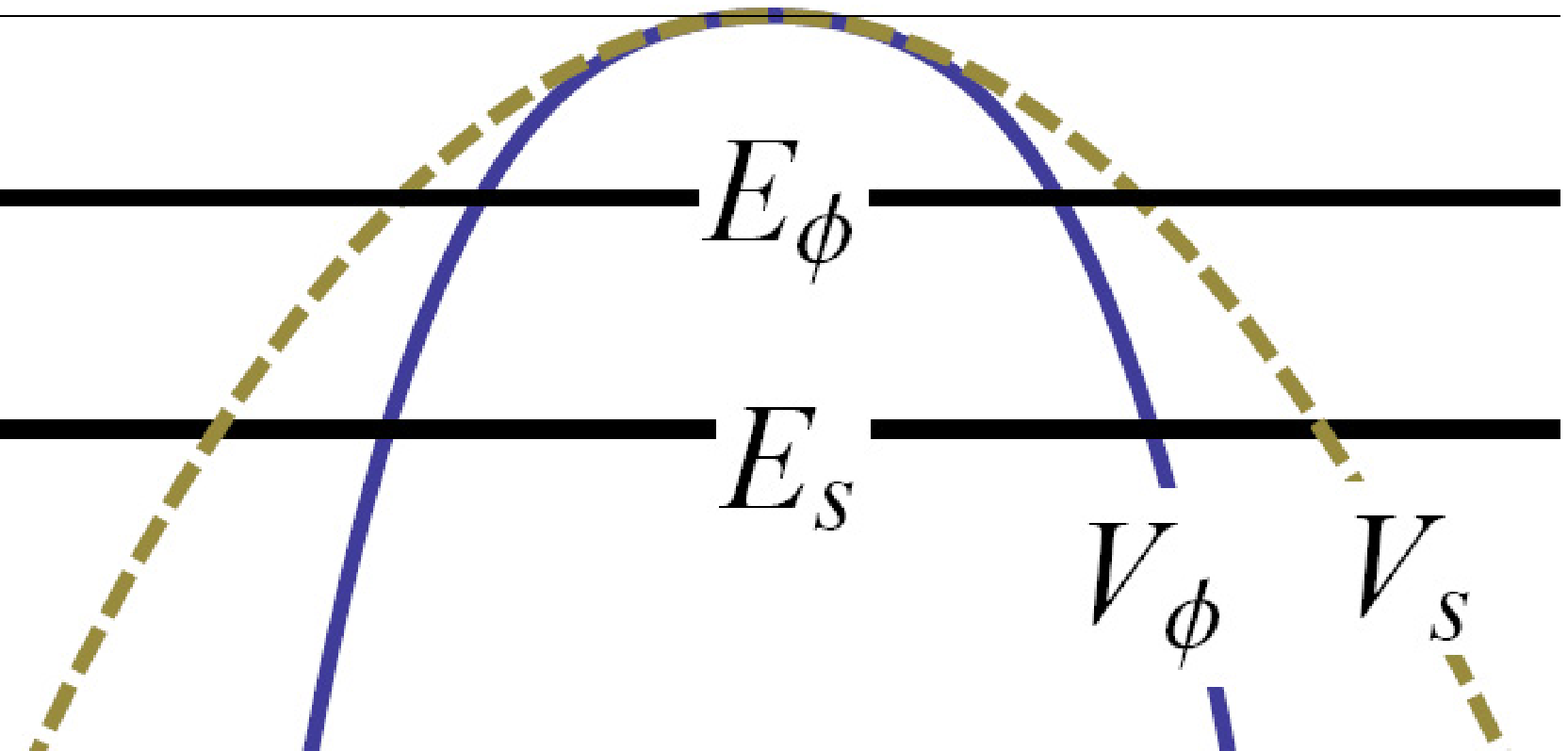}%
}%
\\
\text{FIG. 29}\\
b>0,\text{ }c<0,\text{ }0>E_{\phi}\geq E_{s}%
\end{array}
$ & $%
\begin{array}
[c]{l}%
\phi=\sqrt{\frac{1-\left\vert K\right\vert T_{\phi}^{2}}{4b}}\frac
{1}{cn\left(  \frac{\tau+\delta}{T_{\phi}}|m_{\phi}\right)  },\
\begin{array}
[c]{l}%
m_{\phi}=\frac{1}{2}\left(  1+\left\vert K\right\vert T_{\phi}^{2}\right)
\leq1\\
T_{\phi}=\left(  K^{2}+16b\left\vert E+\rho_{r}\right\vert \right)  ^{-1/4}%
\end{array}
\\
s=\sqrt{\frac{1-\left\vert K\right\vert T_{s}^{2}}{4\left\vert c\right\vert }%
}\frac{1}{cn\left(  \frac{\tau}{T_{s}}|m_{s}\right)  },\
\begin{array}
[c]{l}%
m_{s}=\frac{1}{2}\left(  1+\left\vert K\right\vert T_{s}^{2}\right)  \leq1\\
T_{s}=\left(  K^{2}+16\left\vert c\right\vert \left\vert E\right\vert \right)
^{-1/4}%
\end{array}
\end{array}
$\\\hline\hline
\end{tabular}


\medskip

\bigskip\newpage

\begin{thebibliography}{99}                                                                                               %


\bibitem {inflation Guth}A. H. Guth, Phys. Rev. D \textbf{23}, 347 (1981).

\bibitem {inflation Linde}A. D. Linde, Phys. Lett. B \textbf{108}, 389 (1982).

\bibitem {inflation Steinhardt}A. Albrecht, P. J. Steinhardt, Phys. Rev.
Lett.\textbf{\ 48}, 1220 (1982).

\bibitem {Mathiazhagan}C. Mathiazhagan and V. B. Johri, Class. Quantum Grav.
\textbf{1}, L 29 (1984).

\bibitem {extendedLa}D. La and P. J. Steinhardt, Phys. Rev. D \textbf{62}, 376 (1989).

\bibitem {hyperextendedAccetta}F. Accetta and P. J. Steinhardt, Phys. Rev.
Lett. \textbf{32}, 2740 (1990).

\bibitem {LindeBD}A. Linde, Phys. Lett. B\textbf{238}, 160 (1990).

\bibitem {ekpyrotic1}J. Khoury, B. A. Ovrut, P. J. Steinhardt and N. Turok,
Phys. Rev. D 64, 123522 (2001) [arXiv:hep-th/0103239].

\bibitem {ekpyrotic2}J. K. Erickson, D. H. Wesley, P. J. Steinhardt and N.
Turok, Phys. Rev. D 69, 063514 (2004) [arXiv:hep-th/0312009].

\bibitem {SN}A.~G.~Riess \textit{et al.} [Supernova Search Team
Collaboration], Astron.\ J.\ \textbf{116}, 1009 (1998); S. Perlmutter
\textit{et al.} [Supernova Cosmology Project Collaboration], Astrophys. J.
\textbf{517}, 565 (1999); C. L. Bennett \textit{et al.}, Astrophys. J. Suppl.
\textbf{148}, 1 (2003); M. Tegmark \textit{et al.} [SDSS Collaboration], Phys.
Rev. D \textbf{69}, 103501 (2004); S. W. Allen, \emph{et al.}, Mon. Not. Roy.
Astron. Soc. \textbf{353}, 457 (2004).

\bibitem {quint}B.~Ratra and P.~J.~E.~Peebles, Phys.\ Rev.\ D \textbf{37},
3406 (1988); C.~Wetterich, Nucl.\ Phys.\ B \textbf{302}, 668 (1988); R.R.
Caldwell, R. Dave, and P.J. Steinhardt, Phys. Rev. Lett. \textbf{80}, 1582
(1998). A.~R.~Liddle and R.~J.~Scherrer, Phys.\ Rev.\ D \textbf{59}, 023509
(1999); I.~Zlatev, L.~M.~Wang and P.~J.~Steinhardt,
Phys.\ Rev.\ Lett.\ \textbf{82}, 896 (1999); Z.~K.~Guo, N.~Ohta and
Y.~Z.~Zhang, Mod.\ Phys.\ Lett.\ A \textbf{22}, 883 (2007);
S.~Dutta and R.~J.~Scherrer,
Phys.\ Rev.\ D \textbf{78}, 123525 (2008);
S.~Dutta and R.~J.~Scherrer,
Phys.\ Rev.\ D \textbf{78}, 083512 (2008);
D.~C.~Dai, S.~Dutta and D.~Stojkovic,
Phys.\ Rev.\ D \textbf{80}, 063522 (2009);
S.~Dutta, S.~D.~H.~Hsu, D.~Reeb and R.~J.~Scherrer,
Phys.\ Rev.\ D \textbf{79}, 103504 (2009); S.~Dutta, E.~N.~Saridakis and
R.~J.~Scherrer, Phys.\ Rev.\ D \textbf{79}, 103005 (2009);
S.~Dutta and R.~J.~Scherrer,
Phys.\ Lett.\ B \textbf{676}, 12 (2009).

\bibitem {f(R)}K. Maeda, Phys. Rev. D\textbf{39}, 3159 (1989).

\bibitem {inflationBC}I. Bars and S-H. Chen, \textquotedblleft The Big Bang
and Inflation United by an Analytic Solution\textquotedblright, Phys. Rev.
\textbf{D83 }(2011) 043522 [arXiv:1004.0752v2].


\bibitem {cyclic BCT}I. Bars, S-H. Chen and N. Turok, \textquotedblleft
Geodesically Complete Analytic Solutions for a Cyclic
Universe.\textquotedblright, Phys.Rev. D84 (2011) 083513 [arXiv:1105.3606].

\bibitem {cyclic IB}I. Bars, \textquotedblleft Geodesically Complete
Universe\textquotedblright, arXiv:1109.5872 [gr-qc].

\bibitem {BCSTletter}I. Bars, S. H. Chen, P. J. Steinhardt, N. Turok,
\textquotedblleft Antigravity and the Big Crunch/Big Bang
Transition\textquotedblright, [arXiv:1112.2470]

\bibitem {BCST}I. Bars, S. H. Chen, P. J. Steinhardt, N. Turok, in preparation.


\bibitem {2TPhaseSpace}For a recent summary and status of 2T-physics, see I.
Bars, \textquotedblleft Gauge Symmetry in Phase Space, Consequences for
Physics and Spacetime\textquotedblright, Int. J. Mod. Phys. \textbf{A25}
(2010) 5235 [arXiv:1004.0668 [hep-th]].


\bibitem {2Tgravity}I. Bars, "Gravity in 2T-Physics", Phys. Rev. \textbf{D77
}(2008) 125027 [arXiv:0804.1585[tep-th]].

\bibitem {2TgravityGeometry}I. Bars, S. H. Chen "Geometry and Symmetry
Structures in 2T Gravity", Phys. Rev. \textbf{D79 }(2009) 085021
[arXiv:0811.2510v2 [hep-th]].

\bibitem {HoravaWitten}P. Ho\v{r}ava and E.Witten, Nucl. Phys. \textbf{B460}
(1996) 506 [arXiv:hep-th/9510209; hep-th/9603142].

\bibitem {RandallSundrum}L. Randall and R. Sundrum, Phys. Rev. Lett.
\textbf{83 }(1999) 4690; ibid. \textbf{83} (1999) 3370.

\bibitem {KOSST}J. Khoury, B. A. Ovrut, N. Seiberg, P. J. Steinhardt, N.
Turok, Phys.Rev. D65 (2002) 086007 [arXiv:hep-th/0108187].

\bibitem {branesMT}P. McFadden and N. Turok, \textquotedblleft Conformal
Symmetry of Brane World Effective Actions\textquotedblright, Phys. Rev.
\textbf{D71} (2005) 021901 [arXiv:hep-th/0409122].

\bibitem {tHooft}G. tHooft. Lecture in conference, \textit{Conformal Nature of
the Universe}, see http://www.pirsa.org/C12027 .

\bibitem {2Tsm}I. Bars, "The Standard Model of Particles and Forces in the
Framework of 2T-physics", Phys.Rev. D74 (2006) 085019 [hep-th/0606045].

\bibitem {higgsCosmo}I. Bars, talk at http://pirsa.org/12050077/, also paper
in preparation.

\bibitem {Abramowitz}M. Abramowitz, I.A. Stegun, "\textit{Handbook of
Mathematical Functions}", Dover (1965), ISBN 0486612724.

\bibitem {bars harmonic oscillator}I. Bars, Phys. Rev. D79 (2009)045009 [arXiv:hep-th/0810.2075]

\bibitem {bars DPF}I. Bars, [arXiv:gr-qc1109.5872]

\bibitem {weinberg}S. Weinberg, \textit{The Quantum Theory of Fields}, Volume
III, Cambridge 2000.


\bibitem {deWit}B. de Wit and A. Van Proeyen, \textquotedblleft Potentials and
Symmetries of General Gauged $\mathcal{N}$=2 Supergravity: Yang-Mills
Models\textquotedblright, Nucl.Phys. \textbf{B245} (1984) 89.


\end{thebibliography}
\end{document}